\documentclass[11pt]{JHEP3}
\usepackage{epsfig}
\usepackage{braket}

\def  \bcen   {\begin{center}}
\def  \ecen   {\end{center}}
\def  \et       {\not\!\! E_T}
\newcommand{\beq}{\begin{eqnarray}}
\newcommand{\eeq}{\end{eqnarray}}
\newcommand{\nn}{\nonumber}
\newcommand{\centeron}[2]{{\setbox0=\hbox{#1}\setbox1=\hbox{#2}\ifdim                                       
\wd1>\wd0\kern.5\wd1\kern-.5\wd0\fi \copy0
\kern-.5\wd0\kern-.5\wd1\copy1\ifdim\wd0>\wd1 \kern.5\wd0\kern-.5\wd1\fi}}
\newcommand{\ltap}{\>\centeron{\raise.35ex\hbox{$<$}}
                               {\lower.65ex\hbox{$\sim$}}\>}
\newcommand{\gtap}{\>\centeron{\raise.35ex\hbox{$>$}}
                               {\lower.65ex\hbox{$\sim$}}\>}

\newcommand\ZZ{\hbox{\zfont Z\kern-.4emZ}}
\font\zfont = cmss10 

\newcommand  {\bjlllet} {$b j \ell^\pm \ell^\mp \ell^\pm \not\!\! E_T$}
\newcommand  {\jjbb}  {$b \bar{b} j j$}
\newcommand  {\bjlet} {$b j \ell^\pm \not\!\! E_T$} 

\title{Heavy Vector-like Top Partners at the LHC and flavour constraints}    

\author{Giacomo Cacciapaglia\\
   King's College London, Department of Physics, 
   Strand, London WC2R 2LS, UK \\
   and Universit\'e de Lyon, France; Universit\'e Lyon 1,\\ 
   CNRS/IN2P3, UMR5822 IPNL, F-69622 Villeurbanne Cedex, France\\
   E-mail: \email{ cacciapaglia@ipnl.in2p3.fr}} 
\author{Aldo Deandrea, Luca Panizzi \\
   Universit\'e de Lyon, France; Universit\'e Lyon 1,\\ 
   CNRS/IN2P3, UMR5822 IPNL, F-69622 Villeurbanne Cedex, France\\
   E-mail: \email{deandrea@ipnl.in2p3.fr, panizzi@ipnl.in2p3.fr}} 
\author{Naveen Gaur\\ 
   Department of Physics, Dyal Singh College (University of Delhi), \\
   Lodi Road, New Delhi - 110003, India \\
   E-mail:  \email{gaur.nav@gmail.com}} 
\author{Daisuke Harada\\ 
   KEK Theory Center, Institute of Particle and Nuclear Studies, KEK,  
   1-1 Oho, Tsukuba, Ibaraki 305-0801, Japan\\    E-mail: \email{dharada@post.kek.jp}} 
\author{Yasuhiro Okada\\ 
   KEK Theory Center, Institute of Particle and Nuclear Studies, KEK,  
   1-1 Oho, Tsukuba, Ibaraki 305-0801, Japan\\ Department of Particle and
   Nuclear Physics, Graduate 
   University for Advanced Studies (Sokendai), 1-1 Oho, Tsukuba, Ibaraki
   305-0801, Japan\\ 
   E-mail: \email{yasuhiro.okada@kek.jp}} 

\abstract{We consider the phenomenology at the Large Hadron Collider
of new heavy vector-like quarks which couple mainly to the third generation
quarks via Yukawa interactions, with special emphasis on non--standard
doublet representations which are less constrained from present
data. We also discuss in detail the flavour limits at tree level and loop level and implications of
a generalised CKM mixing matrix to these cases.} 

\preprint{ KEK-TH-1470 , LYCEN 2011-07, KCL-PH-TH/2011-15}

\begin{document}

\section{Introduction \label{sec:intro}}
Heavy vector-like fermions are not present in the Standard Model (SM), however they naturally arise near the 
electroweak scale in many extensions of new physics. Flavour hierarchies 
among SM masses for fermions and their mixings are most often generated through their 
dynamical mixing with vector-like fermions. These particles are indeed present in many 
models of new physics, like for example extra dimensional, Little Higgs and 
dynamical models. More recently the possibility of gauging the Standard Model flavour 
group requiring anomaly cancellation with the addition of new vector-like fermions was 
studied with explicit model constructions which show the possibility of a relatively low new 
gauge flavour bosons scale~ \cite{Grinstein:2010ve}. Moreover, in this case the vector-like quarks play an important 
role both for anomaly cancellation and for the mechanism of generation of fermion masses 
(see~\cite{Guadagnoli:2011id} for an example with left-right symmetry). The mixing of vector-like quarks with 
the other three generations and in particular the top quark is also a common feature in 
Little Higgs~\cite{lhmod} and composite Higgs models~\cite{modHF} based on top condensation~\cite{tcmod}.
The Large Hadron Collider (LHC) is collecting data that allows to test and discover this 
sector quite soon. Typical cross-sections for pair and single production of heavy vector-like 
quarks are fairly large and within short term reach for the LHC experiments in the few hundred GeV mass range, as 
it will be discussed in detail in the following. 
Assuming some properties, the existing collider and precision data place limits on the new heavy quarks and set the lowest mass scale for these states. 
Direct searches, for example, give mass constraints in the 
300 GeV range, assuming a charged current decay chain~\cite{pdg}. 
In this work, we will focus on a particular example: namely, new coloured quarks that are doublets under the 
SM SU(2) symmetry, but with non-standard hypercharge.
The resulting Yukawa couplings are less constrained by precision measurements than in other 
cases~\cite{Cacciapaglia:2010vn}.
Moreover, a fermion with such quantum numbers is present in models of composite Higgs boson with extended custodial 
symmetry~\cite{Agashe:2006at}, engineered to protect the left-handed $Zbb$ coupling from large corrections.
In this model, the gauge symmetry is extended to SU(2)$_L \times$ SU(2)$_R$ in the bulk, and the left-handed top and 
bottom doublet comes in a bi-doublet of the two SU(2).
The bi-doublet naturally contains a doublet with non-standard hypercharge. 
Moreover, the boundary conditions are such that the resonances from the non-standard doublet are parameterically 
lighter that the others~\cite{Panico:2011pw}, therefore one can approximate the phenomenology of such models by the addition of a massive 
doublet only, after integrating out the other more massive resonances.

The paper is organised as follows. In section~\ref{sec:model} we describe 
the effective Yukawa interactions we consider for this study. We assume the presence of 
the SM quarks and of the standard Higgs doublet and couple them to the vector-like 
quarks. The Cabibbo--Kobayashi--Maskawa (CKM) flavour structure and the couplings 
to the standard gauge bosons and Higgs particle are discussed in detail. In section~\ref{sec:bounds} we 
discuss the electroweak precision tests and the tree level bounds 
which define the accessible parameter space. In section~\ref{sec:loop} we explore the loop effects in 
the Kaon and B meson sector together with loop effects on the couplings of the Higgs boson 
to gluons and photons. In section~ \ref{sec:lhc} we study the LHC phenomenology for the production 
and the decay of vector-like quarks with a detailed simulation of various decay modes 
obtained from single production of a heavy vector-like top partner. Finally we give our 
conclusions. Two appendices contain more detailed material concerning the expansion of 
the CKM matrix and the notations used for the meson mixing formalism. 

\section{The effective interactions} \label{sec:model}

We assume that the new fermions interact with the SM fermions via
Yukawa interactions, therefore the quantum  
numbers of the new fermions with respect to the weak SU(2)$_L\times$
U(1)$_Y$ gauge group are limited by the  
requirement of an interaction with the Higgs doublet and one of the SM
fermions. A complete survey of all the possibilities is given in
\cite{Cacciapaglia:2010vn} (see also \cite{delAguila:2000aa}, \cite{delAguila:2000rc} for previous studies). 
Here we are interested mainly in the case of a new 
coloured fermion that transforms like a doublet of SU(2) 
with hypercharge $7/6$, because the bounds are milder in this case. We will therefore introduce the notation in this specific case. 
In the quark sector, a SM family contains a doublet $q_L = \{ u_L, d_L \}^T = (2,1/6)$
and two singlets $u_R = (1,2/3)$ and  
$d_R = (1,-1/3)$, that couple to each other via the Higgs $H = (2,1/2)$. Here and in the following, $L$ and $R$ label respectively the left-handed or right-handed chirality of the field.
The SM Yukawa couplings are:
\beq \label{eq:SMyuk}
\mathcal{L}_{\rm Yukawa} = - y_u\, \bar q_L H^c u_R - y_d\, \bar q_L H
d_R + h.c.\,;
\eeq
the up-type quarks (top), therefore, couple to the charge-conjugate of the Higgs boson.
If we extend the SM with a new fermion $\psi = (2, 7/6) = \{X, U\}^T$, it is possible to write down a Yukawa coupling between the new fermion and the up-type singlets via the Higgs boson $H$.
The Yukawa sector that we will consider here is
\beq
\mathcal{L}_{\rm Yukawa} & = & - y_u\, \bar q_L H^c u_R - \lambda \,
\bar \psi_L H u_R - M\, \bar \psi_L \psi_R + h.c. \nonumber \\ 
  & = & - \frac{y_u v}{\sqrt{2}}\, \bar u_L u_R - \frac{\lambda
v}{\sqrt{2}}\, \bar U_L u_R - M\,  ( \bar U_L U_R + \bar X_L X_R  )+
h.c. \,. 
\eeq
The new massive fermion contains a quark $X$ with electric charge $5/3$, and a fourth up quark which, 
however, has different gauge couplings with respect to the SM up quarks.
The Higgs vacuum expectation value (VEV) $v$ induces a mass mixing with the up quarks, 
therefore the mass eigenstates will be a mixture of the SM and new fermions.
The new Yukawa couplings $\lambda$ connecting the heavy quarks with the SM
ones will generate flavour mixing between  
the new states and the SM quarks in the up sector, while the down sector is left untouched.
We will first review the simple limit of just one flavour mixing
\cite{Cacciapaglia:2010vn}, in which the new state only  
mixes with the third generation: this is an interesting and physical
limit because of the simpler  
notation and the fact that mixing with the light generations are much more tightly constrained. 
We will then give the more general parameterisation of the
full three flavour mixing, which will show a very peculiar 
flavour structure, which is not flavour minimal and still can be
consistent with present bounds with non-negligible couplings. 

\subsection{One flavour mixing \label{sec:mix1}}

In our previous paper~\cite{Cacciapaglia:2010vn}, we found that the most general mass terms, including the 
Yukawa interactions, is
\beq
\mathcal{L}_{\rm mass} = -\tilde{m}_t\, \bar u_L u_R - x\, \bar U_L u_R - M\, \bar U_L U_R + h.c.\, .
\eeq
where $\tilde{m}_t = \frac{y_u v}{\sqrt{2}}$ and $x = \frac{\lambda v}{\sqrt{2}}$ can be chosen real thanks to 
a phase redefinition of $U$. It can be diagonalised by
\begin{eqnarray}
\left(
\begin{array}{cc}
\cos \theta^{L}_{u} & -\sin \theta^{L}_{u} \\
\sin \theta^{L}_{u} & \cos \theta^{L}_{u} \\
\end{array}
\right)
\left(
\begin{array}{cc}
\tilde{m}_t & 0 \\
x & M \\
\end{array}
\right)
\left(
\begin{array}{cc}
\cos \theta^{R}_{u} & \sin \theta^{R}_{u} \\
-\sin \theta^{R}_{u} & \cos \theta^{R}_{u} \\
\end{array}
\right) =
\left(
\begin{array}{cc}
m_{t} & 0 \\
0 & m_{t^{\prime}} \\
\end{array}
\right)\, ;
\end{eqnarray}
and the relations between parameters and masses and mixing angles are~\cite{Cacciapaglia:2010vn}
\begin{eqnarray}
\tilde{m}_t^2 = \frac{y_u^2 v^2}{2} = m_{t}^{2} \left(1 + \frac{x^{2}}{M^{2} - m_{t}^{2}} \right)\, , & \qquad
& m_{t^{\prime}}^{2} = M^{2} \left(1 + \frac{x^{2}}{M^{2} - m_{t}^{2}} \right)\, , \nonumber \\
\sin \theta^{R}_{u} = \frac{M x}{\sqrt{(M^{2} - m_{t}^{2})^{2} + M^{2}x^{2}}}\, , & \qquad &
\sin \theta^{L}_{u} = \frac{m_t}{M} \sin \theta^{R}_{u}\, . \label{eq:mix1a}
\end{eqnarray}

We can however use those relations to express all parameters in terms of physical observables, like for example $m_t$, $m_{t'}$ and $\theta^R_u$:
\beq
M = m_X &=& \sqrt{m_{t'}^2 + m_t^2 \tan^2 \theta^R_u} \cos \theta^R_u\,, \\
\tilde{m}_t &=& \frac{m_t m_{t'}}{\sqrt{m_{t'}^2 + m_t^2 \tan^2 \theta^R_u}} \frac{1}{\cos \theta^R_u} =  \frac{m_t m_{t'}}{m_X}\,,\\
x &=& \frac{m_{t'}^2 - m_t^2}{\sqrt{m_{t'}^2 + m_t^2 \tan^2 \theta^R_u}} \sin \theta^R_u =  \frac{m_{t'}^2 - m_t^2}{m_X} \sin \theta^R_u \cos \theta^R_u\,,\\
\sin \theta^L_u &=& \frac{m_t}{\sqrt{m_{t'}^2 + m_t^2 \tan^2 \theta^R_u}} \tan \theta^R_u = \frac{m_t}{m_X} \sin \theta^R_u\,.
\eeq
In the rest of the paper we will use $m_{t'}$ and $\theta^R_u$ as the two free parameters of the model, and the previous relations allows to calculate the fundamental parameters $x$ and $M$.

\subsection{Three flavour mixing}
\label{sec:mixing}

If we consider the three families of quarks in the SM, the Yukawa couplings become matrices in flavour space, while $\lambda$ is a vector. With explicit flavour indices $i,j = 1,2,3$, the Yukawa interactions are:
\beq
\mathcal{L}_{\rm yuk} = - y_u^{i,j}\, \bar{Q}_L^i H^c u_R^j - y_d^{i,j}\, \bar{Q}_L^i H d_R^j  - \lambda^{j}\, \bar{\psi}_L H u_R^j \,.
\eeq
In the following we will describe a convenient method to diagonalise the resulting mass terms.
First, we can use the SM flavour symmetry to rewrite the standard Yukawa couplings as:
\beq
y_u^{i,j} = \mbox{diag} (y_u, y_c, y_t)\,, \qquad
y_d^{i,j} = \tilde{V}_{CKM} \cdot  \mbox{diag} (y_d, y_s, y_b)\,.
\eeq
The matrix $\tilde{V}_{CKM}$ is the misalignment between left-handed ups and downs and it would correspond to the Cabibbo-Kobayashi-Maskawa matrix in the absence of the new fermion.
The tilde signifies that this is not the case, as we will see.
After the Higgs develops a VEV, $\langle H \rangle = v/\sqrt{2}$, the mass matrix for the fermions can be written (in this basis) as
\beq
\mathcal{L}_{\rm {mass}} &=& - (\bar{d}_L, \bar{s}_L, \bar{b}_L)\cdot\tilde{V}_{CKM}\cdot \left( \begin{array}{ccc}
\tilde{m}_d & & \\ & \tilde{m}_s & \\ & & \tilde{m}_b
\end{array} \right) \cdot \left( \begin{array}{c} d_R \\ s_R \\ b_R \end{array} \right)  \nonumber \\
  &-&   (\bar{u}_L, \bar{c}_L, \bar{t}_L, \bar{U}_L) \cdot \left( \begin{array}{cccc}
\tilde{m}_u & & & 0\\ & \tilde{m}_c & & 0\\ & & \tilde{m}_t & 0\\ x_1 & x_2 & x_3 & M
\end{array} \right) \cdot \left( \begin{array}{c} u_R \\ c_R \\ t_R \\ U_R\end{array} \right) - M \, \bar{X}_L X_R + h.c.
\eeq
where $\tilde{m}_x = \frac{y_x v}{\sqrt{2}}$ and $x_i = \frac{\lambda^i v}{\sqrt{2}}$.
For the down type quarks, the tilded masses are equal to the physical masses, as in the SM.
For the up type quarks, there will be corrections coming from the mixing with the heavy $U$.
Using the phase of $U$, we can show that $M$ and one of the $x$, say $x_3$, can be made real, while two physical 
phases are still present on $x_1$ and $x_2$. This will be important for CP violation.
In the following we will take $x_3$ real, so that the formulas in the previous section can be used in this case without any change.

The mass matrix for the up quarks $M_u$ can be diagonalised by two unitary 4 by 4 matrices:
\beq
M_u = V_L \cdot \left( \begin{array}{cccc}
m_u & & & \\ & m_c & & \\ & & m_t & \\ & & & m_{t'}
\end{array} \right) \cdot V^\dagger_R\,.
\label{eq:mu}
\eeq
$V_L$, that describes the mixing in the left-handed sector, is the unitary matrix that diagonalises the unitary matrix
\beq
M_u \cdot M_u^\dagger = \left( \begin{array}{cccc}
\tilde{m}_u^2 & 0 & 0 & x_1^* \tilde{m}_u \\
0 & \tilde{m}_c^2 & 0 & x_2^* \tilde{m}_c \\
0 & 0 & \tilde{m}_t^2 & x_3 \tilde{m}_t \\
x_1 \tilde{m}_u & x_2 \tilde{m}_c & x_3 \tilde{m}_t & M^2 + |x_1|^2 + |x_2|^2 + x_3^2
\end{array} \right)\,.
\eeq
From this matrix we can see two things: a SM quark $q = u, c, t$ is exactly massless in the limit $\tilde{m}_q \to 0$, therefore the masses of the light quarks are always proportional to $\tilde{m}_q$; the mixing angles with the heavy $t'$ are suppressed by $\tilde{m}_q/M \sim m_q/m_{t'}$.

On the other hand, $V_R$, that describes the mixing in the right-handed sector, diagonalises
\beq
M_u^\dagger \cdot M_u = \left( \begin{array}{cccc}
\tilde{m}_u^2 + |x_1|^2 & x_1^* x_2 & x_1^* x_3 & x_1^* M \\
x_2^* x_1 & \tilde{m}_c^2 + |x_2|^2& x_2^* x_3 & x_2^* M \\
x_3 x_1 & x_3 x_2 & \tilde{m}_t^2 + x_3^2& x_3 M \\
x_1 M & x_2 M & x_3 M & M^2
\end{array} \right)\,.
\eeq
Here we see that the mixing is only suppressed by the $x$'s couplings and that it is present also in the limit of massless SM quarks $\tilde{m}_q \to 0$.
Therefore, it will be the couplings of the right-handed quarks to pose the most serious flavour constraints.

\subsection{Mixing matrices}

The two mixing matrices $V_L$ and $V_R$ can be calculated in an approximate way: we can assume that the 
parameters relative to the mass and mixing of the light two generations are small (therefore, $\tilde{m}_u$, 
$\tilde{m}_c$, $x_1$ and $x_2$ are of order $\epsilon$ compared to $\tilde{m}_t$, $M$ and $x_3$).
We can then calculate the matrices in an expansion in $\epsilon$.
A good trick is to first diagonalise the top-heavy fermion sector with the formulas in section \ref{sec:mix1}; then, 
we can diagonalise the resulting matrix in the given expansion.

For the right-handed matrix, the result, up to order $\epsilon^2$, is:
\beq
V_R^{ij} = \left( \begin{array}{cccc}
1-\frac{|x_1|^2}{2 m_X^2}  &  - \frac{x^*_1 x_2 m_c^2}{(m_c^2 - m_u^2) m_X^2}  &  - \frac{x^*_1 \sin \theta_R}{m_X}  \\
\frac{x_1 x^*_2 m_u^2}{(m_c^2 - m_u^2) m_X^2}  & 1-\frac{|x_2|^2}{2 m_X^2}  & - \frac{x^*_2 \sin \theta_R}{m_X}   \\
0   &  0   &  \cos \theta_R + \frac{(m_{t'}^2 + m_t^2) (|x_1|^2 + |x_2|^2) \cos \theta_R \sin^2 \theta_R}{2 (m_{t'}^2 - m_t^2) m_X^2}  \end{array} \right)
\eeq
for $i,j = 1,2,3$; and
\beq
& V_R^{14} =  \frac{x^*_1 \cos \theta_R}{m_X}\,, \quad V_R^{24} = \frac{x^*_2 \cos \theta_R}{m_X}\,, \quad V_R^{34} =  \sin \theta_R -  \frac{(m_{t'}^2 + m_t^2) (|x_1|^2 + |x_2|^2) \cos^2 \theta_R \sin \theta_R}{2 (m_{t'}^2 - m_t^2) m_X^2}\,, & \nonumber \\
& V_R^{41} = - \frac{x_1}{m_X} \,, \quad V_R^{42} =  - \frac{x_2}{m_X}\,, \quad V_R^{43} =  - \sin \theta_R  + \frac{(|x_1|^2 + |x_2|^2) (3 m_{t'}^2 - m_t^2 + (m_{t'}^2 + m_t^2) \cos 2 \theta_R) \sin \theta_R}{4 (m_{t'}^2 - m_t^2) m_X^2}\,, & \nonumber \\
& 
V_R^{44} =  \cos \theta_R -   \frac{(|x_1|^2 + |x_2|^2) (m_{t'}^2 - 3 m_t^2 + (m_{t'}^2 + m_t^2) \cos 2 \theta_R) \cos \theta_R}{4 (m_{t'}^2 - m_t^2) m_X^2}\,. &
\eeq

The masses, at order $\epsilon^2$, are
\beq
m_u^2 &=& \tilde{m}_u^2\,, \quad m_c^2 = \tilde{m}_c^2\,,  \nonumber \\
m_t^2 &=& m_{1,t}^2 + \frac{(|x_1|^2+|x_2|^2) m_{1,t}^2}{m_X^2} \sin^2 \theta_R\,, \\
m_{t'}^2 &=& m_{1,t'}^2 + \frac{(|x_1|^2+|x_2|^2) m_{1,t'}^2}{m_X^2} \cos^2 \theta_R\,, \nonumber
\eeq
where $m_{1,t}$ and $m_{1,t'}$ are the top and top prime masses in the 1 generation case, Eq.~(\ref{eq:mix1a}).

A similar procedure can be followed for $V_L$:
\beq
V_L^{ij} = \left( \begin{array}{cccc}
1 &  - \frac{x^*_1 x_2 m_c m_u}{(m_c^2 - m_u^2) m_X^2}  &  - \frac{x^*_1 m_u \sin \theta_L}{m_t^2}  \\
 \frac{x_1 x^*_2 m_c m_u}{(m_c^2 - m_u^2) m_X^2}  & 1  & - \frac{x^*_2 m_c \sin \theta_L}{m_t^2}  \\
\frac{x_1 (m_{t'}^2-m_t^2) m_u \cos \theta_L \sin \theta_L}{m_{t'}^2 m_t^2}   &  \frac{x_2 (m_{t'}^2-m_t^2) m_c \cos 
\theta_L \sin \theta_L}{m_{t'}^2 m_t^2}    &  \cos \theta_L +  \frac{(|x_1|^2 + |x_2|^2) \cos \theta_L \sin^2 \theta_L}
{(m_{t'}^2 - m_t^2)}  \end{array} \right)\nonumber \\
\eeq
with $i,j = 1,2,3$; and
\beq
& V_L^{14} =  \frac{x^*_1 m_u \cos \theta_L}{m_{t'}^2}\,, \quad V_L^{24} = \frac{x^*_2 m_c \cos \theta_L}{m_{t'}^2}\,, 
\quad V_L^{34} =  \sin \theta_L -  \frac{(|x_1|^2 + |x_2|^2) \cos^2 \theta_L \sin \theta_L}{(m_{t'}^2 - m_t^2)}\,, & 
\nonumber \\
& V_L^{41} = - \frac{x_1 m_u}{m_X^2} \,, \quad V_L^{42} =  - \frac{x_2 m_c}{m_X^2}\,, 
\quad V_L^{43} =  - \sin \theta_L  + \frac{(|x_1|^2 + |x_2|^2)  \cos^2 \theta_L  \sin \theta_L}{m_{t'}^2 - m_t^2}\,, 
& \nonumber \\
& V_L^{44} =  \cos \theta_L +   \frac{(|x_1|^2 + |x_2|^2) \cos \theta_L \sin^2 \theta_L}{m_{t'}^2 - m_t^2} &
\eeq
with mass eigenstates
\beq
m_u^2 &=& \tilde{m}_u^2\,, \quad m_c^2 = \tilde{m}_c^2\,, \nonumber \\
m_t^2 &=& m_{1,t}^2 + (|x_1|^2+|x_2|^2) \sin^2 \theta_L\,, \\
m_{t'}^2 &=& m_{1,t'}^2 + (|x_1|^2+|x_2|^2) \cos^2 \theta_L\,. \nonumber
\eeq
The expressions for the masses coincide with the ones found from the right-handed mixing once the relations between $\theta_L$ and $\theta_R$ are taken into account.

\subsection{Charged gauge boson: $W^\pm$ couplings and CKM matrix}

In the basis we discussed above (gauge interaction basis), the couplings of the $W^\pm$ bosons is given 
by a 3 by 4 flavour matrix in the form:
\beq
\mathcal{L}_{W^\pm} = \frac{g}{\sqrt{2}}\, \left( \bar{u}_L, \bar{c}_L, \bar{t}_L, \bar{U}_L \right)\cdot \left( \begin{array}{ccc}
1 & & \\ & 1 & \\ & & 1 \\ 0 & 0 & 0 
\end{array} \right) \cdot \gamma^\mu \left( \begin{array}{c}
d_L \\ s_L \\ b_L
\end{array} \right) W^+_\mu + h.c.\,,
\eeq
where the new fermion $U$ does not couple to the down-type quarks.
In the mass eigenstate basis, the mixing in the left-handed sector will generate couplings with the heavy $t'$, and, in matricial form,  the coupling can be written as
\beq
g_{WL}^{Ij} = \frac{g}{\sqrt{2}} V_{CKM}^{Ij} =
\frac{g}{\sqrt{2}}\, V_L^\dagger \cdot \left( \begin{array}{ccc}
 & & \\ & \tilde{V}_{CKM} & \\ & &  \\ 0 & 0 & 0 
\end{array} \right) \,.
\label{eq:wcoupling}
\eeq
The upper 3 by 3 block, that describes the couplings of the standard quarks, is given by the matrix
\beq
V_{3,CKM}^{ij} = V_{CKM}^{ij} = \sum_{l=1}^3 V_L^{*,li} \tilde{V}_{CKM}^{lj}\,, \label{eq:VCKMdef}
\eeq
that corresponds to the CKM matrix as measured by tree level SM processes.
From this equation we can see that the measured CKM matrix is not equal to the misalignment between the two standard Yukawa matrices; however, this is approximately true because the mixing angles in $V_L$ are small.
The couplings of the heavy top to the SM down quarks is described by a 3-component vector
\beq
V^{t'j}_{CKM} =  \sum_{l=1}^3 V_L^{*,l4} \tilde{V}_{CKM}^{lj}\,.
\eeq
Note that in this model the 3 by 3 CKM matrix is not unitary.
In fact
\beq
\left[V_{3,CKM}\cdot V^\dagger_{3,CKM}\right]^{ij} & = & \delta^{ij} -  V_L^{*,4i} V_L^{4j}\,,  \label{eq:CKMuni1}\\
\left[V_{3,CKM}^\dagger \cdot V_{3,CKM}\right]^{ij} &=& \delta^{ij} - V_{CKM}^{*,t'i} V_{CKM}^{t'j}\,; \label{eq:CKMuni2}
\eeq
where we have used the unitarity of $\tilde{V}_{CKM}$ and $V_L$.
The violation of unitarity are proportional to the matrix elements $V_L^{4i}$ and to the couplings of the $t'$ to the $W$.

\subsection{Neutral gauge boson: $Z^0$ couplings and FCNCs}

In this model, the couplings of the $Z^0$ will also develop off-diagonal terms, absent in the SM, therefore giving rise 
to tree-level flavour changing neutral currents (FCNC).
Note however that the FCNCs will uniquely involve up-type quarks, therefore the strong bounds from Kaon and B 
meson mixing are avoided. Nevertheless, as we will see, the measurements in the D meson sector are yet constraining.
Let's first recall that the couplings of the $Z$ to SM up quarks are proportional to $\frac{g}{c_W} \left( \frac{1}{2} - 
\frac{2}{3} s_W^2\right)$ for a left-handed up, and $\frac{g}{c_W} \left( - \frac{2}{3} s_W^2 \right)$ for a right-handed up 
(where as usual $s_W$ and $c_W$ indicate respectively the sine and cosine of the Weinberg angle); while for the 
new $U$ quark the couplings are proportional to 
\beq
\frac{g}{c_W} \left( -\frac{1}{2} - \frac{2}{3} s_W^2 \right)
\eeq
for both left- and right-handed components.
In terms of interaction eigenstates, the left-handed couplings of the $Z$ can therefore be written as
\beq
\mathcal{L}_{Z} = \frac{g}{c_W}\, \left( \bar{u}_L, \bar{c}_L, \bar{t}_L, \bar{U}_L\right)\cdot \left[ \left( \frac{1}{2} - \frac{2}{3} s_W^2 \right) \left( \begin{array}{cccc}
1&&&\\ &1&&\\ &&1&\\&&&1
\end{array} \right) - \left( \begin{array}{cccc}
&&&\\ &&&\\ &&&\\&&&1
\end{array} \right)\right] \gamma^\mu \cdot \left( \begin{array}{c}
u_L\\c_L\\t_L\\U_L \end{array} \right) Z_\mu\,. \nonumber
\eeq
In the mass eigenstate basis, the coupling becomes
\beq
g_{ZL}^{IJ} = \frac{g}{c_W} \left( \frac{1}{2} - \frac{2}{3} s_W^2 \right) \delta^{IJ} - \frac{g}{c_W} V_L^{*,4I} V_L^{4J}\,;
\eeq
where $I,J = 1,2,3,4$.
Again, the flavour violation is governed by $V_L^{4i}$ elements!
Analogously for the right-handed couplings we obtain
\beq
g_{ZR}^{IJ} = \frac{g}{c_W} \left(- \frac{2}{3} s_W^2 \right) \delta^{IJ} - \frac{1}{2} \frac{g}{c_W} V_R^{*,4I} V_R^{4J}\,.
\eeq

\subsection{Higgs boson couplings}

Finally, for the Higgs, in the interaction basis
\beq
\mathcal{L}_{H} = \frac{1}{v}\, \left( \bar{u}_L, \bar{c}_L, \bar{t}_L, \bar{U}_L\right)\cdot \left[ \left( \begin{array}{cccc}
\tilde{m}_u & & & 0\\ & \tilde{m}_c & & 0\\ & & \tilde{m}_t & 0\\ x_1 & x_2 & x_3 & M
\end{array} \right) - M \left( \begin{array}{cccc}
&&&\\ &&&\\ &&&\\&&&1
\end{array} \right)\right]  \cdot \left( \begin{array}{c}
u_R\\c_R\\t_R\\U_R \end{array} \right) h\,.
\eeq
The first term reproduces the mass matrix, therefore it will be diagonal in the mass eigenstate basis: the flavour 
violation is generated by the second term with an entry only in the 44 term, similar to the $Z$ couplings.
In the mass eigenstate basis the coupling reads :
\beq
C^{IJ} =\frac{1}{v} \left( \begin{array}{cccc}
m_u&&&\\ &m_c&&\\&&m_t&\\&&&m_{t'} \end{array} \right) - \frac{M}{v} V_L^{*,4I} V_R^{4J}\,.
\label{eq:cptp}
\eeq

\section{Tree level bounds}
\label{sec:bounds}

The inclusion of a non-standard doublet of quarks generates a richer flavour structure than the SM one: we have seen a 
non-unitary CKM matrix and flavour violating couplings of the $Z$ and Higgs in the up sector.
In this section we discuss tree level bounds on the new parameters, mainly coming from electroweak precision tests at 
accelerators, low energy observables as the weak charge of atoms, 
and flavour bounds coming from D meson processes. 
Due to the absence of modifications of the coupling of the down quarks to the $Z$ and Higgs, therefore the absence of 
FCNCs in the down sector, the precise measurements in the Kaon and B meson sector do not pose very tight bounds.
Such bounds may arise at loop level, via the modification of the CKM matrix and the new CP violating phases, and 
these effects will be briefly discussed in the next section.
\begin{figure}[tb]
\begin{center}
\includegraphics[width=0.49\textwidth]{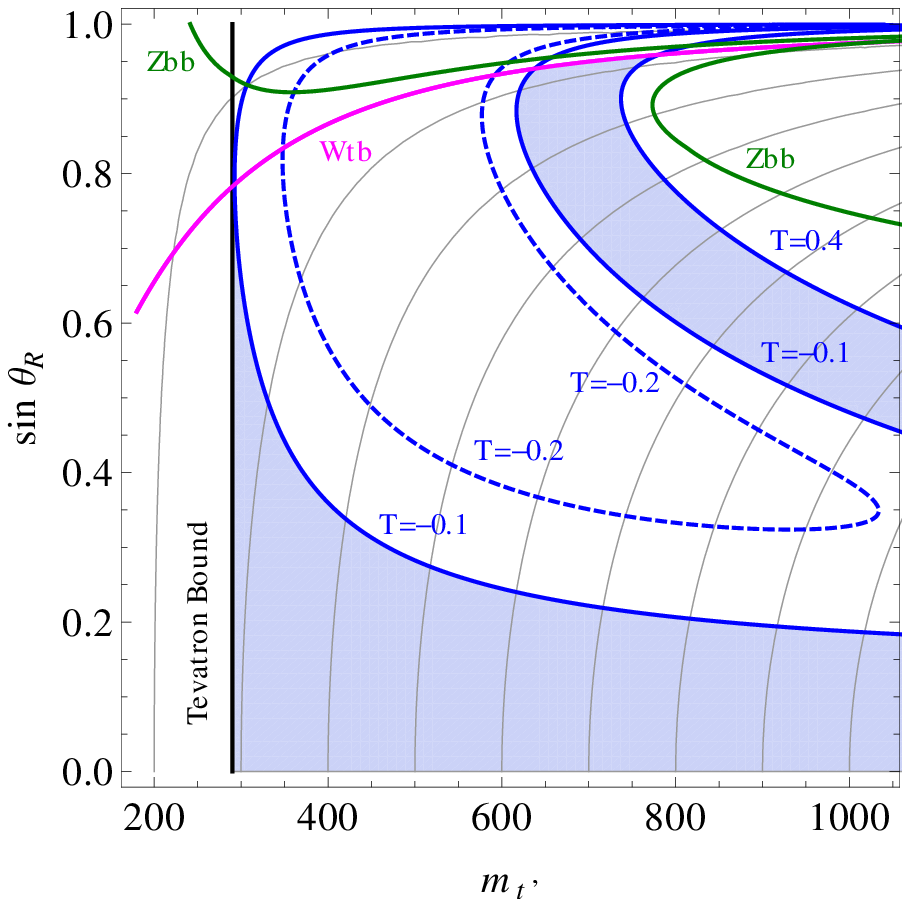} \includegraphics[width=0.49\textwidth]{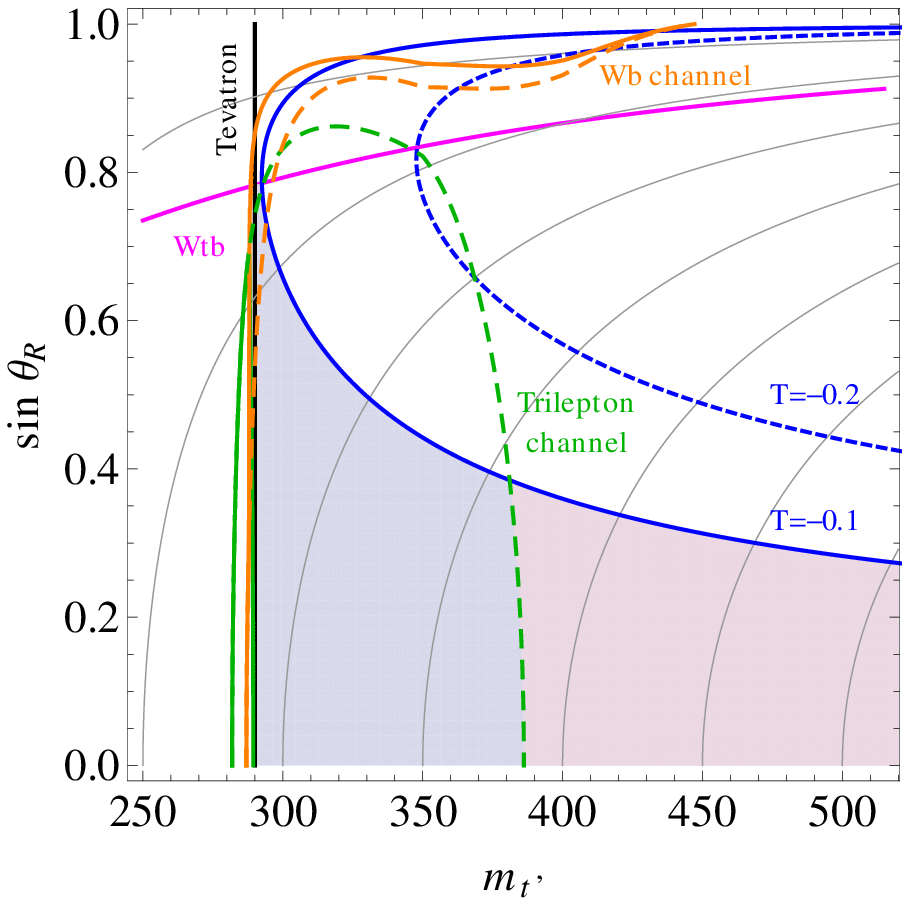}
\end{center}
\caption{Bounds on the top mixing as a function of $m_{t'}$ and $\sin \theta^u_R$. In magenta $Wtb$, in green $Zbb$, in black the 
direct bound from Tevatron, in blue the bound from the $T$ parameter. The grey lines correspond to fixed values of 
$M = m_X$. The light blue areas are the allowed regions. 
On the right panel, we show the bounds from CMS in the trilepton channel and $Wb$ channel. The dashed lines correspond to the case of a decoupled Higgs, the solid lines to $m_h = 115$ GeV. The coloured areas are the allowed regions.} 
\label{fig:bounds}
\end{figure}

\subsection{Precision electroweak measurements}

In~\cite{Cacciapaglia:2010vn} we imposed the bounds on the top sector from precisely measured quantities in the SM.
For earlier bounds see for example \cite{Alwall:2006bx}, \cite{delAguila:2008iz}.
The deviations on the $Z b \bar{b}$ coupling at tree level are absent in this specific case. We have used the exact formulas in 
\cite{hep-ph/9602438} to check that the one-loop contributions to the $Z b \bar{b}$ coupling are present but small for our non-standard 
doublet.  Note that this is true for our specific non-standard 
doublet,  while for other cases, as for example a singlet, these bounds are relevant.  
We studied the  deviations in the $W t b$ 
couplings allowing at most a 20\% deviation from 1, electroweak precision tests (in particular the $T$ parameter) 
where we allow $-0.1 < T < 0.4$ (here we assume $m_h = 120$ GeV and vanishing $S$ and $U$), finally the direct 
bound from Tevatron. The latter bound is nominally at $311$ GeV, however the experiments only look at the $t' \to W b$ 
decay mode, therefore we pose the bound until the branching ratio drops significantly below one: in other words, the 
bound is valid for $m_{t'} < 290$ GeV.
In the left panel of Figure~\ref{fig:bounds}, we show the allowed regions in the $m_{t'}$--$\sin \theta_R^u$ parameter space.
Note that the upper region on the right corresponds to value of the $\lambda$ Yukawa coupling close to the non--perturbative 
region, therefore non--perturbative effects should be considered and the formulas we use 
are not reliable in the upper right corner of the left panel in Figure~\ref{fig:bounds}.

\subsection{Direct bounds at the LHC (CMS)}

The CMS collaboration recently published new bounds on a heavy $t'$ with the 2011 data collected by the LHC.
They consider both the channel $W b$ and the channel $Z t$, in both cases assuming pair production and 100\% branching ratios.
Here we will use their results to pose bounds on our case, where all decay channels are open.

For the $t Z$ channel~\cite{CMStZ}, they select events with a pair of leptons (electrons or muons) near the $Z$ mass (the window for the dilepton invariant mass is set between $60$ and $120$ GeV), one additional lepton and at least two additional jets.
Their analysis is limited to a counting experiment and the bound is extracted by comparing the expected number of events from backgrounds and signal to the observed one: the preliminary results, based on an integrated luminosity of $191$ pb$^{-1}$ shows that there are no events in the dataset.
In our case, many different decay chains can give rise to events that pass the event selection used by CMS:
\beq
p p & \to & (t' \to Z t) (\bar{t}' \to Z \bar{t}) \to b \bar{b}\, Z\, Z\, W^+ W^- \nonumber \\
    & \to & (t' \to Z t) (\bar{t}' \to W^+ \bar{b}) \to b \bar{b}\, Z\, W^+ W^- \nonumber \\
    & \to & (t' \to W^- b ) (\bar{t}' \to Z \bar{t}) \to b \bar{b}\, Z\, W^+ W^- \nonumber \\
    & \to & (t' \to Z t) (\bar{t}' \to h \bar{t}) \to b \bar{b}\, b \bar{b}\, Z\, W^+ W^- \nonumber \\
    & \to & (t' \to h t) (\bar{t}' \to Z \bar{t}) \to b \bar{b}\, b \bar{b}\, Z\, W^+ W^- \nonumber 
 \eeq
where one of the $Z$'s and one of the $W$'s decay leptonically.
In extracting the bound, we consider all these channels together, under the assumptions that the efficiency of the cuts is the same: this is not necessarily true, because the kinematics of the event are different, and we plan to refine this analysis in a future work.
The final result is shown on the right panel in Figure~\ref{fig:bounds}: the dark green area is excluded for a Higgs mass of $m_h = 115$ GeV, while the lighter green one is excluded in the decoupled Higgs limit, where the absence of the $t H$ decay mode boosts the $t Z$ branching.

The analysis in the $W b$~\cite{CMSbW} is more refined because of the presence of an irreducible background from $t \bar{t}$: the selection criteria are the presence of an isolated lepton (electron or muon), a missing $p_T$ greater than $20$ GeV, at least 4 jets where one or more are tagged as b-jets.
To distinguish the $t \bar{t}$ background, the collaboration reconstructs the invariant mass of the top or $t'$ under the assumption of the topology
\beq
p p \to  (t/t' \to W^+ b) (\bar{t}/\bar{t}' \to W^- \bar{b}) \to b \bar{b}\, l^{\pm} \nu\, j j \,, \nonumber
\eeq
and fits the distribution of backgrounds and signal with a free mass for the $t'$.
In the preliminary results, an integrated luminosity of $573$ pb$^{-1}$ for electrons and $821$ pb$^{-1}$ has been considered.
In our case different decay chains can give rise to events that pass the selection rules of this analysis.
However, the reconstruction algorithm will not work on events with decays to $Z t$ and $h t$, where two extra jets or missing energy is coming from the decay of the $Z$ and $h$.
A more detailed study would be required to take into account such events.
To be conservative, therefore, we will limit ourselves to rescale the limit by the branching ratio BR($t' \to W b$): the result is shown in Figure~\ref{fig:bounds}, where the dark (light) grey area correspond to a Higgs mass of $m_h = 115$ GeV ($1000$ GeV - decoupled limit).
Note that for low $\sin \theta_R$, the bound is $m_{t'} \gtrsim 290$ GeV, which is the same bound we assumed for the Tevatron data and given by the suppression in the $W b$ branching ratio.

The plot also shows that the $Zt$ or trilepton channel is much more promising than the more traditional $W b$ one, also taking into account the more limited luminosity of the preliminary data.
Note also that a trilepton signature will also arise the from single production of $t'$, as we will discuss in Section~\ref{sec:lhc}, and that such events are not taken into account in extracting this bound.

\subsection{$D_0$--$\bar{D}_0$ mixing and $D_0 \to l^+ l^-$ decays}

In our model, FCNCs will only affect the up quark sector, therefore the most severe bounds will come from the 
$D$ meson sector.
The most recent experimental measurements in the $D_0$--$\bar{D}_0$ mixing (with no CP violation) are~
\cite{Golowich:2009ii}
\beq
x_D = \frac{\Delta m_D}{\Gamma_D} = 0.0100^{+0.0024}_{-0.0026}\,, \qquad y_D = \frac{\Delta \Gamma_D}{2 \Gamma_D} = 0.0076^{+0.0017}_{-0.0018}\,.
\eeq
In the SM, $x_D$ is dominated by long distance effects, and its value cannot be reliably estimated. Therefore, it is 
possible that the new physics saturates the experimental value and this situation will provide the most conservative 
bound on the new contributions.

Another potential bound may arise from the dilepton $\Delta C = 1$ decays: these channels have not been measured, 
and the present experimental limits are~\cite{pdg}:
\beq
BR (D_0 \to e^+ e^-)_{\rm exp}  < 1.2 \times 10^{-6}\,, \qquad BR (D_0 \to \mu^+ \mu^-)_{\rm exp} < 1.3 \times 10^{-6}\,.
\eeq
Contrary to the $\Delta C = 2$ contributions to the mixing, the branching ratio can be calculated and, in the SM, one obtains
\beq
 BR (D_0 \to \mu^+ \mu^-)_{\rm SM} \sim 3 \times 10^{-13}\,,
 \eeq
while the decay into electrons is suppressed by the electron mass square.
In both cases, therefore, the SM prediction lies well below the experimental limit.

In our case, the contribution of new physics is given in both cases by a tree level exchange of a $Z$ boson in the s-channel:
\beq
c \bar{u} \to Z^* \to u \bar{c}\,, \quad \mbox{and} \quad c \bar{u} \to Z^* \to l^+ l^-\,.
\eeq 
We will only consider the right-handed couplings, because the left-handed ones are suppressed by extra powers of the light quark masses.
The contribution of the new physics can be written as:
\beq
\delta x_D &=& \frac{f_D^2 m_D B_D}{2 m_Z^2 \Gamma_D} \frac{2}{3} r(m_c, m_Z) (g_{ZR}^{uc})^2\,, \\
\delta BR (D_0 \to l^+ l^-) &=& \frac{f_D^2 m_D m_l^2}{32 \pi m_z^4 \Gamma_D} \sqrt{1-\frac{4 m_l^2}{m_D^2}} \frac{\pi \alpha}{\cos^2 \theta_W \sin^2 \theta_W} (g_{ZR}^{uc})^2\,,
\eeq
where the explanation of the formula and the definitions and numerical values of the various parameters can be found in~\cite{Golowich:2009ii}.
If we assume that the new physics saturates the value of $x_D$, we can predict the value for the branching ratio; this value will become an upper bound in the case in which the mixing is dominated by the SM:
\beq
BR (D_0 \to l^+ l^-)_{\rm NPh} = \frac{3}{32} \frac{\alpha}{\cos^2 \theta_W \sin^2 \theta_W} \frac{x_D}{B_D r(m_c, m_Z)} \frac{m_l^2}{m_Z^2}   \sqrt{1-\frac{4 m_l^2}{m_D^2}}\,.
\eeq
Plugging in numbers, at one sigma, we find
\beq
BR (D_0 \to \mu^+ \mu^-)_{\rm NPh}  = (0.8 \pm 0.2) \times 10^{-10}\,,
\eeq
therefore we are still far from the experimental bound.
For electrons, the branching is suppressed by a factor $(m_e/m_\mu)^2$.

The most significant constraint from new physics, therefore, comes from $x_D$.
In our case,
\beq
(g_{ZR}^{uc})^2 =  \frac{\pi \alpha}{\cos^2 \theta_W \sin^2 \theta_W} |V_R^{41}|^2 |V_R^{42}|^2\,.
\eeq
To fit the experimental value would require
\beq
 |V_R^{41}|\, |V_R^{42}| = 2.44^{+0.28}_{-0.34} \times 10^{-4}\,.
 \eeq
This implies a bound at 3$\sigma$~\footnote{This bound roughly agrees with the results in~\cite{Gedalia:2009kh}, where effective operators are considered. The relevant parameter, as defined in~\cite{Gedalia:2009kh}, is $|\tilde{z}_1| < 5.7 \times 10^{-7} \left( \frac{\Lambda}{1 \mbox{TeV}} \right)^2$, where $\Lambda$ is the scale suppressing the operator ($\Lambda = m_Z$ in our case). Converting the bound in our parameterisation, we obtain $|V_R^{41}|\, |V_R^{42}| < 2.0 \times 10^{-4}$.}
\beq
 |V_R^{41}|\, |V_R^{42}| < 3.2 \times 10^{-4}\,. \label{eq:boundD0D0bar}
\eeq
As we will see in the following, this is the most stringent bound on flavour violation.

\subsection{Rare decays of $D$ mesons}

Many rare decays of the $D$ mesons via FCNC have been tested and have strong bounds from experiments such 
as CLEO and BES. In the SM they are loop generated and often dominated by long distance effects while, in our model, 
they may receive tree level contributions from the $Z$ exchange.
A list of the experimental limits~\cite{pdg}, standard model predictions from short and long distance~\cite{Fajfer:2003vi} 
calculations and the possible contribution in our case is presented in Table~\ref{tab:rareDdecays}. 
\begin{table}[tb]
\begin{tabular}{|c|cc|c|c|}
\hline
   mode & SM short & SM long & our model & exp. limit \\
\hline
\hline
$D^+ \to \pi^+ e^+ e^-$ & $9.4 \times 10^{-9}$ & $1.0 \times 10^{-6}$ & $c \to u l^+ l^-$ & $7.4\times 10^{-6}$ \\
$D^+ \to \pi^+ \mu^+ \mu^-$ & $9.4 \times 10^{-9}$ & $1.0 \times 10^{-6}$ & $c \to u l^+ l^-$ & $3.9\times 10^{-6}$ \\
$D^+ \to \rho^+ \mu^+ \mu^-$ & $4.8 \times 10^{-9}$  & $1.5\div 1.8 \times 10^{-6}$ & $c \to u l^+ l^-$ & $5.6\times 10^{-4}$ \\
 \hline
$D_0 \to \gamma \gamma$ &  &  &  1-loop  & $2.7\times 10^{-5}$ \\
$D_0 \to \pi^0 e^+ e^-$ & $1.9 \times 10^{-9}$ & $2.1\times 10^{-7}$ & $c \to u l^+ l^-$ & $4.5\times 10^{-5}$ \\
$D_0 \to \pi^0 \mu^+ \mu^-$ & $1.9 \times 10^{-9}$ & $2.1\times 10^{-7}$ & $c \to u l^+ l^-$ & $1.8\times 10^{-4}$ \\
$D_0 \to \eta e^+ e^-$ & $2.5 \times 10^{-10}$  & $4.9 \times 10^{-8}$ & $c \to u l^+ l^-$ & $1.1\times 10^{-4}$ \\
$D_0 \to \eta \mu^+ \mu^-$ & $2.5 \times 10^{-10}$  & $4.9 \times 10^{-8}$ & $c \to u l^+ l^-$ & $5.3\times 10^{-4}$ \\
$D_0 \to \pi^+ \pi^- e^+ e^-$ &  &  & $c \to u l^+ l^-$ & $3.73\times 10^{-4}$ \\
$D_0 \to \rho^0 e^+ e^-$ &  &  & $c \to u l^+ l^-$ & $1.0\times 10^{-4}$ \\
$D_0 \to \pi^+ \pi^- \mu^+ \mu^-$ &  &  & $c \to u l^+ l^-$ & $3.0\times 10^{-5}$ \\
$D_0 \to \rho^0 \mu^+ \mu^-$ & $9.7\times 10^{-10}$  & $3.5\div 4.7\times 10^{-7}$  & $c \to u l^+ l^-$ & $2.2\times 10^{-5}$ \\
$D_0 \to \omega e^+ e^-$ &  &  & $c \to u l^+ l^-$ & $1.8\times 10^{-4}$ \\
$D_0 \to \omega \mu^+ \mu^-$ & $9.1\times 10^{-10}$ & $3.3\div 4.5 \times 10^{-7}$  & $c \to u l^+ l^-$ & $8.3\times 10^{-4}$ \\
$D_0 \to K^+ K^- e^+ e^-$ &  &  & $c \to u l^+ l^-$ & $3.15\times 10^{-4}$ \\
$D_0 \to \phi e^+ e^-$ &  &  &  & $5.2\times 10^{-5}$ \\
$D_0 \to K^+ K^- \mu^+ \mu^-$ &  &  & $c \to u l^+ l^-$ & $3.3\times 10^{-5}$ \\
$D_0 \to \phi \mu^+ \mu^-$ & $0$ & $6.5\div 9 \times 10^{-8}$  &  & $3.1\times 10^{-5}$ \\
$D_0 \to \pi^+ K^- e^+ e^-$ &  &  &  & $3.85\times 10^{-4}$ \\
$D_0 \to \pi^+ K^- \mu^+ \mu^-$ &  &  &  & $3.59\times 10^{-4}$ \\
$D_0 \to \pi^+ \pi^- \pi^0 e^+ e^-$ &  &  & $c \to u l^+ l^-$ & $8.1\times 10^{-4}$ \\
\hline
$D_s^+ \to K^+ \mu^+ \mu^-$ & $9.0 \times 10^{-10}$ & $4.3 \times 10^{-8}$ & $c \to u l^+ l^-$ & $3.6\times 10^{-5}$ \\
$D_s^+ \to K^+ e^+ e^-$ & $9.0 \times 10^{-10}$ & $4.3 \times 10^{-8}$  & $c \to u l^+ l^-$ & $1.6\times 10^{-3}$ \\
$D_s^+ \to K^*(892)^+ \mu^+ \mu^-$ & $1.6 \times 10^{-9}$ & $5\div 7 \times 10^{-7}$ & $c \to u l^+ l^-$ & $1.4\times 10^{-3}$ \\
\hline
 \end{tabular} 
 \caption{Rare decays of D mesons with SM short and long distance contributions, the decay channel within 
 the vector-like model and the corresponding experimental figures.} \label{tab:rareDdecays}
 \end{table} 
The short distance SM result is dominated by the operator $\mathcal{O}_9$~\cite{Fajfer:2001sa}
\beq
\mathcal{L} = - \frac{4 G_F}{\sqrt{2}} V^*_{cs} V_{us} \frac{\alpha}{4 \pi} c_9\; \bar{u} \gamma^\mu P_L c\, \bar{l} \gamma_\mu l\,,
\eeq
where $c_9 (m_W) \sim \frac{4}{9} \log \frac{m_s}{m_d} = 1.34$, while the contribution of the penguins, enhanced by QCD corrections is small.

In our model, the right-handed couplings of the $Z$ generate the operators $\mathcal{O}'_9$ and $\mathcal{O}'_{10}$~\cite{Fajfer:2001sa}, with coefficients
\beq
\mathcal{L} =  \frac{G_F}{\sqrt{2}} V^{*41}_R V^{42}_R\; \bar{u} \gamma^\mu P_R c\, \left\{ (1-4 \sin^2 \theta_W)\, \bar{l} \gamma_\mu l -  \bar{l} \gamma_\mu \gamma_5 l \right\} \,.
\eeq
Those operators contribute to the short distance term, they do not significantly affect the long distance one because the pair of leptons comes from an heavy $Z$ boson that do not mix significantly to light vector mesons.
The main difference between our contributions and the SM one is that in our case right-handed quarks are involved, and the form factors are not precisely known.
Here we will simply assume that the form factor for right-handed and left-handed quarks are the same to extract an order of magnitude estimate of the contribution.
Under such assumption, we can calculate the ratio $\xi_D$ between the new physics and SM contribution (where the form factors cancel out) and obtain
\beq
\xi_D = \frac{2 \pi^2 (1-4 \sin^2 \theta_W + 8 \sin^4 \theta_W)}{\alpha^2 |V_{cs}|^2 |V_{us}|^2 c_9^2} |V_R^{41}|^2 |V_R^{42}|^2 = 0.5 \times 10^{6} \times  |V_R^{41}|^2 |V_R^{42}|^2\,.
\eeq
This enhancement factor allows some of the bounds to be relevant for our model.
However, imposing the $D_0$--$\bar{D}_0$ mixing bound, we obtain that
\beq
\xi_D < 0.05\,,
\eeq
therefore those processes should not impose further bounds.

\subsection{Top quark FCNC rare decays: $t \to Z c, Z u$}

FCNCs mediated by the $Z$ boson can also affect the physics of the top quark.
The only observable effect is in rare decays of the top quark: $t \to Z c$ and $t \to Z u$, which are very suppressed and 
induced by loops in the SM.
The present bound on the FCNC decays of the top is~\cite{Margaroli:2011zr}:
\beq
\frac{\Gamma (t \to Z c) + \Gamma (t \to Z u)}{\Gamma (t \to W b)} < 3.3 \%\,,
\eeq
which involves the partial width in $W b$ that has not been measured yet.
In our case, there is a tree level vertex of the $Z$ with top and a light quark.
The widths can be written as~\cite{Carvalho:2007yi,Beneke:2000hk} (neglecting the up and charm masses)
\beq
\Gamma (t \to Z q) &=& (g_{ZR}^{tq})^2 \frac{m_t^3}{8 \pi m_Z^2} f \left(\frac{m_Z}{m_t} \right) \,, \\
\Gamma (t \to W b) &=& \frac{G_F m_t^2}{8 \pi \sqrt{2}} |V_{tb}|^2  f \left(\frac{m_W}{m_t} \right)\,;
\eeq
where
\beq
f(x) = \left(1-x^2 \right)^2 \left(1+2 x^2 \right).
\eeq
In our model,
\beq
\Gamma (t \to Z u) = \frac{G_F m_t^3}{16 \pi \sqrt{2}} |V_R^{43}|^2 |V_R^{41}|^2 f \left(\frac{m_Z}{m_t} \right)\,, 
\eeq
and similarly for $t \to Z c$ with $V_R^{41} \to V_R^{42}$.

The bound, therefore, can be written as
\beq
\frac{|V_R^{43}|^2 (|V_R^{41}|^2 + |V_R^{42}|^2)}{2 |V_{tb}|^2} \frac{f (m_Z/m_t)}{f(m_W/m_t)} < 3.3 \%\,,
\eeq
and
\beq
|V_R^{43}| \sqrt{|V_R^{41}|^2 + |V_R^{42}|^2} < 0.28 |V_{tb}|\,. \label{eq:boundtop}
\eeq
Note that this bound depends on $|V_R^{43}| \sim \sin \theta^u_R$, and on the value of $V_{tb}$ which can deviate from 
one and therefore make the bound stronger. Even though this bound is much milder that the one from the $D_0$ 
mesons, the different dependence on the two mixing angles $V_R^{41}$ and $V_R^{42}$ makes it relevant. These 
rare decay modes of the top quark may be relevant for LHC with high statistics and can be studied in detail at an 
electron-positron linear collider.

\subsection{Atomic Parity Violation}

Another strong bound comes from the modifications of the couplings of the up with the $Z^0$ boson: experiments can measure the weak charge $Q_W$ of the nucleus.
The most precise test is from atomic parity violation in Cesium $^{133}$Cs~\cite{pdg}: 
\beq
\left. Q_W (\mbox{$^{133}$Cs})\right|_{\rm exp.} = -73.20\pm 0.35\,, \qquad \left. Q_W (\mbox{$^{133}$Cs})\right|_{\rm SM} = -73.15 \pm 0.02\,.
\eeq

In general, the weak charge can be written as~\cite{Deandrea:1997wk}
\beq
Q_W = (2Z+N) (\tilde{g}_{ZL}^u + \tilde{g}_{ZR}^u) + (Z+2N) (\tilde{g}_{ZL}^d + \tilde{g}_{ZR}^d)\,,
\eeq
where the $\tilde{g}$'s are the left- and right-handed couplings of the $Z^0$ divided by a factor $\frac{g}{2 \cos \theta_W}$, and $N$ and $Z$ are the number of neutrons and protons in the nucleus.
In our case, the only large deviation appears in 
\beq
\tilde{g}_{ZR}^u = - \frac{4}{3} \sin^2 \theta_W + \delta g_{ZR}\,, \qquad \delta g_{ZR} = - |V_R^{41}|^2\,.
\eeq
The correction to the weak charge is therefore
\beq
\delta Q_W (\mbox{Cs}) = - (2Z+N) |V_R^{41}|^2 = - 188 |V_R^{41}|^2\,,
\eeq
where the numerical value corresponds to $^{133}$Cs ($Z=55$ and $N=78$).
As the SM prediction is in very good agreement with the experimental value and its error is very small, we can directly compare the new physics contribution to the experimental error.
At 3 sigma $|\delta Q_W| < (-73.15+0.06) - (-73.20-1.05) = 1.16$, therefore
\beq
|V_R^{41}| < 7.8 \cdot 10^{-2}\,. \label{eq:boundAPV}
\eeq

Atomic parity violation has also been measured for the Thallium $^{204}$ Tl, however the bound is milder~\footnote{This case is more favourable because of the larger atomic number ($Z=81$ and $N=123$), however the experimental precision is lower that in Cesium~\cite{pdg}: $$\left. Q_W (\mbox{Tl})\right|_{\rm exp.} = -116.4\pm 3.6\,, \qquad \left. Q_W (\mbox{Tl})\right|_{\rm SM} = -116.76 \pm 0.04\,;$$ 
for which the correction is
$\delta Q_W (\mbox{Tl})= - (2Z+N) |V_R^{41}|^2 = - 285 |V_R^{41}|^2\,,
$
and the bound
$ |V_R^{41}| < 0.19\,$.}

\subsection{Measurement of the charm couplings at LEP1}

The couplings of the charm quark to the $Z$ boson have been precisely measured at LEP1~\cite{LEP1}. 
Under the hypothesis of universal lepton couplings, the result of the fit is
\beq
g_{ZL}^c = 0.3453 \pm 0.0036\,, \qquad g_{ZR}^c = -0.1580 \pm 0.0051\,.
\eeq
In our case, the largest correction occurs on the right-handed couplings, and requiring that the new contribution is smaller that 3 standard deviations yields
\beq
\frac{g}{2 \cos \theta_W} |V_R^{42}|^2 < 0.015\; \Rightarrow |V_R^{42}| < 0.2\,. \label{eq:boundLEP1}
\eeq
This bound is stronger that the bound from top decays in Eq.~\ref{eq:boundtop}.
Note also that no significant bound can be extracted from the couplings of the up quarks, which are measured with a much worse precision, while the couplings of the top have not been measured.
A summary of all the relevant bounds on $|V_R^{14}|$ and $|V_R^{24}|$ is presented in Table~\ref{tab:sumbounds}.

\begin{table}[tb]
\begin{center}
\begin{tabular}{|l|c|}
\hline
$D_0$--$\bar{D}_0$ mixing & $|V_R^{41}||V_R^{42}| < 3.2 \times 10^{-4}$ \\
\hline
APV in Cs & $|V_R^{41}| < 7.8 \times 10^{-2}$ \\
\hline
LEP1, charm couplings & $|V_R^{42}| < 0.2$ \\
\hline
Tevatron: $t \to Z c, Z u$ & $|V_R^{43}| \sqrt{|V_R^{41}|^2 + |V_R^{42}|^2} < 0.28 |V_{tb}|$\\
\hline
$D$ meson decays & none\\
\hline
\end{tabular}
\caption{Bounds on $|V_R^{41}|$ and $|V_R^{42}|$ in order of importance.} 
\label{tab:sumbounds}
\end{center}
\end{table}

\subsection{CKM Matrix}
\label{subsec:CKMmatrix}

The CKM matrix has been measured very precisely, both at tree and one-loop level.
Even though in our case the corrections are proportional to the suppressed mixing angles in the left-handed sector, it is worth studying the detailed structure of the modified CKM matrix to ensure that no further bounds arise.
In the following we will limit ourselves to tree level measurements, because loop effects will also include the effect of the heavy top $t'$ and the flavour violating $Z$ boson couplings (a brief discussion of loop effect is present in next section).
The modified CKM matrix is given in Eq.~(\ref{eq:VCKMdef}): as $V_L$ is close to the identity matrix, we can assume that $\tilde{V}_{CKM}$ is approximately equal to the measured matrix, while $V_L$ generates small corrections. 
For a study discussing the CKM structure in the case of an iso-singlet see \cite{Botella:2008qm}.
In this section, we will estimate how small such corrections can be.

We first observe that both $\tilde{V}_{CKM}$ and $V_L$ are hierarchical matrices, as the off diagonal entries of the latter are suppressed by ratios of quark masses.
Therefore, we can use the Cabibbo angle $\lambda = \sin \theta_{12}$ as an expansion parameter for both (knowing that $\frac{m_c}{m_t} \sim \mathcal{O} (\lambda^3)$ and $\frac{m_u}{m_t} \sim \mathcal{O} (\lambda^7)$).
More details about this expansion are presented in Appendix~\ref{app:CKM}.
On top of the suppression from the light quark masses, off diagonal $V_L$ terms are also suppressed by the new yukawa couplings $x_i$, which are constrained to be small as we discusses previously.
One can relate the elements of $V_L$ that enter in the definition of $V_{CKM}$ to $V_R^{41}$ and $V_R^{42}$, and use the bounds in Eq.s~(\ref{eq:boundD0D0bar}), (\ref{eq:boundtop}) and (\ref{eq:boundAPV}) to find an upper bound.
For instance
\beq
V_L^{12} = - \frac{m_c m_u}{m_c^2 - m_u^2} \frac{x_1 x_2^*}{m_X^2} = - \frac{m_c m_u}{m_c^2 - m_u^2} \, V_R^{*,41} V_R^{42} = - V_L^{*,21}\,.
\eeq
Therefore, $|V_L^{12}| = |V_L^{21}|$ is suppressed by a power $\lambda^4$ from the quark mass ratio, but also by the product $|V_R^{41}| |V_R^{42}| < 3.2 \cdot 10^{-4}$:
\beq
|V_L^{12}|  < 8.3 \cdot 10^{-7}\,.
\eeq
Therefore, $V_L^{12}$ is always highly suppressed!
Regarding $V_L^{13}$ and $V_L^{31}$, we can write
\beq
V_L^{13} = - \frac{x_1^* m_u}{m_t^2} \sin \theta_L = - \frac{m_u}{m_t} \frac{x_1^* \sin \theta_R}{m_X} = \frac{m_u}{m_t} V_R^{*,41} \sin \theta_R\,.
\eeq
Besides the $\lambda^7$ suppression from the quark masses, this element is proportional to $|V_R^{42}| < 7.8 \cdot 10^{-2}$ (from APV in Cesium); therefore, considering $\sin \theta_R < 0.3$ as a conservative bound, we find
\beq
|V_L^{13}|  < 4.2 \cdot 10^{-7}\,.
\eeq
The same is true for
\beq
V_L^{31} = \frac{m_{t'}^2 - m_t^2}{m_{t'}^2} \cos \theta_L\, (- V_L^{*,13})\,.
\eeq
Finally, 
\beq
V_L^{23} = - \frac{x_2^* m_c}{m_t^2} \sin \theta_L = - \frac{m_c}{m_t} \frac{x_2^* \sin \theta_R}{m_X} = \frac{m_c}{m_t} V_R^{*,42} \sin \theta_R\,.
\eeq
This element is proportional to $V_R^{42}$, therefore it is maximal when the bound from the charm couplings in Eq.~\ref{eq:boundLEP1} is saturated:
\beq
|V_L^{23}|  < 2.3 \cdot 10^{-3} \times |V_R^{42}| \sim 4.6 \cdot 10^{-4}\,.
\eeq
The same is true for
\beq
V_L^{32} = \frac{m_{t'}^2 - m_t^2}{m_{t'}^2} \cos \theta_L\, (- V_L^{*,23})\,.
\eeq

The final comment is about $V_L^{33}$: this modifies the top couplings and therefore may seriously affect the loop processes that are Glashow-Iliopoulos-Maiani (GIM) suppressed in the SM.
The tree level bound coming from modifications of $V_{tb}$ have already been considered 
in~\cite{Cacciapaglia:2010vn}, and are less competitive than bounds from the $T$ parameter.
We can estimate
\beq
V_L^{33} \sim \cos \theta_L = 1-\frac{1}{2} \frac{m_t^2}{m_X^2} \sin^2 \theta_R
\eeq
The bounds from the top sector typically imply $m_X > 2 m_t$ and $\sin \theta_R < 0.3$, therefore
\beq
1-V_L^{33} < 0.01\,.
\eeq

\subsubsection*{A numerical example}

We have seen that corrections to the CKM matrix are typically very small.
To be more concrete, here we will focus on two numerical examples 
to illustrate the relevance of the corrections.
First, we parameterise the two phases as $x_1 = |x_1| e^{i \beta_1}$ and $x_2 = |x_2| e^{i \beta_2}$.
As a benchmark mass for the $t'$ we consider $m_{t'} = 350$ GeV, and a conservative bound on the mixing 
$\sin \theta_R = 0.3$.
As we want a maximal flavour violating effect, we will saturate all bound, in particular the strongest bound from $D_0$--$\bar{D}_0$ mixing $|V_R^{41}| |V_R^{42}| = 3.2 \cdot 10^{-4}$.
We distinguish two cases: in case A, the mixing with the up is maximal, therefore from the APV bound we fix 
$|V_R^{41}| = 0.078$; in case B, we require maximal mixing with the charm, which is bounded from the charm couplings 
$|V_R^{42}| = 0.2$.
In these numerical examples only, for simplicity, we also set the phase in $\tilde{V}_{CKM}$ to zero, while the absolute 
values of the entries are taken from the tree level measurements in the SM~\cite{pdg}.
The final results are listed in Table~\ref{tab:CKMnum}.

\begin{table}[tb]
{\small
\begin{tabular}{|l|c|c|c|}
\hline
  & SM & Case A & Case B \\
\hline
\hline
$V_{CKM}^{ud}$ & $0.974$ & $+ 1.9 \cdot 10^{-7}\, e^{i (\beta_2-\beta_1)} + 2.6 \cdot 10^{-9}\, e^{-i \beta_1}$ & $+ 1.9 \cdot 10^{-7}\, e^{i (\beta_2-\beta_1)} + 5.4 \cdot 10^{-11}\, e^{-i \beta_1}$ \\
$V_{CKM}^{us}$ & $0.225$  & $+ 8.5 \cdot 10^{-7} e^{i (\beta_2-\beta_1)} + 1.2 \cdot 10^{-8}\, e^{-i \beta_1}$  & $+ 8.5 \cdot 10^{-7} e^{i (\beta_2-\beta_1)} + 2.5 \cdot 10^{-10}\, e^{-i \beta_1}$\\
$V_{CKM}^{ub}$ & $0.0039$  & $+ 3.4 \cdot 10^{-8}\, e^{i (\beta_2-\beta_1)} + 2.7 \cdot 10^{-7}  e^{-i \beta_1}$ & $+ 3.4 \cdot 10^{-8}\, e^{i (\beta_2-\beta_1)} + 5.6 \cdot 10^{-9}  e^{-i \beta_1}$\\
\hline
$V_{CKM}^{cd}$ & $0.23$  & $- 8.1 \cdot 10^{-7} e^{- i (\beta_2-\beta_1)}+ 5.8 \cdot 10^{-8}\, e^{-i \beta_2}$ &  $+ 2.8 \cdot 10^{-6}\, e^{-i \beta_2}- 8.1 \cdot 10^{-7} e^{- i (\beta_2-\beta_1)}$ \\
$V_{CKM}^{cs}$ & $1.023$ & $- 1.9 \cdot 10^{-7} e^{- i (\beta_2-\beta_1)}  + 2.6 \cdot 10^{-7}  e^{-i \beta_2}$ &  $ + 1.3 \cdot 10^{-5}  e^{-i \beta_2} - 1.9 \cdot 10^{-7} e^{- i (\beta_2-\beta_1)}$ \\
$V_{CKM}^{cb}$ & $0.041$  & $+ 6.1 \cdot 10^{-6} \, e^{-i \beta_2} - 3.2 \cdot 10^{-9} e^{- i (\beta_2-\beta_1)}$ & $+ 3 \cdot 10^{-4} \, e^{-i \beta_2} - 3.2 \cdot 10^{-9} e^{- i (\beta_2-\beta_1)}$ \\ 
\hline
$V_{CKM}^{td}$ & $0.008$ & $- 1 \cdot 10^{-4}  - 2.1 \cdot 10^{-6}\,e^{i \beta_2}- 4.1 \cdot 10^{-7} \, e^{i \beta_1}$ & $- 0.9 \cdot 10^{-4}  - 1.0 \cdot 10^{-4}\,e^{i \beta_2} - 8.4 \cdot 10^{-9} \, e^{i \beta_1}$ \\
$V_{CKM}^{ts}$ & $0.038$  & $-4.6 \cdot 10^{-4} - 1. \cdot 10^{-5}\,  e^{i \beta_2} - 1. \cdot 10^{-7}\, e^{i \beta_1} $  & $- 4.2 \cdot 10^{-4} - 4.7 \cdot 10^{-4}\,  e^{i \beta_2} - 1.9 \cdot 10^{-9}\, e^{i \beta_1}$\\
$V_{CKM}^{tb}$ & $0.88$  & $-0.01  - 3.8 \cdot 10^{-7}\, e^{i \beta_2}  - 1.6 \cdot 10^{-9}\, e^{i \beta_1}$ & $ - 0.009  - 1.9 \cdot 10^{-5}\, e^{i \beta_2}  - 3.3 \cdot 10^{-11}\, e^{i \beta_1}$ \\
\hline
$V_{CKM}^{t'd}$ &  & $0.0013 + 3.4 \cdot 10^{-6}\, e^{i \beta_2} + 6.5 \cdot 10^{-7}\,e^{i \beta_1}$ & $0.001 + 1.7 \cdot 10^{-4}\, e^{i \beta_2} + 1.3 \cdot 10^{-8}\,e^{i \beta_1}$ \\
$V_{CKM}^{t's}$ &  & $0.006 + 1.5 \cdot 10^{-5}\, e^{i \beta_2} +  1.5 \cdot 10^{-7}\,  e^{i \beta_1}$ & $ 0.006 + 7.4 \cdot 10^{-4}\, e^{i \beta_2} +  3.1 \cdot 10^{-9}\,  e^{i \beta_1}$ \\
$V_{CKM}^{t'b}$ &  & $0.14 + 6.1 \cdot 10^{-7} \, e^{i \beta_2}  + 2.6 \cdot 10^{-9}\, e^{i \beta_1}$ &  $0.13 + 3.0 \cdot 10^{-5} \, e^{i \beta_2}  + 5.3 \cdot 10^{-11}\, e^{i \beta_1}$ \\
\hline
\end{tabular} }
\caption{Numerical values of the corrections to the CKM matrix for $m_{t'} = 350$ GeV, $\sin \theta_R = 0.3$,  
$|V_R^{41}| |V_R^{42}| = 3.2 \cdot 10^{-4}$ and: in case A, $|V_R^{41}| = 0.078$; in case B, $|V_R^{42}| = 0.2$.} 
\label{tab:CKMnum}
\end{table}

\section{Loop constraints and couplings}
\label{sec:loop}

So far, we have discussed tree level bounds from precision and flavour observables coming mainly from the right-handed mixing of up type quarks, while the left handed mixings are suppressed by the SM quark masses.
In particular, the CKM matrix is only affected by the left-handed mixings and no significant tree level bounds arise.
The situation is quite different at loop level: in fact, loops involving the modified CKM couplings, will correct the SM results.
In some cases, where a precise cancellation occurs in the SM loops, even small corrections may give rise to observable effects.
Another important constraint may come from CP violating effects, which arise in the SM from a single phase.
In our case, there are two extra phases in the game.
Moreover, even if the extra phases are small, the mixing with the $t'$ will change the phase structure of the CKM matrix and new CP violating effects may arise.
In the following we will focus on the two most constraining cases: $\Delta F = 2$ mixings in the Kaon and B meson sectors.

\subsection{Kaon sector: $K^{0} - \bar{K}^{0}$ mixing}

\begin{figure}[tb]
\begin{center}
\epsfig{file=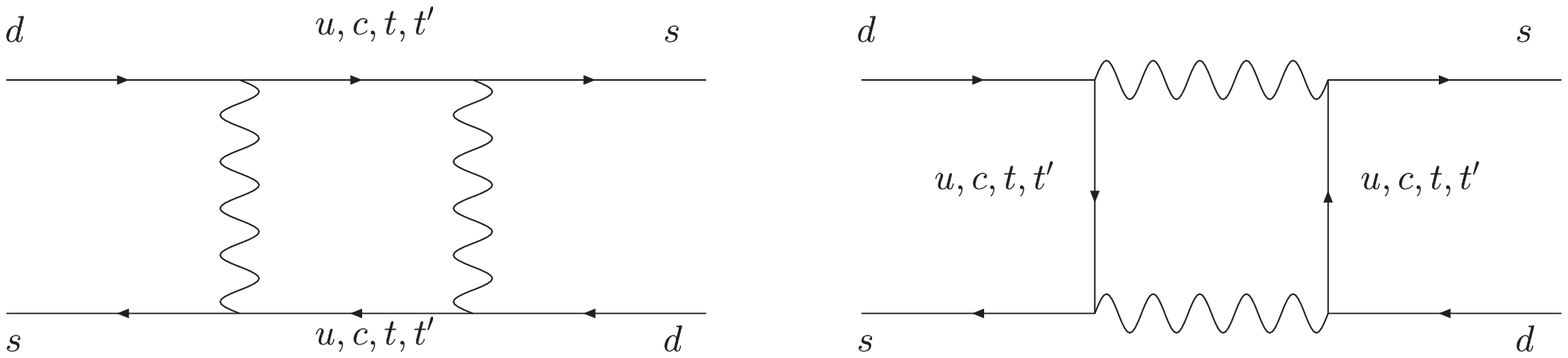,width=.75\textwidth}
\caption{Box diagrams contributing to $K^{0}-\bar{K}^{0}$ mixing.}
\label{k0k0barmixing}
\end{center}
\end{figure}

The $K^{0} - \bar{K}^{0}$ mixing can be calculated in our model with extra non-standard doublet by a simple 
generalisation of the SM formulas, by including the effect of the $t'$ quark as shown in Figure~\ref{k0k0barmixing}.
In fact, loops involving the flavour violating couplings of the $Z^0$ are irrelevant because they only involve up-type 
quarks. The general formalism to describe such mixing is presented in Appendix~\ref{app:mesonmixing}: here we will 
limit ourselves to report the most relevant results. In the non-SM doublet case, the non-diagonal mass matrix element 
$M_{12}$ between $K^0$ and $\bar{K}^0$ mesons can be written as
\beq
M_{12} = \frac{1}{3} f_{K}^{2}B_{K}m_{K}
\frac{G_{F}}{\sqrt{2}}\frac{\alpha}{4\pi\sin^{2}\theta_{W}} \sum_{i,j=c,t,t^{\prime}} \eta_{ij}\xi^{i}\xi^{j} E(x_{i},x_{j}) \,,
\eeq
where the Inami-Lim functions $E$~\cite{Inami:1980fz} of the mass ratios $x_i = m_i^2/m_W^2$ are defined in the Appendix, and the coefficients $\xi^i$ contain the CKM mixings generalised to include $t'$: $\xi^{i}=V^{*,is}_{CKM}V^{id}_{CKM}$, with $i = u, c, t, t'$.
The factors $\eta_{ij}$ encode the QCD loop corrections to the electroweak loop diagrams.
The condition $V_{CKM}^\dagger \cdot V_{CKM} = 1$ in Eq.~(\ref{eq:CKMuni2}) implies the quadrangle condition $\xi^{u} + \xi^{c} + \xi^{t} + \xi^{t^{\prime}}=0$, therefore one can eliminate the contribution of the up quark loop as in the SM calculation.
The functions $E$ are strongly dependent on the value of the masses, and in particular they grow with the mass of the fermions in the loop: to quantify this statement, we list the values of the relevant functions for $m_{t'} = 350, 500, 1000$ GeV:
\beq
E(x_{c},x_{c}) = 2.5 \times 10^{-4}\,, \quad &
E(x_{c},x_{t}) = 2.2 \times 10^{-3}\,, \quad &
E(x_{c},x_{t^{\prime}}) = (2.4, 2.5, 2.6) \times 10^{-3}\,, \nonumber\\
E(x_{t},x_{t}) = 2.5\,, \quad&
E(x_{t},x_{t^{\prime}}) = 4.0, 4.8, 6.4\,, \quad &
E(x_{t^{\prime}},x_{t^{\prime}}) = 7.5, 13, 44\,, 
\eeq
while the function depending on $x_u$ are negligible.
Here we are interested in two observables: the mass difference $\Delta m_K$ between the two mass eigenstates 
and the CP violating parameter $\epsilon_K$, given by
\begin{eqnarray}
\Delta m_{K} \equiv m_{K_{L}} - m_{K_{S}} &=& 2  Re M_{12} \simeq 2 |M_{12}|\,, \\
\epsilon_{K} & \simeq& \frac{e^{i\pi/4}}{\sqrt{2}\Delta m_{K}} Im M_{12}\,.
\end{eqnarray}
They are to be compared with the experimental results \cite{pdg} :
\beq
\Delta m_{K}|_{{\rm exp}} = (3.483 \pm 0.006) \times 10^{-15} {\rm GeV}\,, \quad \mbox{and} \;\; |\epsilon_{K}|_{\rm exp} = (2.233 \pm 0.015) \times 10^{-3}\,. \nonumber
\eeq
The real part of $M_{12}$, that is related to $\Delta m_K$, is dominated by the charm contribution: in fact, the suppression from the function $E$ is compensated by the suppression of the CKM mixing for top and $t'$, therefore
\begin{eqnarray}
\Delta m_{K} \sim
1.75791 \times 10^{-10}
\Bigl|
(V_{cs}^{*}V_{cd})^{2}E(x_{c},x_{c})
\Bigr|
\sim 2.1 \times 10^{-15} \,\,{\rm GeV}\,.
\end{eqnarray}
The corrections to $\xi^c$ from the new mixing are very small, therefore no significant modification occurs.
The situation is rather different for $\epsilon_K$.
In this case, in the SM the imaginary part of $\xi^c$ is of the same order as the imaginary part of $\xi^t$, 
therefore $\epsilon_K$ is dominated by the top contribution.
Effects of the new physics enter either via modifications of the top couplings and via the new phases, so we can expect 
large modifications in this case. In addition to $\xi^{i}$, the proportion of the short and long distance contributions in measured values are also important.
Compared with the short distance contributions, in the real part of $M_{12}$ the long distance contributions are sizable effects, but in the imaginary part of $M_{12}$ they are negligible. In other words, $\epsilon_{K}$ is more sensitive to the new physics effects than the mass difference $\Delta m_{K}$.
The non-diagonal mass matrix element $M_{12}$ of top- and $t^{\prime}$-mediated box diagrams in the vector-like model can be approximately written as
\beq
\frac{M_{12}}{C_{K}} &=& (V_{ts}^{*}V_{td})^{2}
\biggl\{
\tilde{E}(x_{t},x_{t}) + 2\frac{m_{t}^{2}\sin^{2}\theta_{R}}{m_{X}^{2}} \left( \tilde{E}(x_{t},x_{t^{\prime}}) - \tilde{E}(x_{t},x_{t}) \right) \nonumber \\
&& + \frac{m_{t}^{4}\sin^{4}\theta_{R}}{m_{X}^{4}} \tilde{E}(x_{t^{\prime}},x_{t^{\prime}}) 
\biggr\} \nonumber \\
&+& (V_{ts}^{*}V_{td}) \frac{m_{c}}{m_{t}}|V_{R}^{42}|\sin\theta_{R} \nonumber \\
&& \times \left\{
e^{i\beta_{2}}(V_{ts}^{*}V_{cd}) \tilde{E}(x_{t},x_{t})
+ e^{-i\beta_{2}}(V_{cs}^{*}V_{td}) \left( \tilde{E}(x_{t},x_{t}) + 2\frac{m_{t}^{2}}{m_{t}^{\prime2}} \tilde{E}(x_{t},x_{t^{\prime}}) \right)
\right\} \,,
\eeq
where $C_{K}=\frac{G_{F}^{2}}{12\pi^{2}}m_{W}^{2}m_{K}f_{K}B_{K}$ and $\tilde{E}(x_{i},x_{j})=\eta_{ij}E(x_{i},x_{j})$. The contributions from $t^{\prime}$-mediated diagrams with heavy $t^{\prime}$ are suppressed by its mass and these effects become sub-leading terms.
The $\beta_{1}$ term is only appear in $\xi^{c}$.
Consequently, the result is practically independent of  
$\beta_1$ while it can be modified slightly when varying $\beta_2$. 

\begin{figure}[htb]
\begin{center}
\epsfig{file=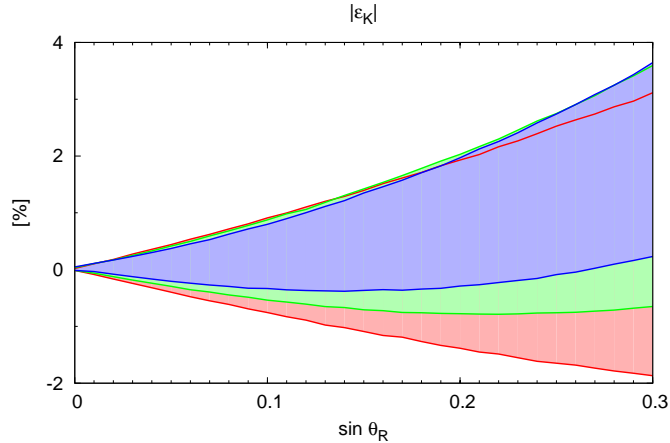,width=0.6\textwidth}
\caption{Deviation of $|\epsilon_{K}|$ in Vector-like model from the SM value is shown as a function of $\sin\theta_{R}$
 for three different values of $m_{t^{\prime}}$ (350 GeV (blue), 500 GeV (green), 1000 GeV (red)). 
We vary the other parameters within the ranges : $|V_R^{41}| \leq 0.078$, $|V_R^{42}| \leq 0.2$, $|V_R^{41}||V_R^{42}| 
\leq 3.2 \times 10^{-4}$, $0 \leq \beta_1 < 2\pi$ and $0 \leq \beta_2 < 2\pi$.}
\label{fig:epstheta}
\end{center}
\end{figure}

To estimate the range of the effect, we study the correction to $\epsilon_K$ numerically.
For the SM CKM part $\tilde{V}_{CKM}$, we use the Wolfenstein parameterisation with $\lambda = 0.2253$,
$A = 0.808$, $\bar{\rho} = 0.132$ and $\bar{\eta} = 0.341$.
In terms of new physics parameters, we take account of tree level bounds which we discussed in previous section. Therefore, we vary new physics parameters within the ranges : $0 \leq \sin\theta_{R} \leq 0.3$, $|V_R^{41}| \leq 0.078$, $|V_R^{42}| \leq 0.2$, $|V_R^{41}||V_R^{42}| \leq 3.2 \times 10^{-4}$, $0 \leq \beta_1 < 2\pi$ and $0 \leq \beta_2 < 2\pi$.
Here the upper bound values of $|V_{R}^{41}|=0.078$ and $|V_{R}^{42}|=0.2$ correspond to case A and B respectively, and the parameter ranges of two new phases $\beta_{1}$ and $\beta_{2}$ in our model are not restricted by the present experimental results.
The dependence of $|\epsilon_{K}|$ in the vector-like model as a function of $\sin\theta_{R}$ is shown in Figure 
\ref{fig:epstheta} for three different values of $m_{t^{\prime}}$ (350 GeV (blue), 500 GeV (green), 1000 GeV (red)).
Deviations are typically at few percent level.
The main source of the theoretical uncertainty is the bag parameter $B_{K}$.
Note that the effect is always below the theoretical uncertainty in the matrix elements.

\subsection{$B$ meson sector: $B_{d}^{0} - \bar{B}_{d}^{0}$ mixing}

The mixing in the $B$ meson sector can be computed in a similar way as the $K^0$--$\bar{K}^0$ mixing.  In this 
case $\Delta M_{B_{q}} = 2 |M_{12}(B_{q})|$, $M_{12}(B_{q})=|M_{12}(B_{q})|e^{i\Phi_{B_{q}}}$
while the experimental value of $\Delta M_{B_{d}}$ is~\cite{pdg}
$$
\Delta M_{B_{d}} = (3.337 \pm 0.033) \times 10^{-13} {\rm GeV} \,.
$$
In the SM, the real and imaginary parts of $M_{12}({B_{d}})$ are given by the following forms
\beq
Re M_{12}({B_{d}}) &=& C_{B_{d}} \eta_{tt}^{B}E(x_{t},x_{t}) A^{2} \left\{ (1-\rho)^{2} - \eta^{2} \right\}\lambda^{6} \,, \\
Im M_{12}({B_{d}}) &=& -2C_{B_{d}}\eta_{tt}^{B}E(x_{t},x_{t}) A^{2} (1-\rho)\eta \lambda^{6} \,,
\eeq
where $C_{B_{d}}=\frac{G_{F}^{2}}{12\pi^{2}}m_{B_{d}}f_{B_{d}}^{2}B_{d}m_{W}^{2}$.
The leading contribution of $M_{12}(B_{d})$ is the top-mediated box diagrams and $Re M_{12}({B_{d}})$ is of the same order of magnitude as $Im M_{12}({B_{d}})$.
To check the leading contribution of $M_{12}(B_{d})$ in the vector-like model, we expand $\xi_{d}^{i}$ in a power of $1/m_{X}^{2}$:
\beq
\frac{M_{12}(B_{d})}{C_{B_{d}}} &=& (V_{tb}^{*}V_{td})^{2}
\biggl\{
\tilde{E}(x_{t},x_{t}) + 2\frac{m_{t}^{2}\sin^{2}\theta_{R}}{m_{X}^{2}} \left( \tilde{E}(x_{t},x_{t^{\prime}}) - \tilde{E}(x_{t},x_{t}) \right) \nonumber \\
&& + \frac{m_{t}^{4}\sin^{4}\theta_{R}}{m_{X}^{4}} \tilde{E}(x_{t^{\prime}},x_{t^{\prime}}) 
\biggr\} \nonumber \\
&-& 2e^{i\beta_{2}}(V_{tb}^{*}V_{td})(V_{tb}^{*}V_{cd}) \frac{m_{c}}{m_{t}}|V_{R}^{42}|\sin\theta_{R} \nonumber \\
&& \times \left\{
\tilde{E}(x_{t},x_{t}) - \frac{m_{t}^{2}}{m_{t}^{\prime2}} \tilde{E}(x_{t},x_{t^{\prime}}) - \frac{m_{t}^{4}\sin^{2}\theta_{R}}{m_{t}^{\prime2}m_{X}^{2}} \tilde{E}(x_{t^{\prime}},x_{t^{\prime}})
\right\} \,,
\eeq
where $\tilde{E}(x_{i},x_{j})=\eta^{B}_{ij}E(x_{i},x_{j})$. The function $E(x_{t^{\prime}},x_{t^{\prime}}) \simeq x_{t^{\prime}}$ grows with the $t^{\prime}$ mass, while the coefficient $(\xi_{d}^{t^{\prime}})^{2}$ is suppressed by $1/m_{t^{\prime}}^{4}$. Consequently, deviations which comes from this effect are no more than $m_{t}^{2}/m_{X}^{2}$.
Similarly, the top-$t^{\prime}$-mediated box diagrams are also suppressed by a power of $m_{t}^{2}/m_{t^{\prime}}^{2}\sim m_{t}^{2}/m_{X}^{2}$. Therefore these new physics effects are sub-leading terms in allowed parameter space.
It is important to check the new physics effect which is caused by the modification of the CKM matrix.
The contributions from the top-mediated diagrams are written as follows:
\beq
Re M_{12}({B_{d}}) &=& C_{B_{d}} \eta_{tt}^{B}E(x_{t},x_{t}) \biggl[
\left( 1-2\frac{m_{t}^{2}\sin^{2}\theta_{R}}{m_{X}^{2}} \right)
A^{2}\{(1-\rho)^{2} - \eta^{2} \}\lambda^{6} \nonumber \\
&& + 2\frac{m_{c}}{m_{t}}|V_{R}^{42}|\sin\theta_{R}
A \{ (1-\rho)\cos\beta_{2} + \eta\sin\beta_{2} \}\lambda^{4}
\biggr] \,, \\
Im M_{12}({B_{d}}) &=& -2C_{B_{d}}\eta_{tt}^{B}E(x_{t},x_{t}) \biggl[
\left( 1-2\frac{m_{t}^{2}\sin^{2}\theta_{R}}{m_{X}^{2}} \right) A^{2} (1-\rho)\eta \lambda^{6} \nonumber \\
&& - \frac{m_{c}}{m_{t}}|V_{R}^{42}|\sin\theta_{R}
A \{ (1-\rho)\sin\beta_{2} - \eta\cos\beta_{2} \}\lambda^{4}
\biggr] \,.
\eeq
The second term is proportional to $\lambda^{7}$ due to $\frac{m_{c}}{m_{t}}\sim{\cal O}(\lambda^{3})$. Therefore, in the $B_{d}^{0} - \bar{B}_{d}^{0}$ mixing, we find that new physics contributions are smaller than the SM ones.

\begin{figure}[htb]
\begin{center}
\epsfig{file=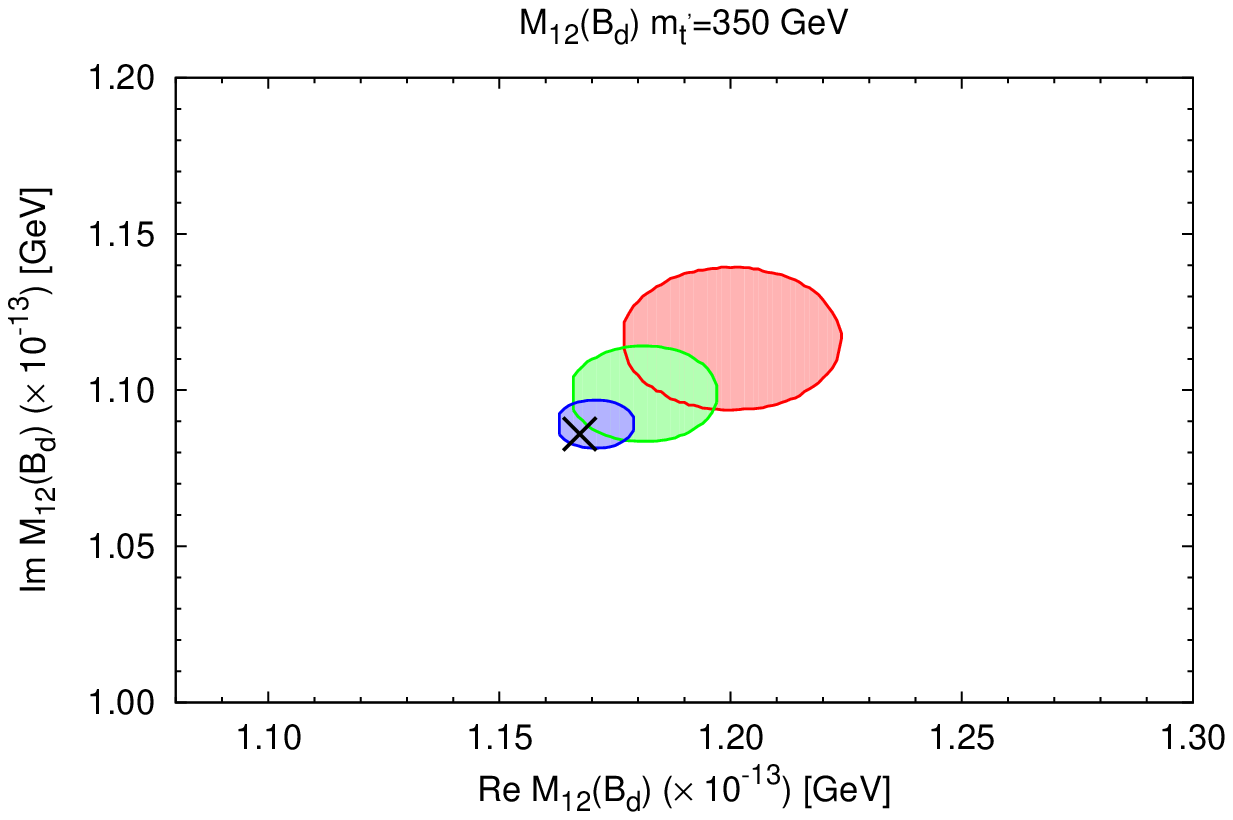,width=.45\textwidth}
\epsfig{file=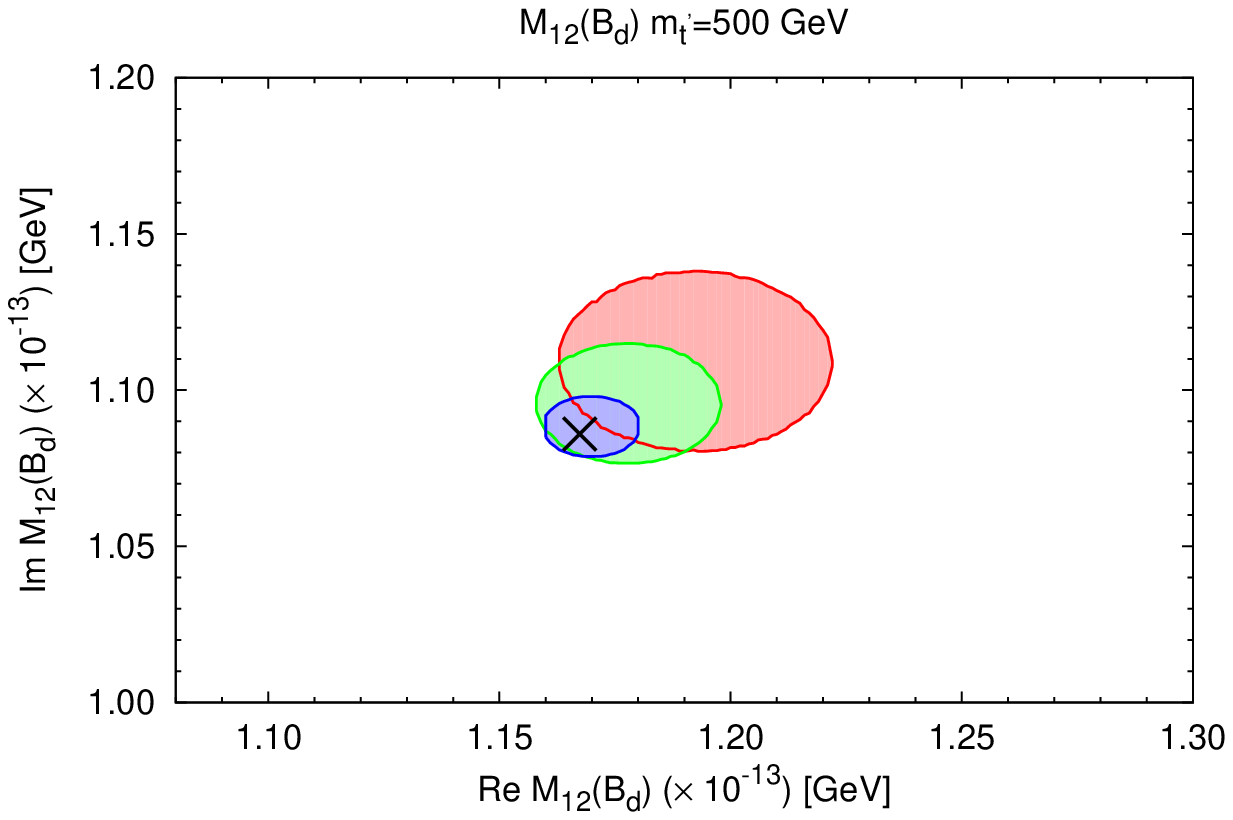,width=.45\textwidth}
\epsfig{file=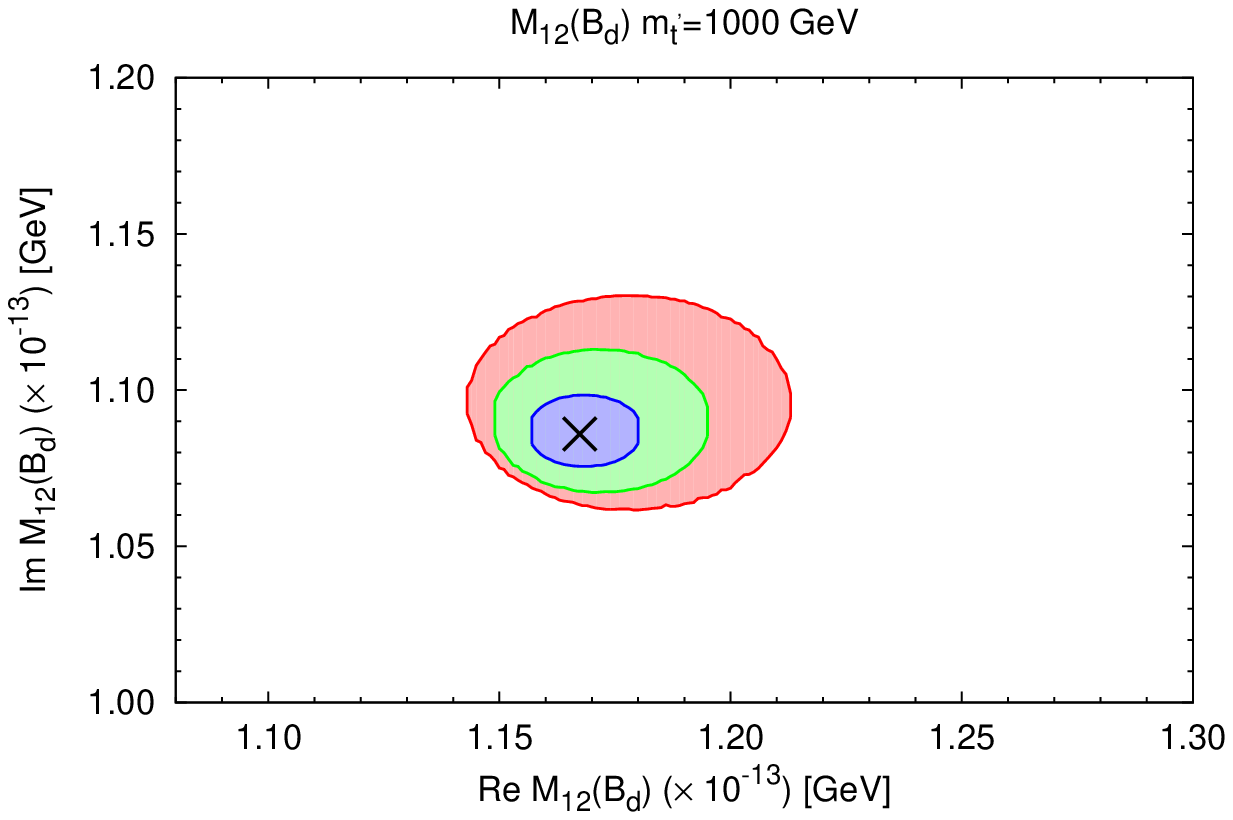,width=.45\textwidth}
\caption{Real and imaginary parts of $M_{12}(B_{d})$ for $m_{t^{\prime}}=350, 500, 1000$ GeV. The Vector-like model is shown for three different values of $\sin\theta_{R}$ (0.1 (blue), 0.2 (green), 0.3 (red)). 
The SM value is shown with a black cross. We set that $|V_R^{41}| \leq 0.078$, $|V_R^{42}| \leq 0.2$, $|V_R^{41}||V_R^{42}| \leq 
3.2 \times 10^{-4}$, $0 \leq \beta_1 < 2\pi$ and $0 \leq \beta_2 < 2\pi$.}
\label{fig:ribd}
\end{center}
\end{figure}

In the $B_d$ system, numerical results are analogous to $K^{0} - \bar{K}^{0}$ mixing as typical difference from the 
standard model value of $\Delta M_{B_{d}}$ is in the few to 10\% range. These values are below 
the theoretical uncertainty in the matrix elements.
In our numerical calculations, we set $\eta_{tt}^{B}=\eta_{tt^{\prime}}^{B}=\eta_{t^{\prime}t^{\prime}}^{B}$. Although this is not accurate, the discrepancy between our numerical results and the results including NLO QCD corrections is small because the top mediated box diagrams are the leading contributions in our model.
We plot the real and imaginary parts of $M_{12}(B_{d})$ for $m_{t^{\prime}}=350, 500, 1000$ GeV in Figure \ref{fig:ribd}. All the effects are typically small deviations with respect to the SM in the $B_d$ system.

\subsection{$B$ meson sector: $B_{s}^{0} - \bar{B}_{s}^{0}$ mixing}
\begin{figure}[htb]
\begin{center}
\epsfig{file=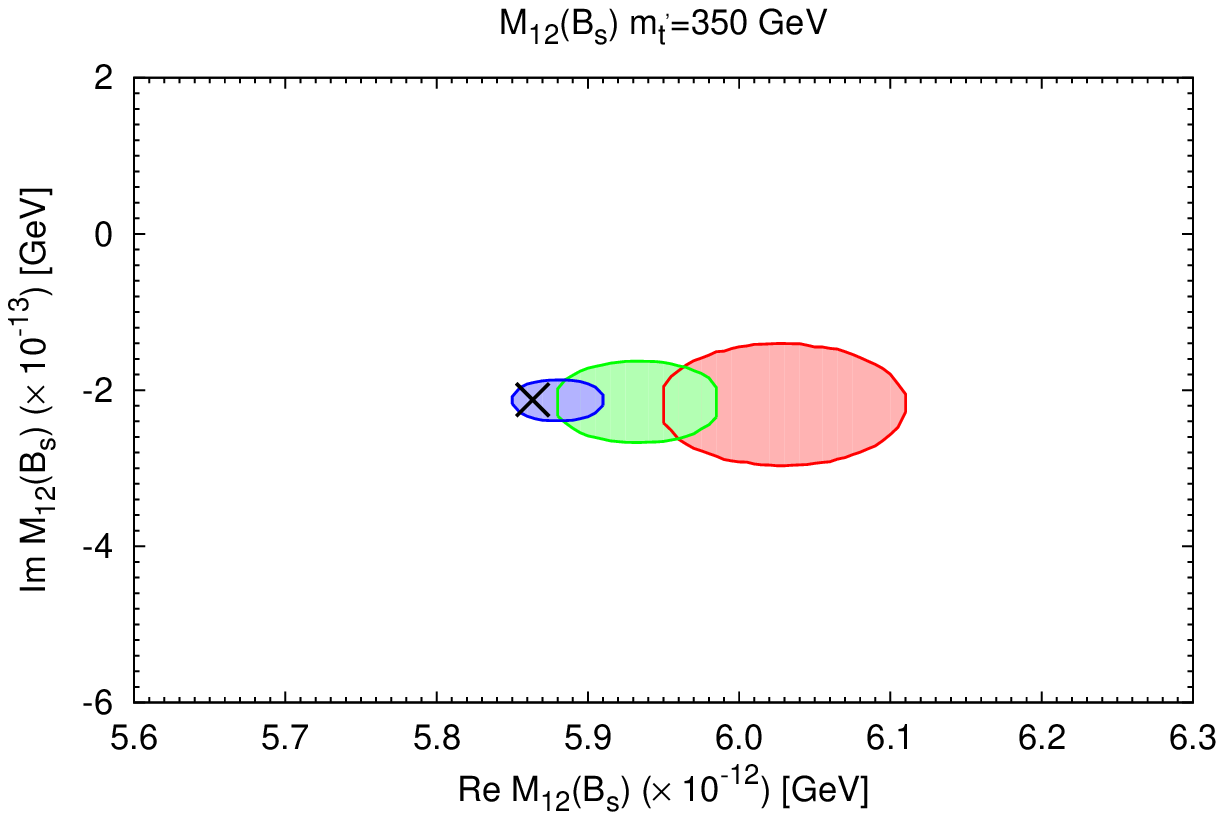,width=.45\textwidth}
\epsfig{file=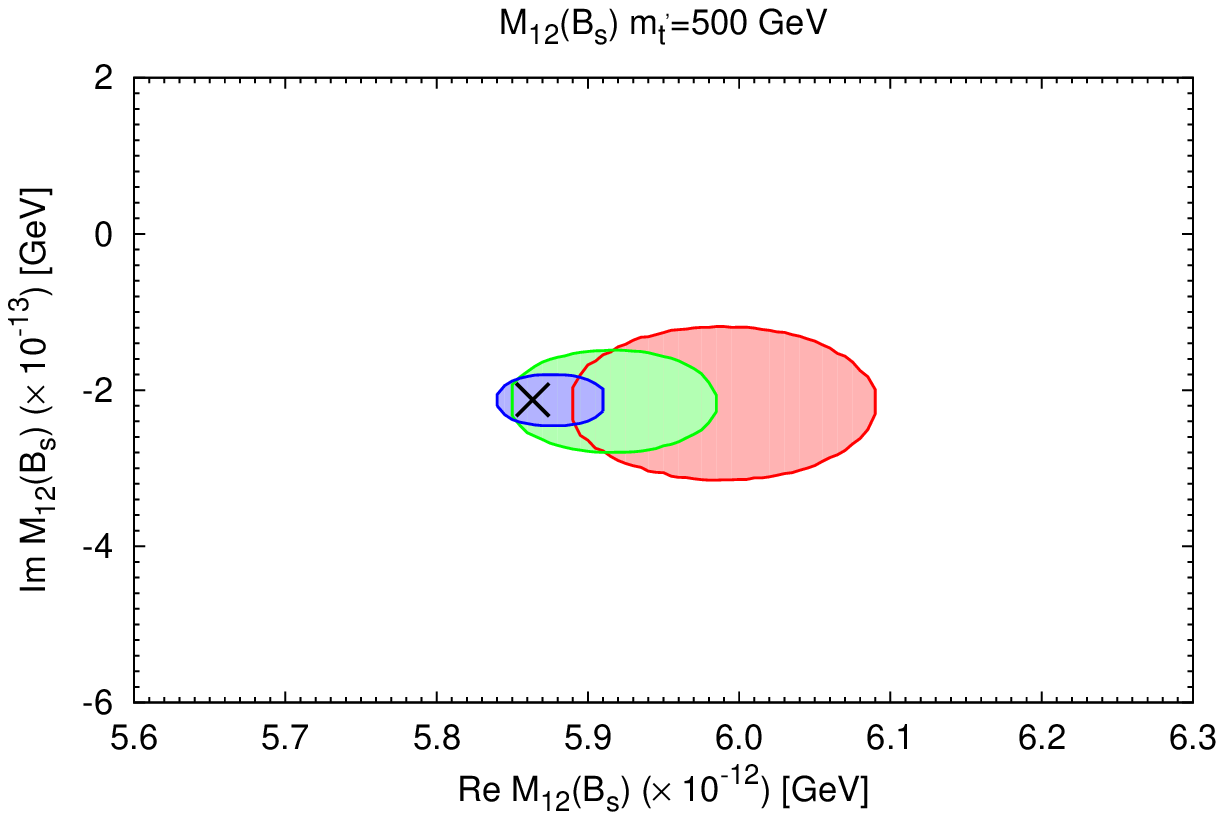,width=.45\textwidth}
\epsfig{file=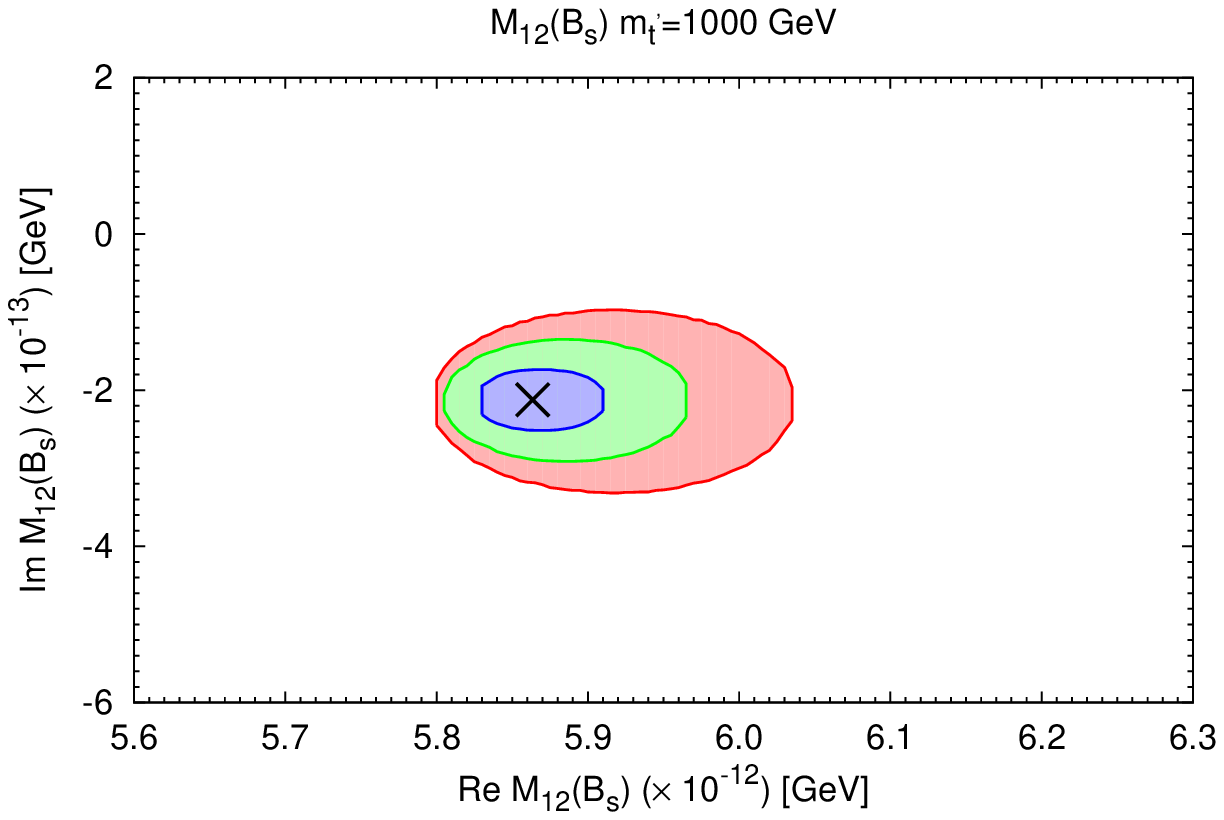,width=.45\textwidth}
\caption{Real and imaginary parts of $M_{12}(B_{s})$ for $m_{t^{\prime}}=350, 500, 1000$ GeV. The Vector-like model is shown for three different values of $\sin\theta_{R}$ (0.1 (blue), 0.2 (green), 0.3 (red)). 
The SM value is shown with a black cross. We vary the other parameters within the ranges $|V_R^{41}| \leq 0.078$, 
$|V_R^{42}| \leq 0.2$, $|V_R^{41}||V_R^{42}| \leq 
3.2 \times 10^{-4}$, $0 \leq \beta_1 < 2\pi$ and $0 \leq \beta_2 < 2\pi$.}
\label{fig:ribs}
\end{center}
\end{figure}

The experimental value of $\Delta M_{B_{s}}$ is
\begin{equation}
\Delta M_{B_{s}} = (1.170 \pm 0.008) \times 10^{-11} {\rm GeV} \,.
\end{equation}
The real and imaginary parts of $M_{12}({B_{s}})$ in the SM can be written as
\beq
Re M_{12}({B_{s}}) &=& C_{B_{s}}\eta_{tt}^{B}E(x_{t},x_{t}) A^{2} \lambda^{4} \,, \\
Im M_{12}({B_{s}}) &=& 2C_{B_{s}}\eta_{tt}^{B}E(x_{t},x_{t}) A^{2}\eta\lambda^{6} \,,
\eeq
where $C_{B_{s}}=\frac{G_{F}^{2}}{12\pi^{2}}m_{B_{s}}f_{B_{s}}^{2}B_{s}m_{W}^{2}$. In comparison with the real part $Re M_{12}({B_{s}})$, the imaginary part $Im M_{12}(B_{s})$ is suppressed by a power of $\lambda^{2}$. This is a remarkable feature of $B_{s}^{0} - \bar{B}_{s}^{0}$ mixing.
Even though the new physics effects to $Re M_{12}({B_{s}})$ are small, there is some possibility of that $Im M_{12}(B_{s})$ is largely shifted by the new CP phases.
In the vector-like model, $M_{12}(B_{s})$ can be written as
\beq
\frac{M_{12}(B_{s})}{C_{B_{s}}} &=& (V_{tb}^{*}V_{ts})^{2}
\biggl\{
\tilde{E}(x_{t},x_{t}) + 2\frac{m_{t}^{2}\sin^{2}\theta_{R}}{m_{X}^{2}} \left( \tilde{E}(x_{t},x_{t^{\prime}}) - \tilde{E}(x_{t},x_{t}) \right) \nonumber \\
&& + \frac{m_{t}^{4}\sin^{4}\theta_{R}}{m_{X}^{4}} \tilde{E}(x_{t^{\prime}},x_{t^{\prime}}) 
\biggr\} \nonumber \\
&-& 2e^{i\beta_{2}}(V_{tb}^{*}V_{ts})(V_{tb}^{*}V_{cs}) \frac{m_{c}}{m_{t}}|V_{R}^{42}|\sin\theta_{R} \nonumber \\
&& \times \left\{
\tilde{E}(x_{t},x_{t}) - \frac{m_{t}^{2}}{m_{t}^{\prime2}} \tilde{E}(x_{t},x_{t^{\prime}}) - \frac{m_{t}^{4}\sin^{2}\theta_{R}}{m_{t}^{\prime2}m_{X}^{2}} \tilde{E}(x_{t^{\prime}},x_{t^{\prime}})
\right\} \,.
\eeq
It is found that the top-mediated box diagrams become leading contribution in the $B_{s}^{0} - \bar{B}_{s}^{0}$ mixing.
The real and imaginary part of $M_{12}(B_{s})$ due to the top-mediated box diagrams can be expressed analogously to the $M_{12}(B_{d})$ as
\beq
Re M_{12}({B_{s}}) &=& C_{B_{s}}\eta_{tt}^{B}E(x_{t},x_{t}) \nonumber \\
&& \times \biggl[
\left( 1-2\frac{m_{t}^{2}\sin^{2}\theta_{R}}{m_{X}^{2}} \right) A^{2} \lambda^{4}
 +\frac{m_{c}}{m_{t}}|V_{R}^{42}|\sin\theta_{R}\cos\beta_{2}A\lambda^{2}
\biggr] \,, \label{eq:Rem12bs} \\
Im M_{12}({B_{s}}) &=& 2C_{B_{s}}\eta_{tt}^{B}E(x_{t},x_{t}) \nonumber \\
&& \times \biggl[
\left( 1-2\frac{m_{t}^{2}\sin^{2}\theta_{R}}{m_{X}^{2}} \right) A^{2}\eta\lambda^{6} +\frac{m_{c}}{m_{t}}|V_{R}^{42}|\sin\theta_{R}\sin\beta_{2} A\lambda^{2}
\biggr] \,. \label{eq:Imm12bs}
\eeq
In the imaginary part of $M_{12}(B_{s})$, this shows that the new physics contributions are sizable effects in comparison with the SM prediction for $|V_{R}^{42}|=0.2$ (case B), $\sin\theta_{R}=0.3$ and $\beta_{2}=\pm \frac{\pi}{2}$. It is found that the $|V_{R}^{41}|$ and $\beta_{1}$ dependences are negligible. This large effect completely vanish when $\beta_{2}=0,\pi$.
In case A, the correction to $Im M_{12}(B_{s})$ cannot be large due to the small value of $|V_{R}^{42}|$ ($|V_{R}^{42}| \leq 3.2\times 10^{-4}/|V_{R}^{41}|$ with $|V_{R}^{41}|=0.078$).
It is also found that the $t^{\prime}$ mass dependence of the imaginary part of $M_{12}(B_{s})$ is quite small, because the second term in Eq.~(\ref{eq:Imm12bs}) is the leading new physics contribution and does not depend on the $t^{\prime}$ mass.
In the real part of $M_{12}(B_{s})$, new physics effect is proportional to ${\cal O}(\lambda^{5})$. Therefore this effect is at least smaller than the SM value by a factor of $\lambda$.
The numerical results of real and imaginary parts of $M_{12}(B_{s})$ for $m_{t^{\prime}}=350, 500 ,1000$ GeV are shown in Figure \ref{fig:ribs}.
It is found that the deviations of $Re M_{12}(B_{s})$ are less than 5\%, while the deviations of $Im M_{12}(B_{s})$ are about 
$\pm50$\% for $\sin\theta_{R}=0.3$. These numerical results agree well with the analytical results in Eq.s~(\ref{eq:Rem12bs}) and (\ref{eq:Imm12bs}).

\begin{figure}[htb]
\begin{center}
\epsfig{file=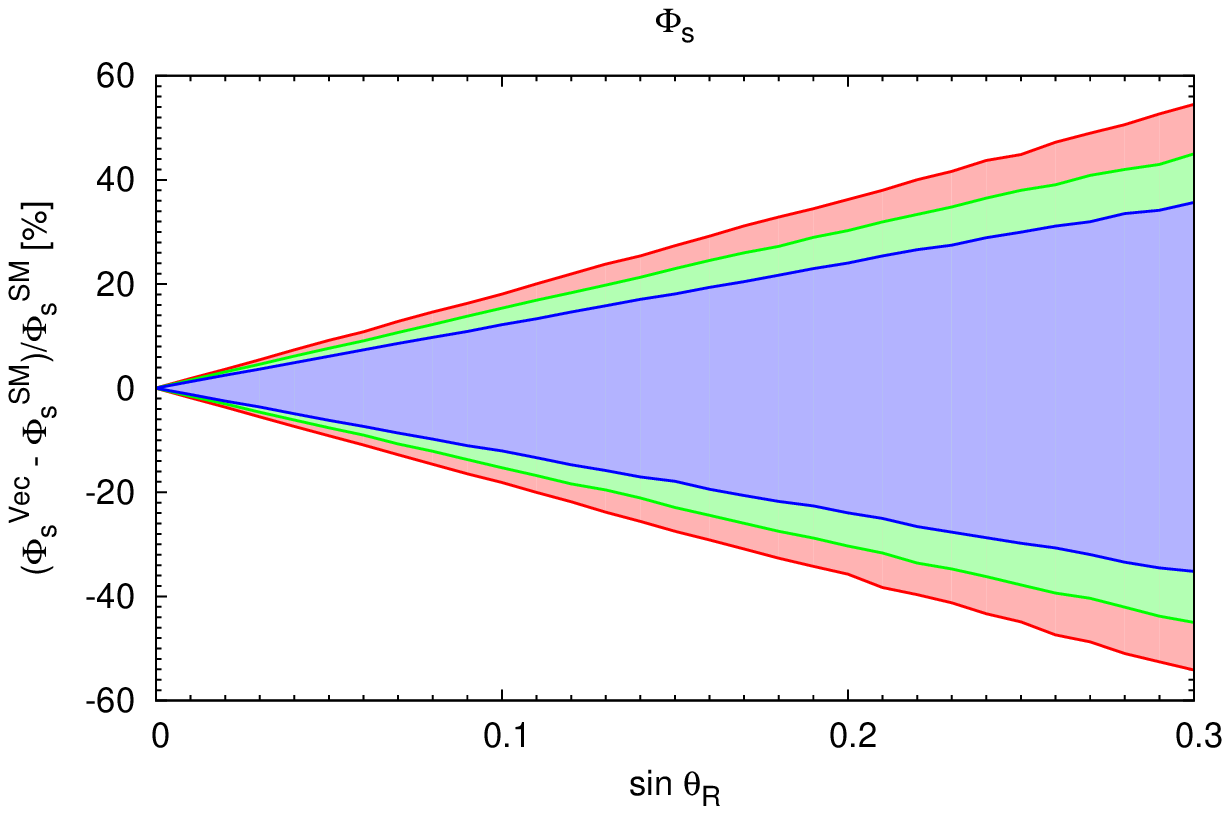,width=0.6\textwidth}
\caption{Deviation of $\Phi_{s}$ in Vector-like model from the SM value is shown as a function of $\sin\theta_{R}$
for three different values of $m_{t^{\prime}}$ (350 GeV (blue), 500 GeV (green), 1000 GeV (red)).
We vary the other parameters within the ranges
$|V_R^{41}| \leq 0.078$, $|V_R^{42}| \leq 0.2$, $|V_R^{41}||V_R^{42}| \leq 3.2 \times 10^{-4}$, $0 \leq \beta_1 < 2\pi$ 
and $0 \leq \beta_2 < 2\pi$.}
\label{fig:phisB}
\end{center}
\end{figure}

In the $B_s$ system deviations of $\Delta M_{B_{s}}$ and the phase  $\Phi_{s}$ with respect to the SM are at few 
percent level for $\beta_{1}=\beta_{2}=0$. 
When $\beta_{1}=\beta_{2}=0$ the phase  $\Phi_{s}$ in the vector-like model is very close to zero in both cases A and B.  
However the situation can be drastically different when $\beta_2$ is non-zero, see Figure \ref{fig:phisB}.
Here we define the deviation of the phase $\Phi_{s}$ in the vector-like model from the SM value as
$\Delta \Phi_{s} = (\Phi_{s}^{{\rm Vec}}-\Phi_{s}^{\rm SM})/\Phi_{s}^{\rm SM}$.
For $\beta_{2}=\frac{\pi}{2}$ $(-\frac{\pi}{2})$, it becomes maximal (minimum) value in case B.
The deviation of the phase $\Delta \Phi_{s}$ can be enhanced by $\sim +40(+60)$\% for $m_{t^{\prime}}=350(1000)$ GeV and $\sin\theta_{R}=0.3$ due to the large deviation of $Im M_{12}(B_{s})$.
For $\beta_{2}=-\frac{\pi}{2}$, it becomes negative deviation $\sim -40(-60)$\%.
A large phase is naturally allowed within our 
vector-like model with a non-standard doublet of quarks. This feature is also present in other models beyond the SM. 
A detailed study of this feature, in particular in relation with the dimuon asymmetry in $B_{s}^{0} - \bar{B}_{s}^{0}$ 
mixing will be described in a separate paper.

\begin{figure}[htb]
\begin{center}
\epsfig{file=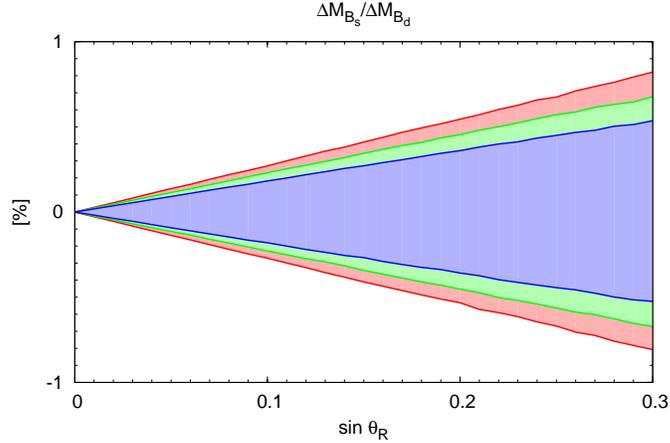,width=0.6\textwidth}
\caption{Deviation of $\left(\frac{\Delta M_{B_{s}}}{\Delta M_{B_{d}}}\right)$ in Vector-like model from the SM value is 
shown as a function of $\sin\theta_{R}$ for three different values of $m_{t^{\prime}}$ (350 GeV (blue), 500 GeV 
(green), 1000 GeV (red)).
We vary the other parameters within the ranges $|V_R^{41}| \leq 0.078$, $|V_R^{42}| \leq 0.2$, $|V_R^{41}||V_R^{42}| 
\leq 3.2 \times 10^{-4}$, $0 \leq \beta_1 < 2\pi$ and $0 \leq \beta_2 < 2\pi$.}
\label{fig:ratiodm}
\end{center}
\end{figure}

We also plot the ratio $\left(\frac{\Delta M_{B_{s}}}{\Delta M_{B_{d}}}\right)$ in Figure \ref{fig:ratiodm}, 
which is less affected by theoretical 
uncertainties. However the deviation with respect to the standard model is small (at most 1\%).

\subsection{Coupling of the Higgs to gluons and photons}

An important effect of new states that couple to the Higgs is to affect the loop-induced couplings to gluons and photons.
Phenomenologically, this is of crucial importance at the LHC, because the main Higgs production mechanism is gluon fusion and, for low masses, the golden discovery channel is in two photons.
Even though the new fermion could be heavy, the effect on the loop may be significant.

We can easily calculate the effect on the loop by using the parameterisation proposed in~\cite{Hgg}: if we neglect the contribution of the light fermions, the only contributions will come from the couplings of the $t'$ and from modifications of the top couplings.
The two parameters can be written as:
\beq
\kappa_{gg} = \kappa_{\gamma \gamma} = \frac{v}{m_t} h_{ht\bar{t}} + \frac{v}{m_{t'}} h_{ht'\bar{t}'} -1\,,
\eeq
where $ h_{hf\bar{f}}$ is the coupling of the Higgs with the fermion $f$.
From Eq.~(\ref{eq:cptp}), we have
\beq
 h_{ht\bar{t}} = \frac{m_t}{v} - \frac{M}{v} V_L^{*,43} V_R^{43}\,, \quad \mbox{and} \quad  h_{ht'\bar{t}'} =  
 \frac{m_{t'}}{v} - \frac{M}{v} V_L^{*,44} V_R^{44}\,.
 \label{eq:ht}
\eeq
Therefore:
\beq
\kappa_{gg} = \kappa_{\gamma \gamma} = 1 -  \frac{M}{m_t} V_L^{*,43} V_R^{43} -  
\frac{M}{m_{t'}} V_L^{*,44} V_R^{44} = 
\frac{(|x_1|^2 + |x_2|^2)}{m_X^2}\,.
\eeq
It is interesting to notice that this result is rather small, being proportional to the mixing with the light generations,
and does not depend significantly on the new coupling of the top $x_3$, as we may have naively expected.
The reason for this is the following: if we take the inverse of Eq.~(\ref{eq:mu}), we obtain the relation
\beq
M_u^{-1} = V_R \cdot \left( \begin{array}{cccc}
\frac{1}{m_u} & & & \\ & \frac{1}{m_c} & & \\ & & \frac{1}{m_t} & \\ & & & \frac{1}{m_{t'}}
\end{array} \right) \cdot V^\dagger_L\,.
\eeq
The component $44$ of this matrix is
\beq
(M_u^{-1})_{44} = \frac{1}{M} = \sum_{i=1}^4 \frac{1}{m_i} V_R^{4i} V_L^{*,4i}\,.
\eeq
In the case of mixing only between top and new fermion, $x_1 = x_2 = 0$, the relation reduces to
\beq
\sum_{i=3,4} \frac{M}{m_i} V_R^{4i} V_L^{*,4i} = 1\,.
\eeq
This relation implies that $\kappa_{gg} = \kappa_{\gamma \gamma}$ vanish in the 2 fermion mixing case.
From Eq.~(\ref{eq:ht}), we can extract an upper bound on the $\kappa$ parameters. 
Using the relation $|V_R^{41}|^2 + |
V_R^{42}|^2 = \frac{|x_1|^2 + |x_2|^2}{m_X^2}$, we 
obtain
\beq
\kappa_\gamma = \kappa_{gg} =(|V_R^{41}|^2 + |V_R^{42}|^2) \lesssim  0.04\,.
\eeq
The upper bound is saturated by the maximal value of $V_R^{42}$: this shows that loop effects are very small and therefore the Higgs phenomenology is  the same as in the standard model.

\section{LHC phenomenology \label{sec:lhc}}

In the following we consider the LHC phenomenology for the decay of a heavy vector-like quark. Few promising decay modes are analysed and their relevance with respect to backgrounds is discussed.
In this section we will present our numerical results for the benchmark values of the parameters in Table~\ref{tab:benchmark}.
The value of the mixing $V_R^{41}$ in particular is fixed to its maximal value allowed by precision measurements, while $V_R^{42}$ maximises the bound from the $D^0$--$\bar{D}^0$ mixing.

\begin{table}[h]
\begin{center}
\begin{tabular}{|c|c|c|c|c|c|} \hline
     & $m_{t'}$ (GeV) & $sin\theta_R$ & $V_R^{41}$ & $V_R^{42}$ & $m_H$ (GeV)\\  
value   & $350$, $500$ & $0.3$ & $0.077$ & $0.0025$ & $120$, $1000$ \\ 
\hline
\end{tabular} \end{center}
\caption{\sl Benchmark point values of the parameters. The ones with two values will be specified in each case.} 
\label{tab:benchmark}
\end{table}

\subsection{Decays}  \label{sec:decay}

\FIGURE{
\hspace*{-1.8cm}
\epsfig{file=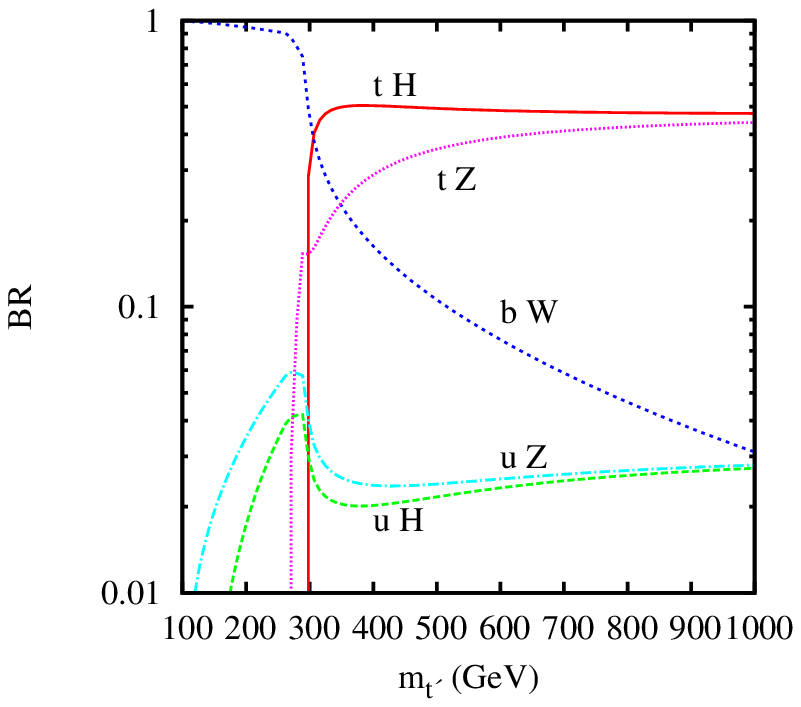,width=0.63\textwidth}
\hspace*{-2.7cm}
\epsfig{file=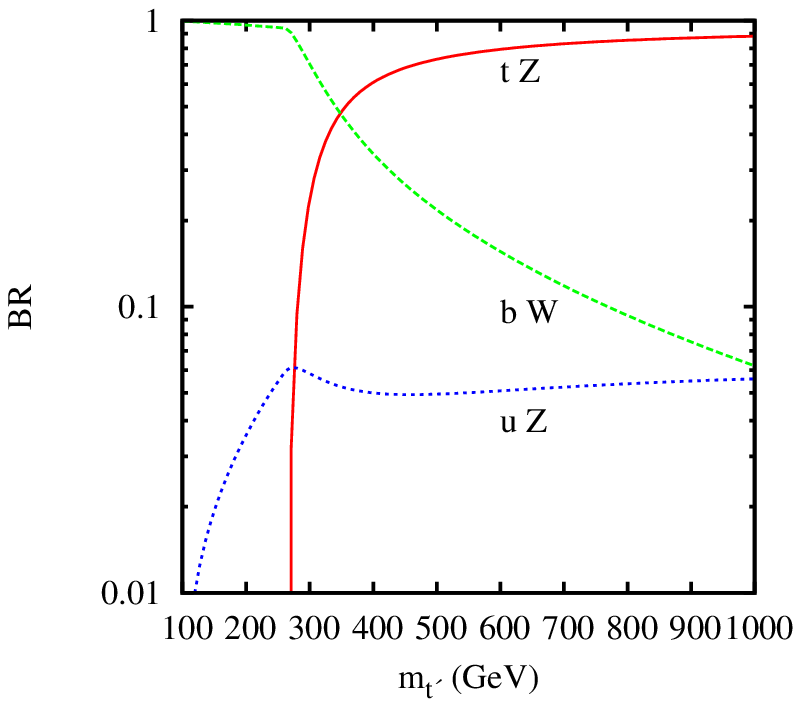,width=0.63\textwidth}
\caption{\sl BR of $t'$ as function of $m_{t'}$ for the benchmark parameters in Table~\ref{tab:benchmark}. The two panels correspond to a Higgs mass $m_h=120$ GeV (left panel) and $m_h$ decoupled from the spectrum
(right panel).}   
\label{fig:br_tp} 
}

The branching ratios for the decay of a heavy vector-like quark are quite
different from those of a sequential fourth family 
and therefore this is an important physical characteristics to
distinguish the two different situations. The phenomenology is also  
novel as vector-like quarks can decay at tree level via neutral currents, to a SM
fermions plus a $Z$/$h$, while a fourth generation quark can only decay via the $W$ boson.  
A detailed and general discussion of the parameter and case dependence is given in appendix B2 of
\cite{Cacciapaglia:2010vn}. Here, we will focus on the  
specific case of the non-standard vector-like doublet containing a new
$t'$. 
In Figure ~\ref{fig:br_tp} we show the branching rations for a light Higgs with $m_h = 120$ GeV and for a decoupled Higgs ($m_H = 1000$ GeV).
All the other parameters are fixed to the benchmark values in Table~\ref{tab:benchmark}.
\FIGURE{
\epsfig{file=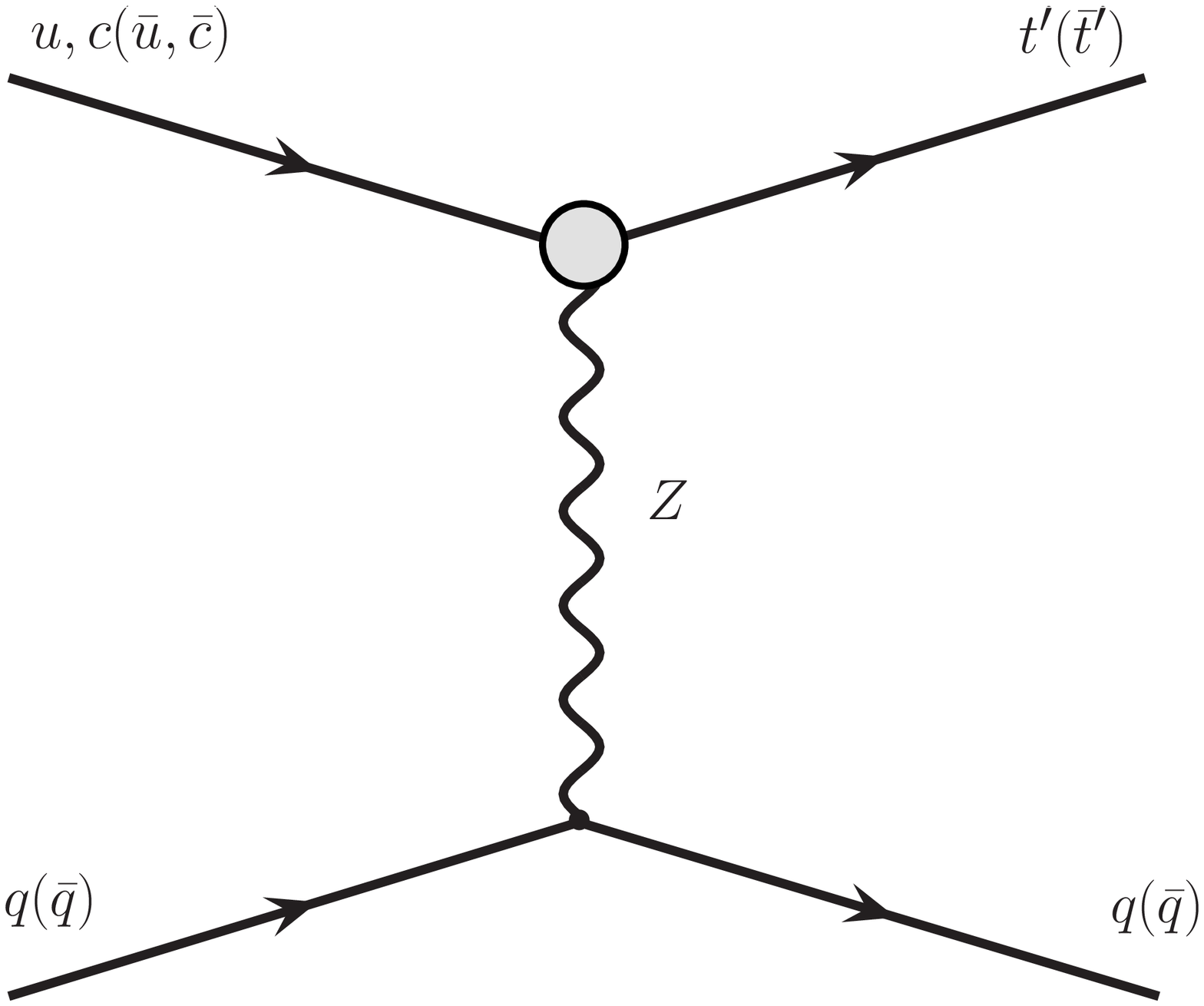,width=0.3\textwidth}
\caption{\sl Feynman diagram for the single production of $t'$ at
  LHC. } 
\label{fig:feynmandiagram1} 
}

For light $t'$ masses, below the $tZ$ and $t h$ thresholds, the decay is dominated by the $Wb$ channel, like for a fourth generation.
However, after the $t h$ channel opens up, it quickly saturates to a BR of 50\%, while $t Z$ more slowly replaces $W b$ for large masses.
This shows that the final states produced by the decays of the vector-like $t'$ are very different from the ones typical of a fourth generation.
The trend is similar in the case of a decoupled Higgs, apart for the absence of the $t h$ channel.

\subsection{Single production \label{sec:sp}}

The single production of $t'$ takes place via flavour violating interactions of the $Z$ and $W$. 
The $W$ mediated channels are similar to the ones giving single top, however the mixing angles depend on the mixing matrix in the left-handed sector and they are very suppressed.
Therefore, the production is dominated by the exchange, in t- and s-channel, of the $Z$ boson depicted in Figure~\ref{fig:feynmandiagram1} where the grey blob represents the flavour violating coupling. Such process is absent in the case of a fourth generation.
The production cross-section is shown on the left panel of
Figure \ref{fig:prod_single_tp}  as
a function of $m_{t'}$ for the benchmark parameters in Table~\ref{tab:benchmark}. 
The dominant contribution comes from the t-channel exchange of the $Z$ with a valence up quark in the initial state, therefore it is directly proportional to the mixing angle $V_R^{41}$.
The benchmark point we chose is characterised by a maximal $V_R^{41}$, therefore the values of the cross-section in the figure are the maximal values in this model.
On the right panel, we also show the pair production cross-section as a function of $m_{t'}$: one important difference is that the single production cross-section decreases more slowly compared to the pair production and, for masses above $450$ GeV it dominates at 7 TeV.
In the following we will study in detail the single production case, which seems more promising at 7 TeV, and leave the pair production case for future investigation.

In Table~\ref{table_lhc:input}, we listed the values of the cross-section and branching ratios in a few benchmark points that we will consider more in detail in the following.
As the general mixing structure is taken into account, decays to first
and second generation quarks are allowed and controlled by the mixing
parameters. 
In the case $\sin \theta_R = 0$, points 1-A and 2-A, the mixing with the top vanishes, therefore the $t'$ can only decay into a light quark plus $Z$/$h$. The decay into a $W$ is suppressed by the light quark masses.

\TABLE{ 
\begin{tabular}{|c|c|c|c|ccc|} \hline
     & $\sin\theta_R$ & ($m_{t'}, m_h$ )  
    & $\sigma$  & \multicolumn{3}{c|}{Branching Ratios of $t'$ } \\  
     &   & (GeV) &  (fb) & $t Z$ & $t h$ & $W b$ \\ \hline
BP 1 & 0.3 & (350, 120)  &  691 & 0.23 & 0.51 & 0.23  \\ 
BP 2 & 0.3 & (350, 1000) &  691 & 0.50  & -   & 0.49  \\ 
BP 3 & 0.3 & (500, 120)  &  254 & 0.36 & 0.50 & 0.10 \\
BP 4 & 0.3 & (500, 1000) &  254 & 0.77 & -  &  0.23 \\
\hline\hline 
     &         &             &      & $u Z$ & $u h$ &  \\ \hline 
BP 1-A  & 0   & (350, 120)  &  436 & 0.13 & 0.87 &  \\
BP 2-A  & 0   & (500, 120)  &  223 & 0.06 & 0.93 &  \\ \hline
\end{tabular}
\caption{\sl Benchmark points, production cross-section ($\sigma$) and
Branching ratios of $t'$ for the results of LHC simulations. } 
\label{table_lhc:input}
}

The production cross-sections and branching ratios have been evaluated with CalcHEP~\cite{Pukhov:2004ca} with
CTEQ6l PDFs and the factorization scale set to be $m_{t'}$ ($2 m_{t'}$ for the pair production). 

\FIGURE{
\epsfig{file=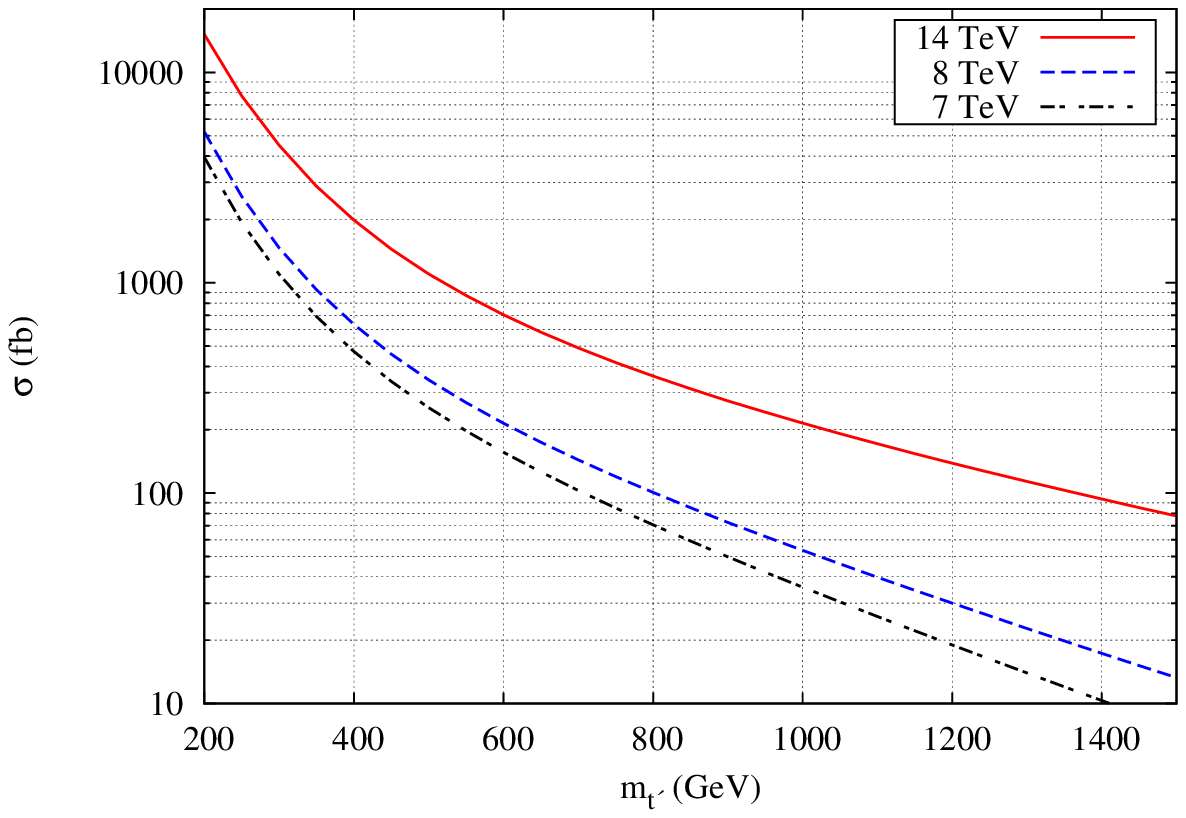,width=0.49\textwidth} \epsfig{file=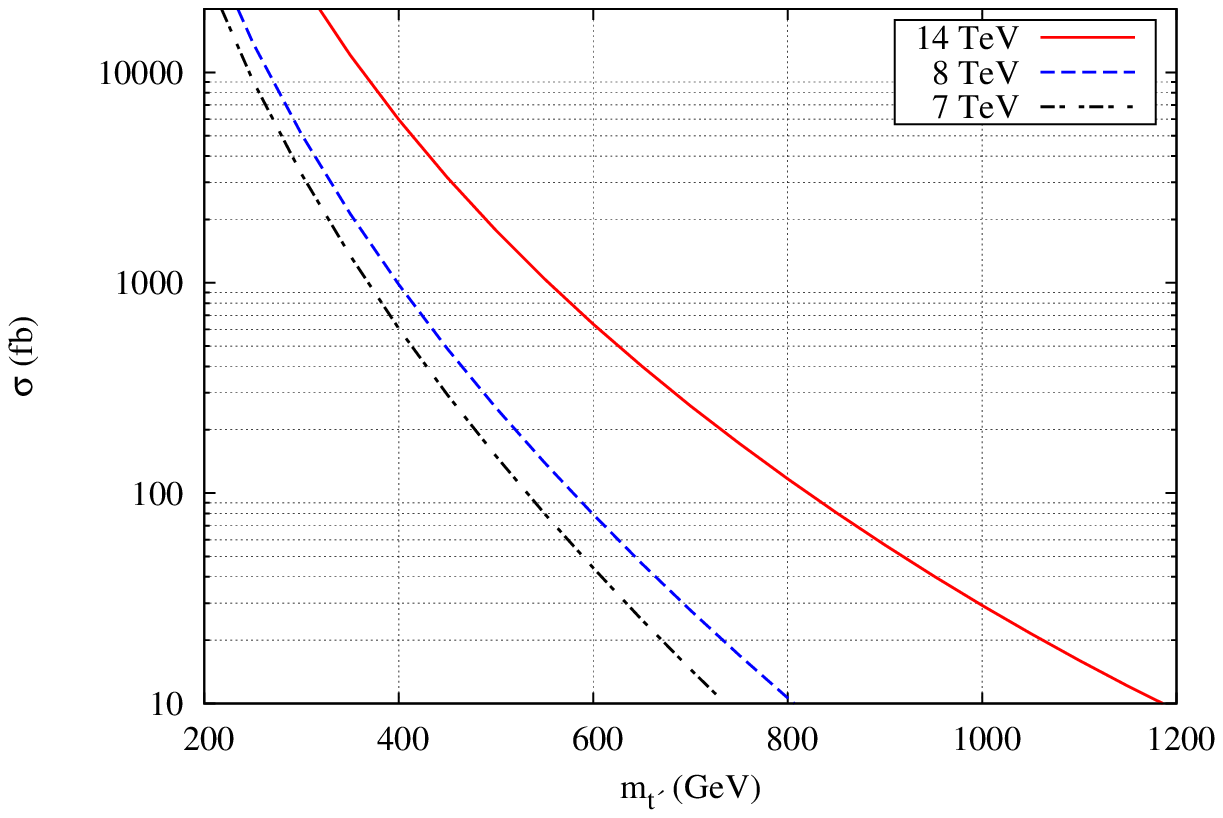,width=0.49\textwidth}
\caption{\sl Production cross-section for a single $p p \to t' j$ (left) and a pair $p p \to t' \bar{t}'$ (right) of heavy quarks as a
function of $m_{t'}$ for the benchmark parameters in Table~\ref{tab:benchmark}. We consider various energy options for the LHC.} 
\label{fig:prod_single_tp} 
}

As shown in Figure \ref{fig:br_tp} the dominant decay mode of $t'$ are $t' \to t h$ (for a light Higgs) and $t' \to t Z$. 
Therefore the possible signatures of single $t'$ production at LHC are:
\begin{enumerate}
\item{}  $p p \to b j \ell^\pm \et$ via the decay chains
$$ t' \to t (\to b \ell^\pm \nu) Z (\to \nu \bar{\nu}) \
  \to b \ell^\pm \et\,, $$
$$ t' \to b W (\to \ell \nu)\
  \to b \ell^\pm \et\,; $$
\item{} $p p \to b \ \ell^\pm \ell^\mp \ell^\pm j \et$  via the decay chain
$$ t' \to t (\to b \ell \nu) Z (\to \ell^+ \ell^-)\,; $$
\item{} $p p \to b \bar{b} b j \ \ell^\pm \et$  via the decay chain 
$$t' \to t H (\to b \bar{b}) \to t (\to b \ell \nu) b \bar{b}
\to b \bar{b} b \ \ell \et\,;$$
\item{} $p p \to b \ \ell^\pm j j j \et $  via the decay chain 
$$ t' \to t (\to b \ell \nu) Z (\to j j)\,. $$
\end{enumerate}
In addition to these channels we have also considered a very
interesting possibility when $\sin\theta_R = 0$. This corresponds to
no-mixing between $t$ and $t'$ and hence the decay channel $t' \to t
Z/h $ is closed and the dominant decay modes of $t'$ are $t' \to u
Z/h$.

In the following sections, we will discuss each of these signatures separately.

\subsubsection{Framework of event generation and
  analysis \label{sss:event} } 

The setup used for signal and background event generation is the
following: 
\begin{itemize}
\item{}{\bf Signal events: } The partonic level signal
  events were generated by CalcHEP 2.5.6 \cite{Pukhov:2004ca}. 
  The partonic level events were
  then interfaced with PYTHIA 6.4.21 \cite{Sjostrand:2006za} via the
  LHE (Les Houches Event) interface in order to include ISR/FSR and
  hadronization.  
\item{}{\bf Background events: } The $t \bar{t}$
  background was simulated using PYTHIA 6.4.21. The single top and $Z$ +
  jets backgrounds were generated using ALPGEN 2.14
  \cite{Mangano:2002ea}. The backgrounds for the signature \bjlllet  were generated
  using Madgraph \cite{Maltoni:2002qb}. 
\end{itemize}

We have used a modified version of ATLFAST \cite{atlfast} to
reconstruct the jets and identify the isolated particles. Finally the
events were analyzed within ROOT framework. For our
analysis we have considered LHC with centre of mass energy of 7 TeV and an integrated luminosity of
10 fb$^{-1}$. In our analysis leptons are intended to be only
electrons and muons ($\ell = e, \mu$).

\subsubsection{Signature $p p \to t' (\to t Z) j \to b \ j \ell^\pm
\et$ } 

This signature originates from two independent decay chains:
\beq
t' & \to&  t (\to b \ell^\pm \nu) Z (\to \nu \bar{\nu}) \
  \to b \ell^\pm \et   \nn \\
t' & \to & b W (\to \ell \nu)\
  \to b \ell^\pm \et \nn
\eeq
The effective cross-section of this process is
\beq
\sigma^{eff} &=& \sigma \times \bigg[ 
   BR(t' \to t Z) \times BR(Z \to \nu \bar{\nu}) \times BR(t \to b W )  
   + BR(t' \to b W) \bigg] \times BR(W \to \ell \nu) 
 \nn  \\
&\approx& \sigma \times \bigg[ BR(t' \to t Z) \times 0.2 \times 0.99    
    + BR(t' \to b W) \bigg] \times 0.2,,
\eeq
where $\sigma$ is the single production cross-section.
This indicates that, despite of $BR(t' \to t Z)$ being larger than
$BR(t' \to b W)$ (refer to Figure \ref{fig:br_tp}), the invisible branching ratio of the $Z$ suppresses the former contribution to the effective cross section.
For light masses $m_{t'} \lesssim 500$ GeV, the \bjlet signature is therefore dominated by the $W b$ decays.
\FIGURE{
\epsfig{file=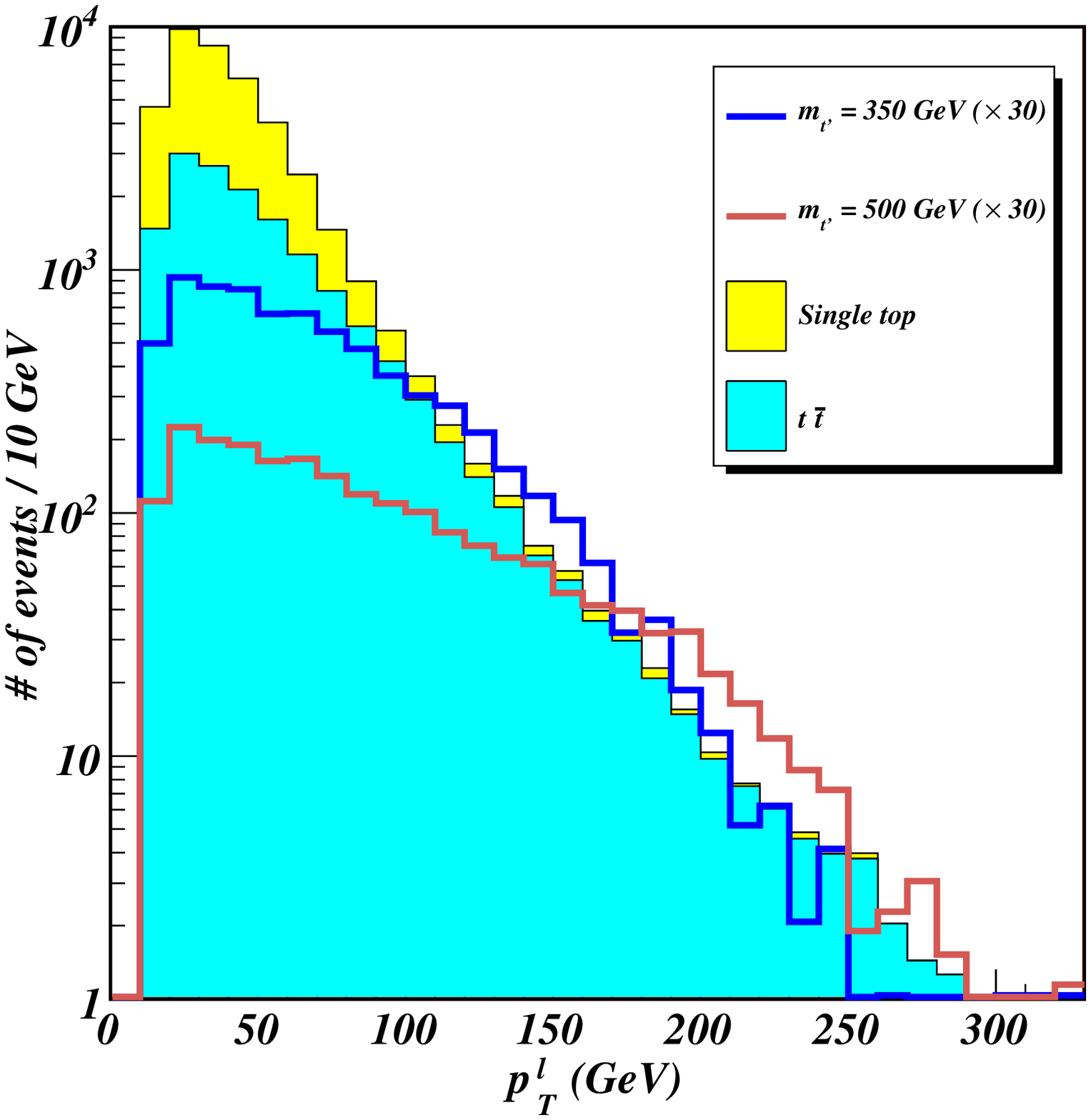,width=0.45\textwidth}
\epsfig{file=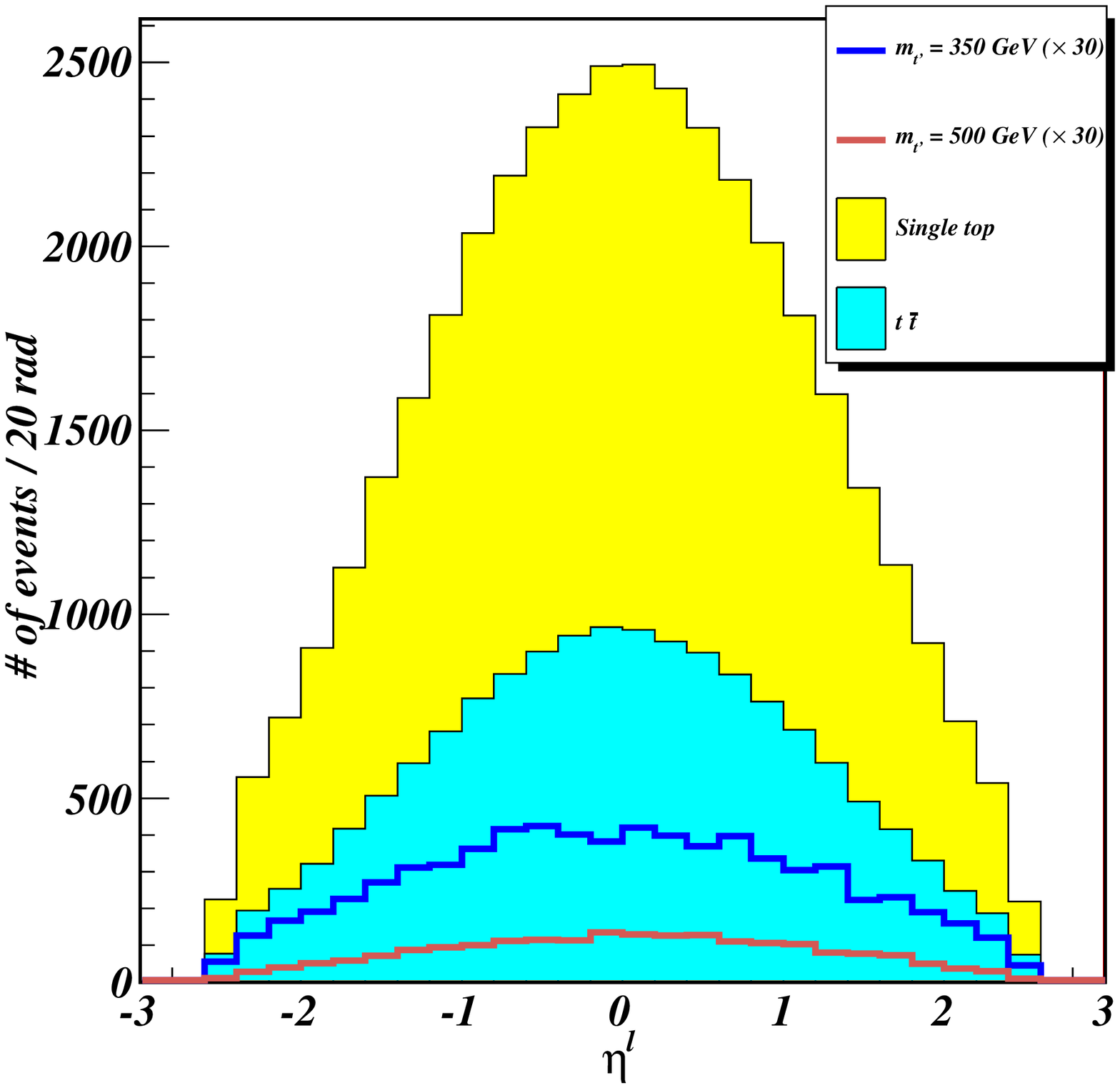,width=0.45\textwidth}
\caption{\sl \underline{\bjlet} : event distributions with respect to the $p_T$ (left) and rapidity (right) of the lepton
at the LHC with centre of mass energy of 7 TeV and luminosity of 10 fb$^{-1}$. }  
\label{fig:1}
}

\FIGURE{
\epsfig{file=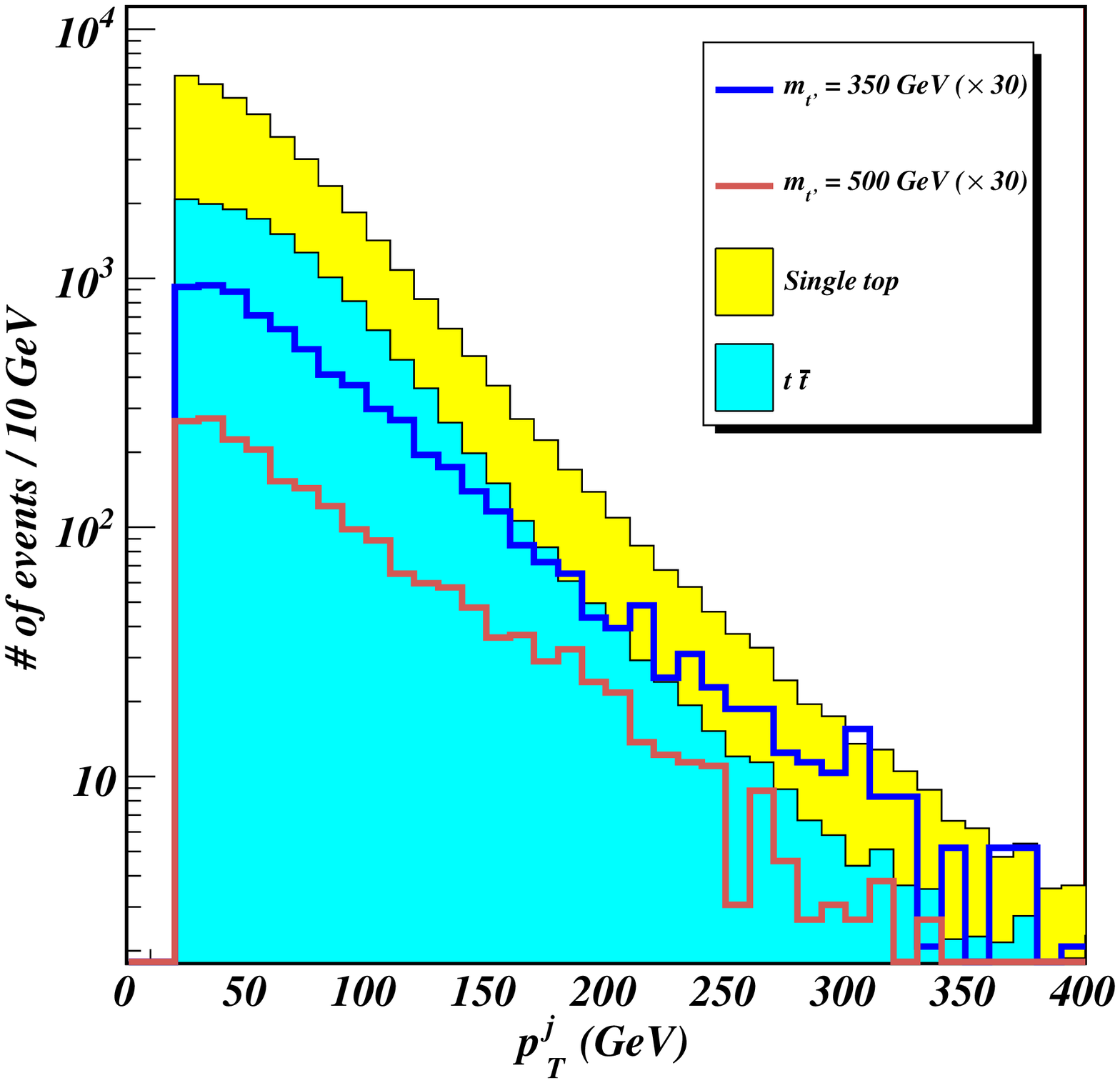,width=0.45\textwidth}
\epsfig{file=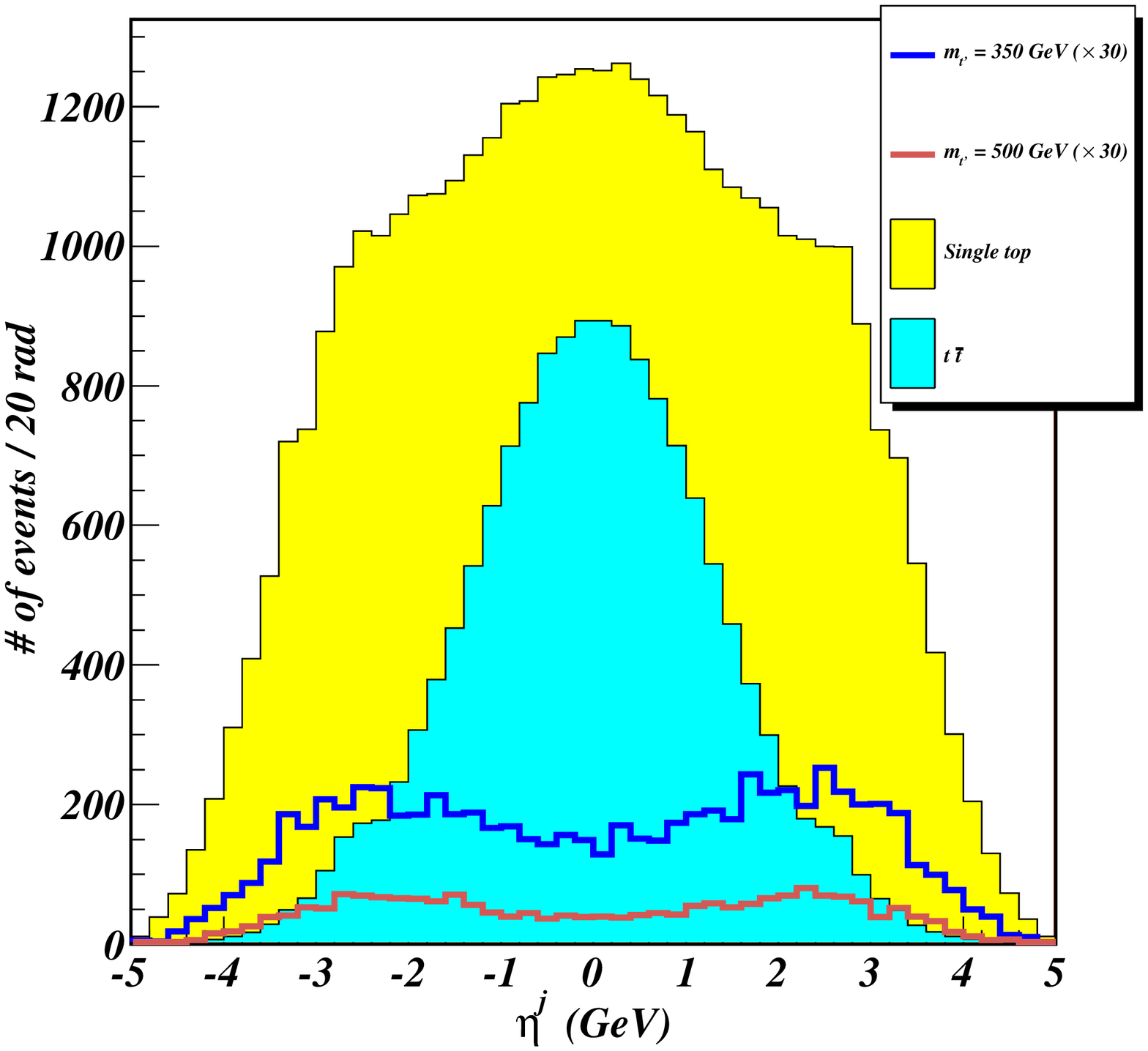,width=0.45\textwidth}
\caption{\sl \underline{\bjlet} : event distributions with respect to the $p_T$ (left) and rapidity (right) of the light
jet at the LHC with centre of mass energy of 7 TeV and luminosity of 10 fb$^{-1}$. } 
\label{fig:2}
}
\FIGURE{
\epsfig{file=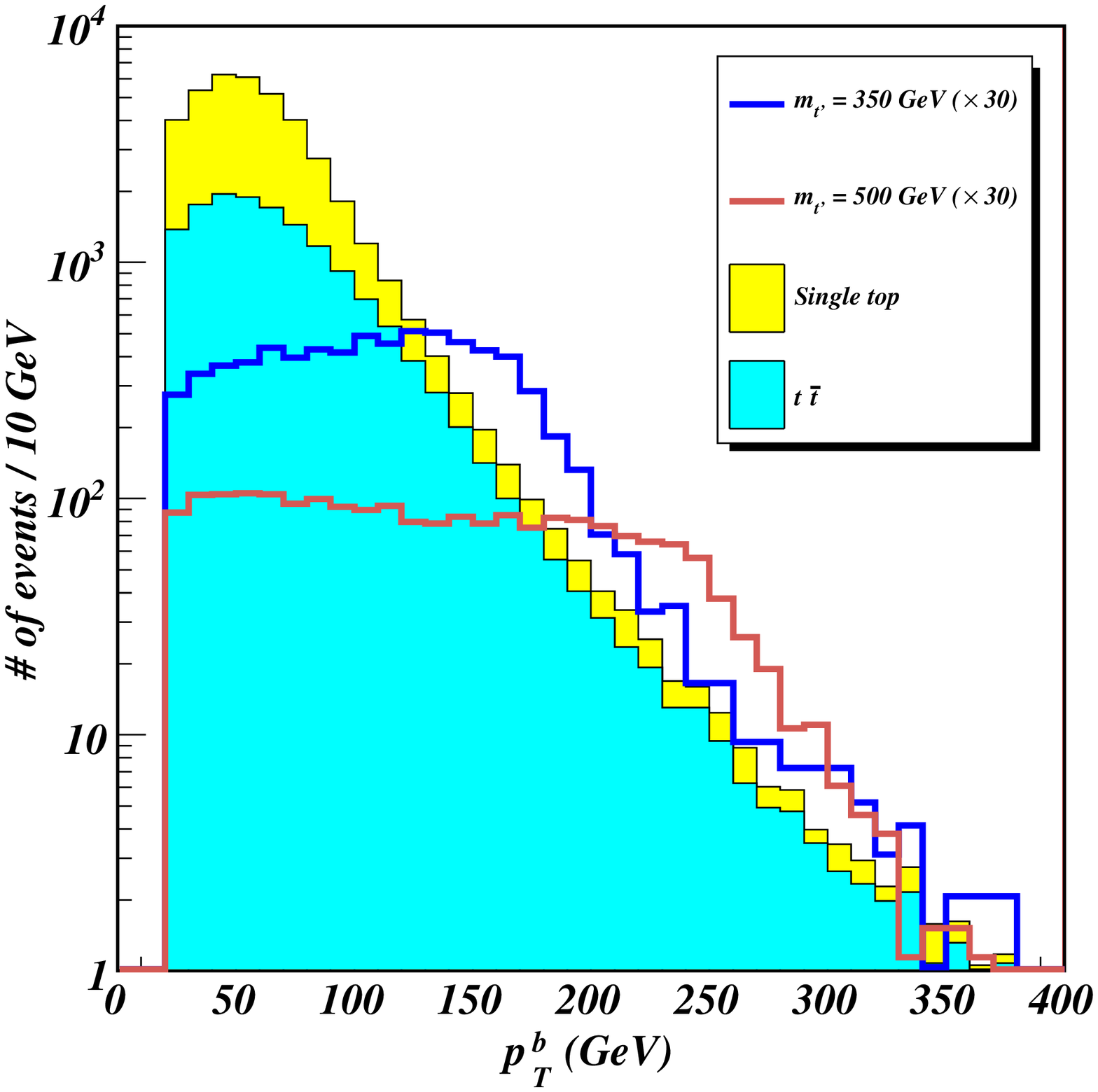,width=0.45\textwidth}
\epsfig{file=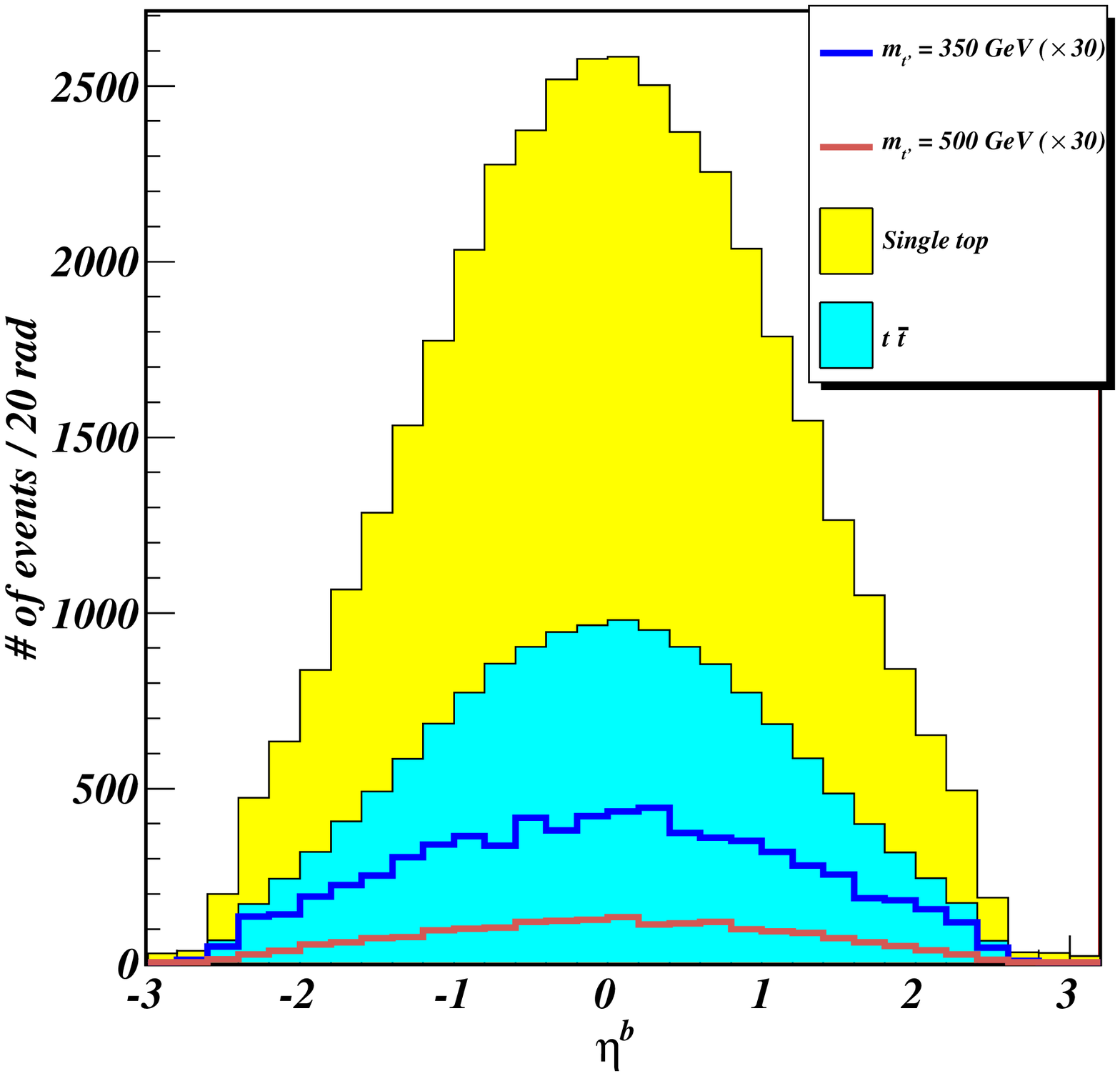,width=0.45\textwidth}
\caption{\sl \underline{\bjlet} : event distributions with respect to the $p_T$ (left) and rapidity (right) of the b-jet at the LHC with centre of mass energy of 7 TeV and luminosity of 10 fb$^{-1}$.} 
\label{fig:3}
}
\FIGURE{
\epsfig{file=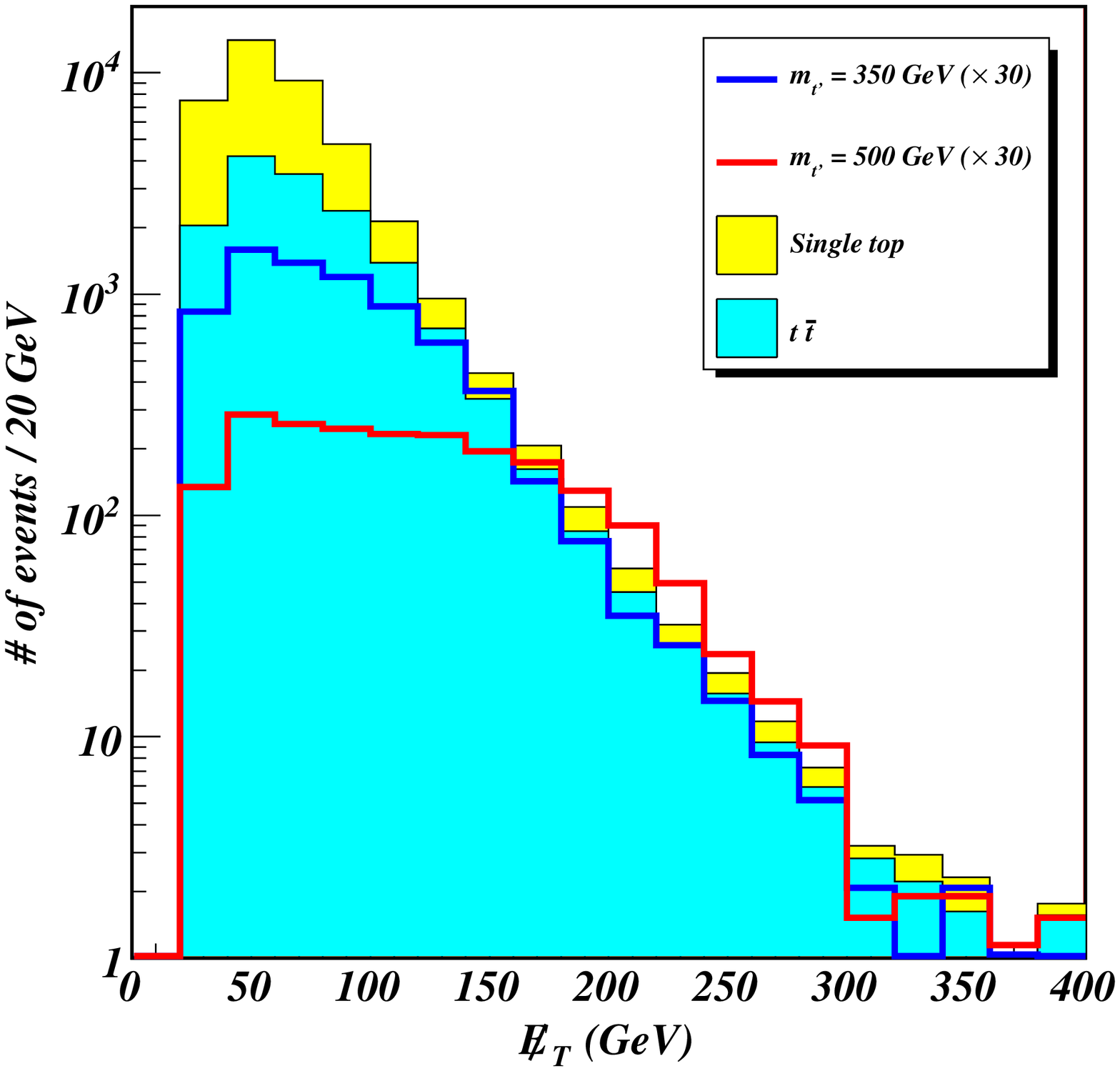,width=0.45\textwidth}
\epsfig{file=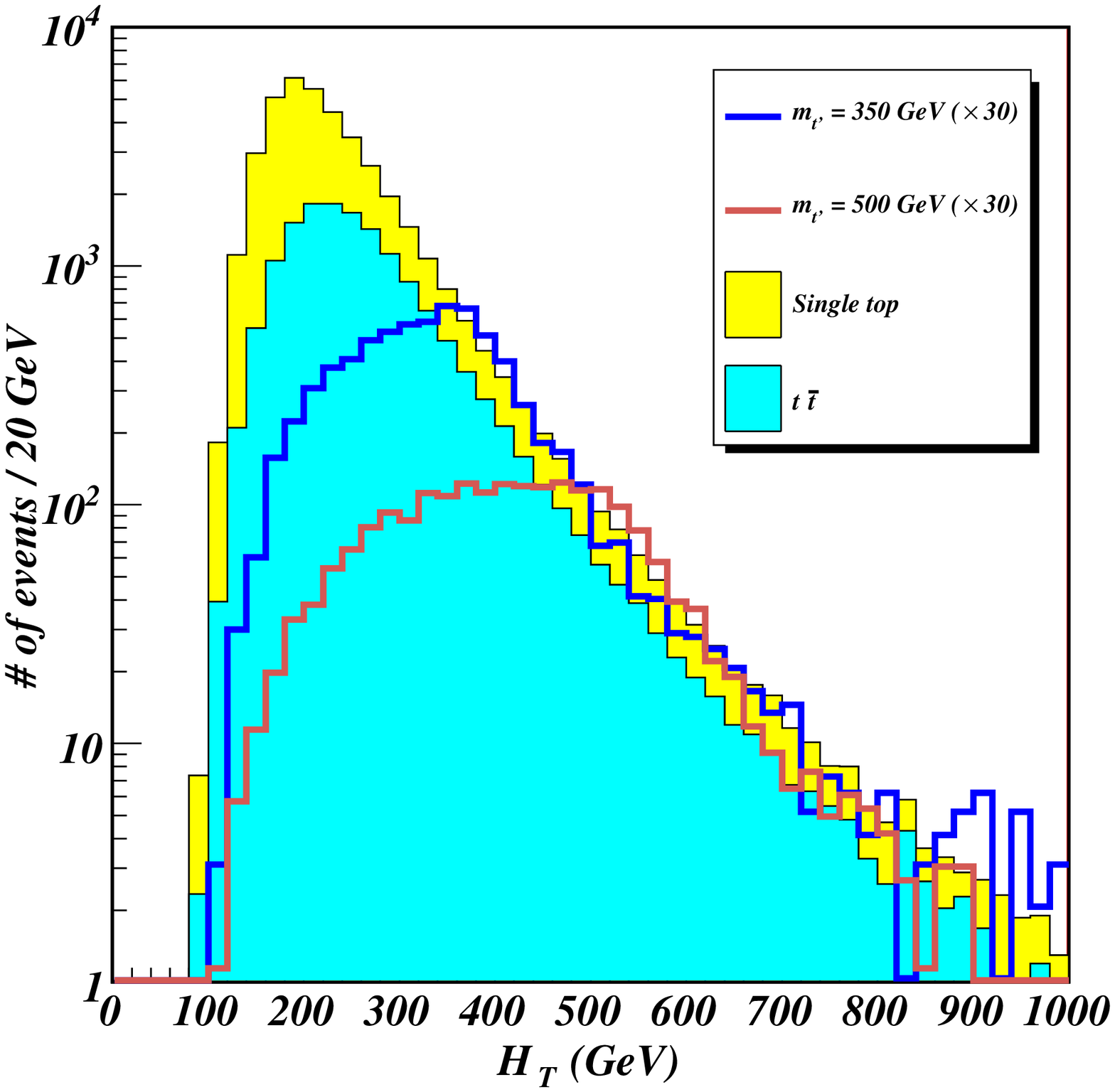,width=0.45\textwidth}
\caption{\sl \underline{\bjlet} : $\et$ (left) and $H_T$ (right) distributions  
at the LHC with centre of mass energy of 7 TeV and luminosity of 10 fb$^{-1}$. }
\label{fig:4}
}
\FIGURE{
\epsfig{file=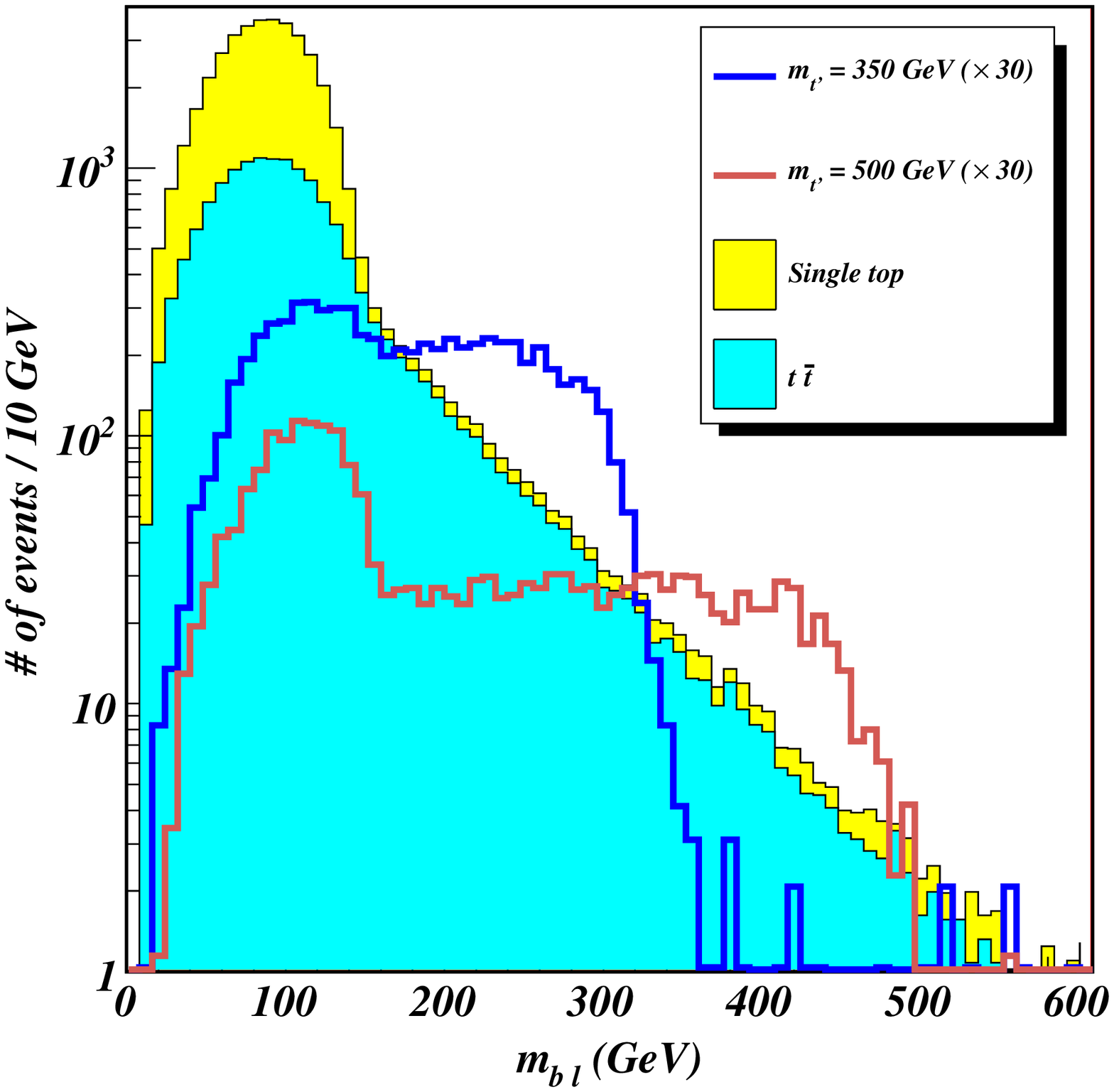,width=0.6\textwidth}
\caption{\sl \underline{\bjlet} : Invariant mass distributions of the bottom $b$ and lepton $l$ 
at the LHC with centre of mass energy of 7 TeV and luminosity of 10 fb$^{-1}$. }
\label{fig:invmassbl}
}

As shown in Figure \ref{fig:br_tp} the above decay modes are the
dominant ones for heavy Higgs masses. Accordingly, to maximise the signal, we have
estimated this signature using a Higgs mass of $m_h = 1000$ GeV. 
For a light Higgs, the branching ratios (and therefore the number of signal events) are approximately halved.
One of the advantages of this process is that there is only one b-jet to be tagged and large missing transverse energy (MET) from the neutrinos could allow us to distinguish the signal from
backgrounds. 

In order to have an idea of the relevance of this channel at 7 TeV LHC we have perfomed a simplified analysis. The {\sl pre-selection} cuts we used are:

\begin{itemize}
\item{} Exactly one isolated lepton with transverse momentum $p_T > 15$ GeV and rapidity
$|\eta| < 2.5$.
\item{} Exactly one b-tagged jet with $p_T > 20$ GeV and $|\eta| < 3$. We assume a b-tagging efficiency of 60\%.
\item{} Exactly one additional light jet with $p_T > 20$ GeV and
$|\eta| < 5$.
\item{} Minimum MET threshold: $\et > 30$ GeV.
\end{itemize}

\noindent The distributions after pre-selection cuts are shown in Figures \ref{fig:1},
\ref{fig:2}, \ref{fig:3}, \ref{fig:4} and \ref{fig:invmassbl}. An important feature
of the distributions is that for signal events the $p_T$ of the light jet is relatively smaller
than of the b-jet. On the other hand, the rapidity of light jet tends to be higher as
compared to the b-jet. 
Another important feature is the distribution of the invariant mass of the bottom quark and the lepton in Figure~\ref{fig:invmassbl}: this quantity has been used in the CMS search for pair production of fourth generation quarks decaying into $W b$ to remove the top background~\cite{CMSbW}.
In fact, for a $b$ and lepton from a top leptonic decay, we would expect the invariant mass to be smaller than the top quark mass.
In Figure~\ref{fig:invmassbl} we can see that the $t \bar{t}$ and single top backgrounds do indeed peak below $m_{top}$.
For the signal, there is also a peak below the top mass coming from the $Z t$ decay modes, while the $W b$ events leak above the top mass.
A cut $m_{bl}>170$ GeV could therefore help reducing the top backgrounds for small $t'$ masses, while for large $t'$ masses the decays are dominated by the $Z t$ mode which is severely removed by this cut.
As it can be seen from Figure \ref{fig:4.5}, the two decay chains $t' \to b W$ and $t'\to t Z$ can also be distinguished from the
$p_T$ of b-jet: the chain $t' \to b W$ typically gives larger $p_T$ to
the b-jet. The MET distribution can also be used for
distinguishing the decay chains. The decay $t' \to b W$ typically
shows larger MET as compared to $t' \to t Z$.

In order to further improve the significance of our results we have
imposed the following {\sl secondary cuts}:  
\begin{itemize}
\item{} $p_T^b > 100$ GeV,
\item{} $m_{bl} > 170$ GeV,
\item{} $\et > 100$ GeV. 
\end{itemize}
The results of signal and backgrounds after the cuts mentioned above are
shown in Table \ref{table:lhc_bjl_btag} and in Figure \ref{fig:invmassbl}. 
The table shows that this mode is
problematic at 7 TeV LHC. The reason is that 
the heavy vector-like quark masses explored here (up to $500$ GeV) have a
typical signature which is too close to the SM single top signal. We
expect that higher mass vector-like quarks will be easier to disentangle
from background because of a larger $pT$ of the b-jet and larger MET, however due to the smaller cross-section these may be
accessible only at higher centre of mass energy of higher luminosities at the LHC.  
\FIGURE{
\epsfig{file=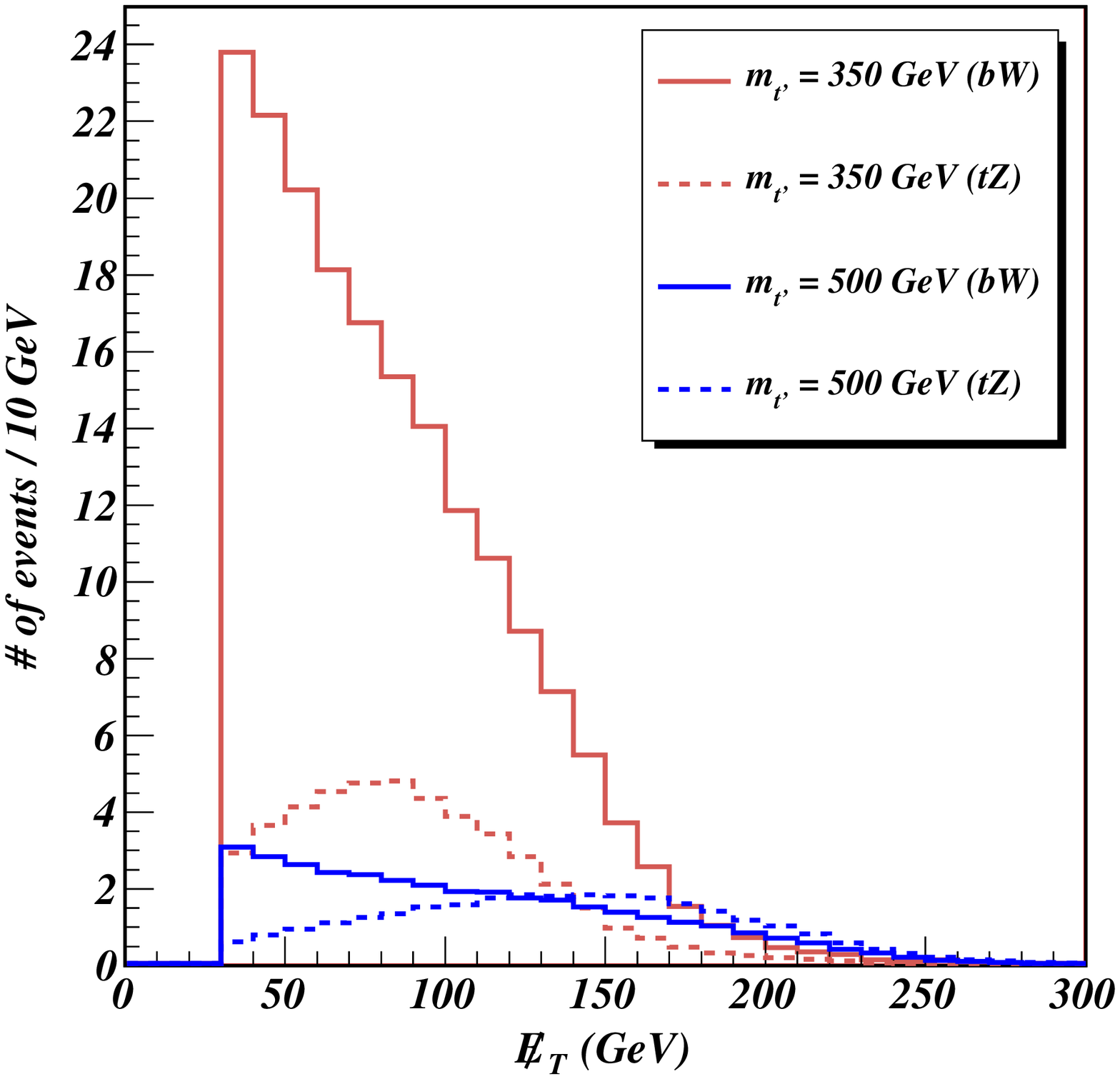,width=0.45\textwidth}
\epsfig{file=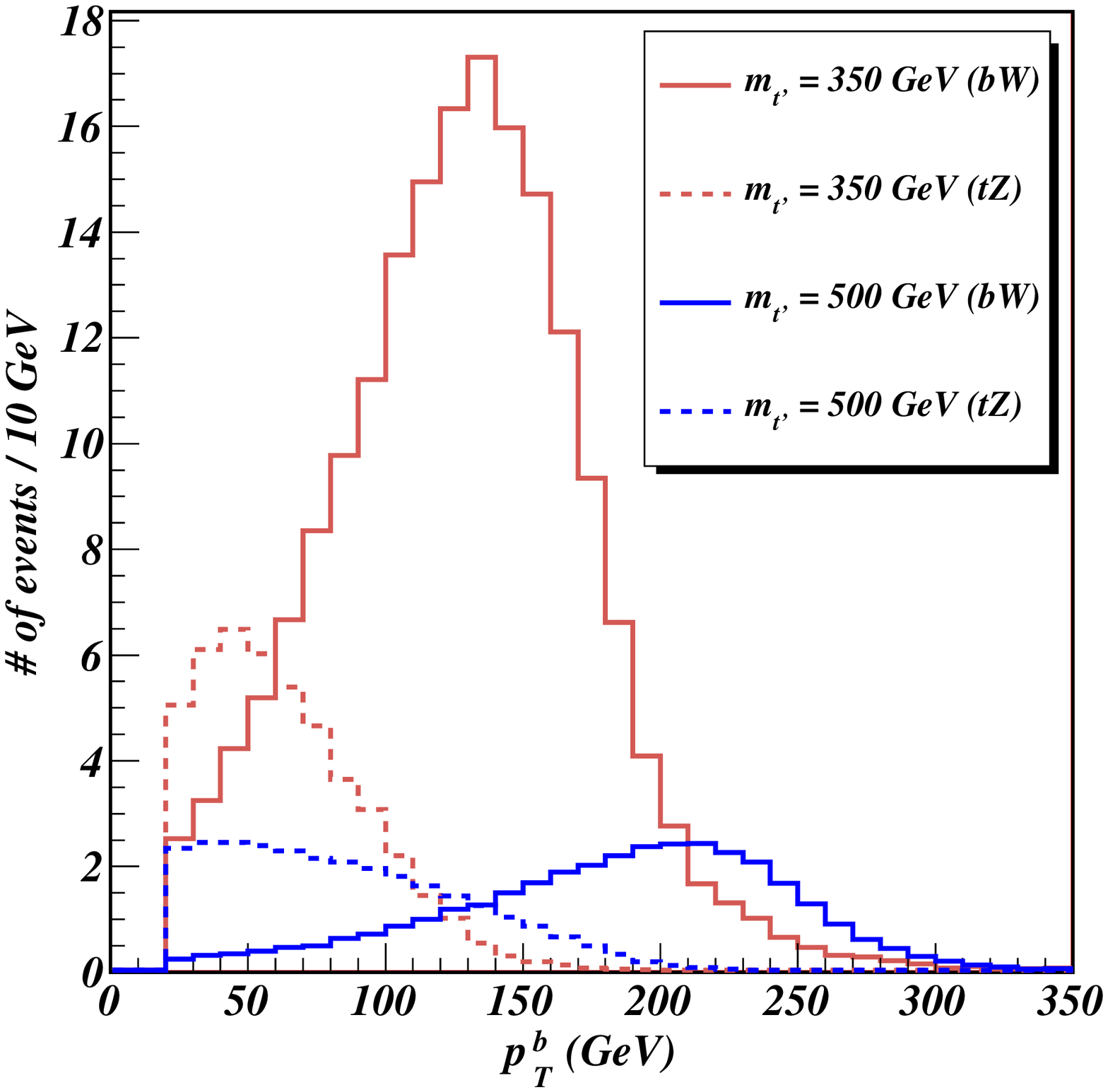,width=0.45\textwidth}
\caption{\sl \underline{\bjlet} : $\et$ (left) and $p_T$ of the b-jet (right) distributions  
at the LHC with centre of mass energy of 7 TeV and luminosity of 10 fb$^{-1}$. 
The contributions of the $W b$ and $Z t$ channels are plotted separately.}
\label{fig:4.5}
}

\TABLE{ 
\begin{tabular}{|ccccc|c|} \hline
Cut                       & BP 2 & BP 4 & BP 1 & BP 3 &  SM   \\ \hline
Pre-selection             & 238.9 &  69.4 & 144.8 &  41.8 & 39486   \\
$S/\sqrt{S+B}$ & 1.20 & 0.35 & 0.73 & 0.21 &  -  \\  \hline
$p_T^b > 100$ GeV         & 137.9 &  52.6 &  98.6 &  25.9 &  4053.6 \\
$S/\sqrt{S+B}$ & 2.13 & 0.82 & 1.53 & 0.40 &  - \\ \hline

$m_{lb} > 170$ GeV         & 117.7 &  31.7 & - &  - &  1951.1 \\
$S/\sqrt{S+B}$ & 2.59 & 0.71 &  &  &  - \\ \hline

Both previous        & 90.8 &  28.0 & - &  - &  992.8 \\
$S/\sqrt{S+B}$ & 2.76 & 0.86 &  &  &  - \\ \hline
 
$\et > 100$ GeV           &  42.3 &  24.8 &  22.5 &  14.1 &   985.3 \\ 
$S/\sqrt{S+B}$ & 1.32 & 0.78 & 0.71 & 0.45 &  - \\ \hline
\end{tabular}
\caption{\sl \underline{\bjlet} : Number of signal events after the cuts at the LHC with 
centre of mass energy of 7 TeV and luminosity of 10 fb$^{-1}$.}
\label{table:lhc_bjl_btag}
}
\subsubsection{Signature $p p \to t' (\to t Z) j \to b \ell^\pm
\ell^\mp \ell^\pm \ j \et$ } 

\begin{figure}[htb]
\begin{center}
\epsfig{file=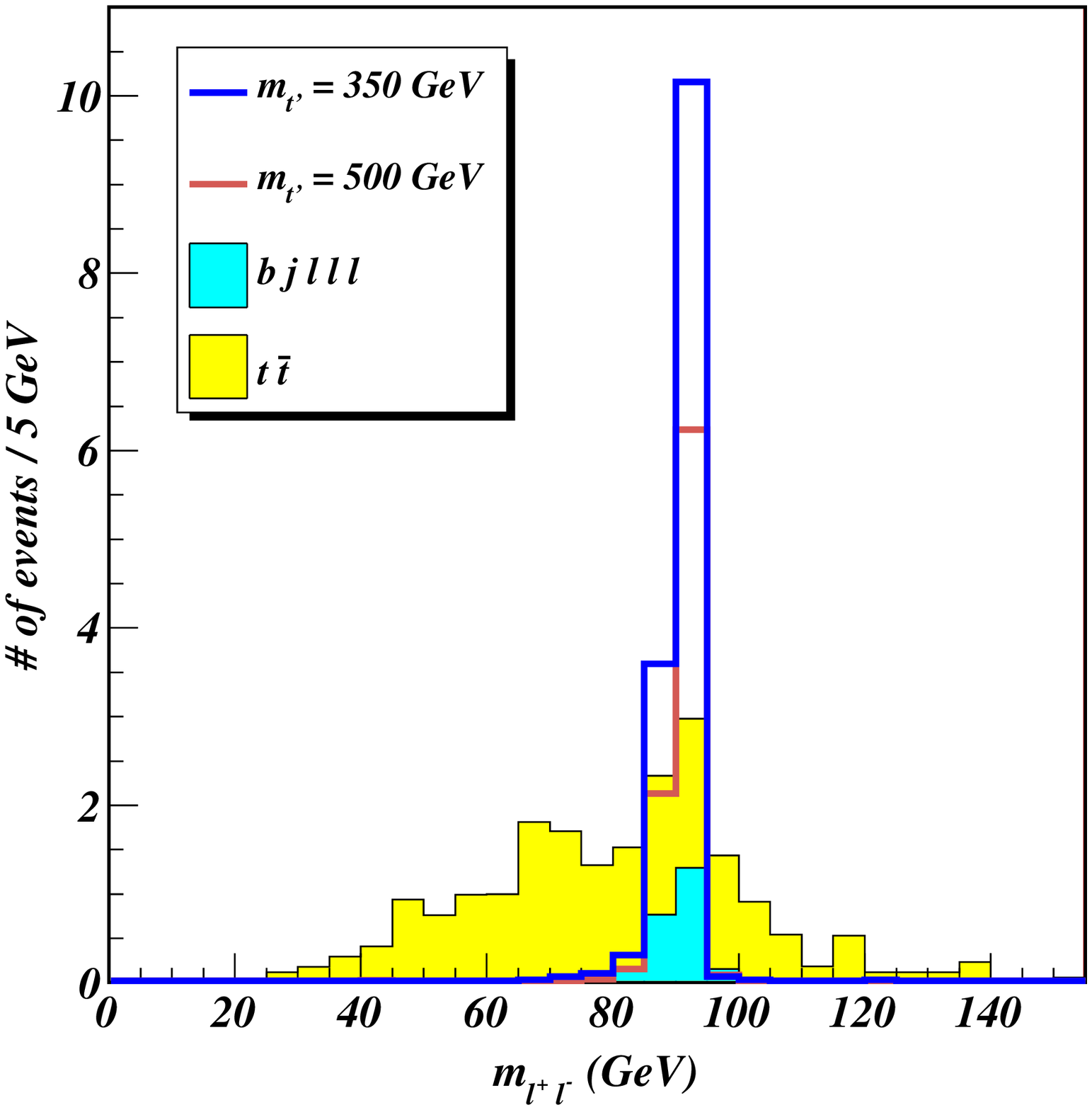,width=0.45\textwidth}
\caption{\sl \underline{\bjlllet} : Opposite sign dilepton invariant mass
($m_{\ell^+ \ell^-}$)  distribution at the LHC with centre of mass energy of 7 TeV
and luminosity of 10 fb$^{-1}$. The correct combination of leptons has been selected by minimising $\chi^2$ as defined in the text.} 
\label{fig:5}
\end{center}
\end{figure}

The trilepton channel is regarded as one one of the golden
discovery channels because of the low SM background and very low probability of faking it with
jets. In our case, only one tagged b-jet is required, hence it does not loose much on
tagging efficiency. This signature is obtained from the leptonic decay
of the Z boson: $Z \to \ell^+ \ell^-$. One can also associate the whole   
MET as originating from the decay of $t$
and hence one can fully reconstruct the top and $t'$. 
 For the analysis we have used the following {\it pre-selection cuts}:
\begin{itemize}
\item{} Exactly three (two of same sign) isolated leptons with
$p_T > 20$ GeV and $|\eta| < 2.5$. 
\item{} Exactly one b-tagged jet with $p_T > 20$ GeV and $|\eta| < 3$. We assume a b-tagging efficiency of 60\%.
\item{} One additional light jet with $p_T > 20$ GeV and $|\eta| < 5$.
\end{itemize}
Requiring three charged lepton with sufficiently high $p_T$ helps in
reducing the $t \bar{t}$ background where one of the lepton comes from
the decay of the b-jet.

We can reconstruct the $Z$-boson in dileptonic decay. We
have three leptons in the final state and hence we can have two
combinations of opposite sign dileptons. The correct combination can
be chosen by selecting the one that minimise the quantity:
$$
\chi^2 = (m_Z - m_{\ell^+ \ell^-})^2\,.
$$
To further reduce the backgrounds we impose that the invariant mass of
the opposite sign dileptons should be close to $m_Z$. We have shown
the opposite sign dilepton distribution in Figure \ref{fig:5}. The
results of signal and background events are given in Table
\ref{table:lhc_bjlll}. Our results show that it is indeed possible to
probe $t'$ in this channel at 7 TeV LHC with a luminosity of 10
fb$^{-1}$.

\TABLE{
\begin{tabular}{|ccccc|c|} \hline
Cut                       & BP 2 & BP 4 & BP 1 & BP 3 &  SM   \\ \hline
Pre-selection             & 14.4 & 8.8  & 6.5  & 4.2  & 20.7  \\ 
$S/\sqrt{S+B}$            & 2.43 & 1.62 & 1.25 & 0.84 & -     \\ \hline 
$|m_{\ell^+ \ell^-} - m_Z| < 5$ GeV & 13.6 & 8.3 & 5.7 & 4    &  3.8 \\ 
$S/\sqrt{S+B}$                      & 3.3  & 2.4 & 1.8 & 1.43 & - \\ \hline 
\end{tabular}
\caption{\sl \underline{\bjlllet} : Number of signal events after the cuts
at the LHC with centre of mass energy of 7 TeV and luminosity of 10 fb$^{-1}$. }  
\label{table:lhc_bjlll}
}

\subsubsection{Signature $p p \to t' (\to t H) j \to b \bar{b} b \ j
\ell^\pm \et$ } 

This signature requires the presence of three tagged b-jets. This
makes the detection of this signature challenging, especially at low
energy and luminosities. The largest backgrounds to this channel might
come from $t \bar{t}$ production where one of the light jets is
misidentified with a b-jet. 

The additional problem in this channel comes because of the presence
of three b-jets and hence, due to combinatorial background problems, it
might be difficult to reconstruct the Higgs boson.

\subsubsection{Signature $p p \to t' (\to t Z) j \to b \ j j j \ell^\pm
\et$ } 

Although this channel will probably have the highest effective
cross-section (when we take BRs into account), one has to face the problem of the $Z$ reconstruction.
It would be possible to
reconstruct the $Z$ invariant 
mass using $Z \to j j$, however the reconstruction might not be very good
because of the presence of an additional light jet in the event which
will give rise to combinatorial backgrounds. 
Even if we were able to reconstruct Z, we may not be able to reduce
backgrounds with $W \to j j$ as this too 
occurrs around the $Z$ mass and the energy resolution with jets is
typically not very good. Such backgrounds could possibly come from
$t\bar{t}$ and it would be quite challenging to reduce them.

\subsubsection{Signature $p p \to t' (\to j H) j
\to j j \bar{b} b$, in the case $\sin\theta^R = 0$.}   

\FIGURE{
\epsfig{file=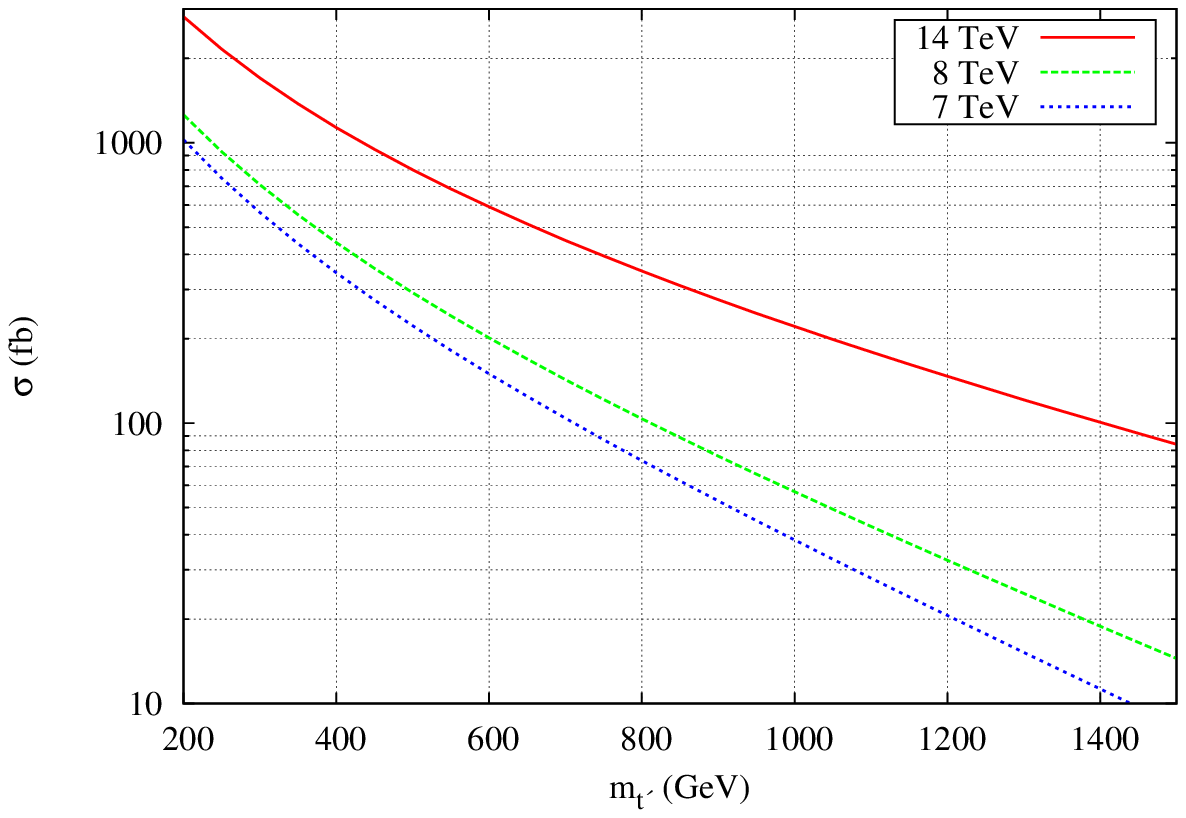,width=0.45\textwidth}
\epsfig{file=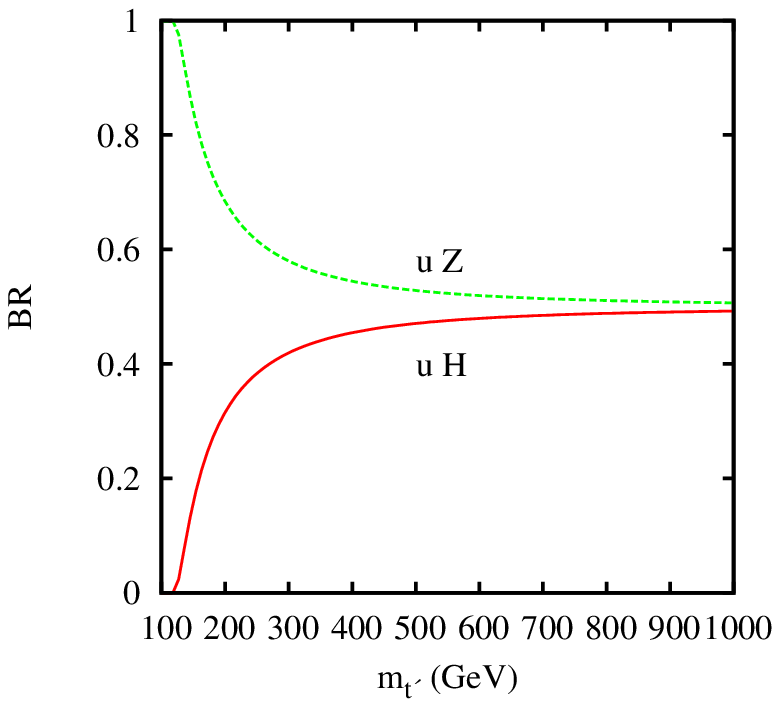,width=0.5\textwidth}
\caption{\sl Single production cross-section (left) and branching ratios (right) of $t'$ as a function of $m_{t'}$ for
$\sin\theta^R = 0$. The remaining parameters are fixed at the benchmark values in Table~\ref{tab:benchmark}.  } 
\label{fig:br_tp_str0} 
}

The single production of $t'$ in association with light jet has also
been considered in \cite{Atre:2011ae} where they considered $t'$ decays
to be:
$$
t' \to W/Z + j
$$
with j=u,d and argued that this is the most general scenario. 
Here we showed that these modes are quite suppressed compared to decays into the third generation, due to the tight bounds from precision flavour measurements that limit the size of the couplings to light quarks.

\FIGURE[t]{
\epsfig{file=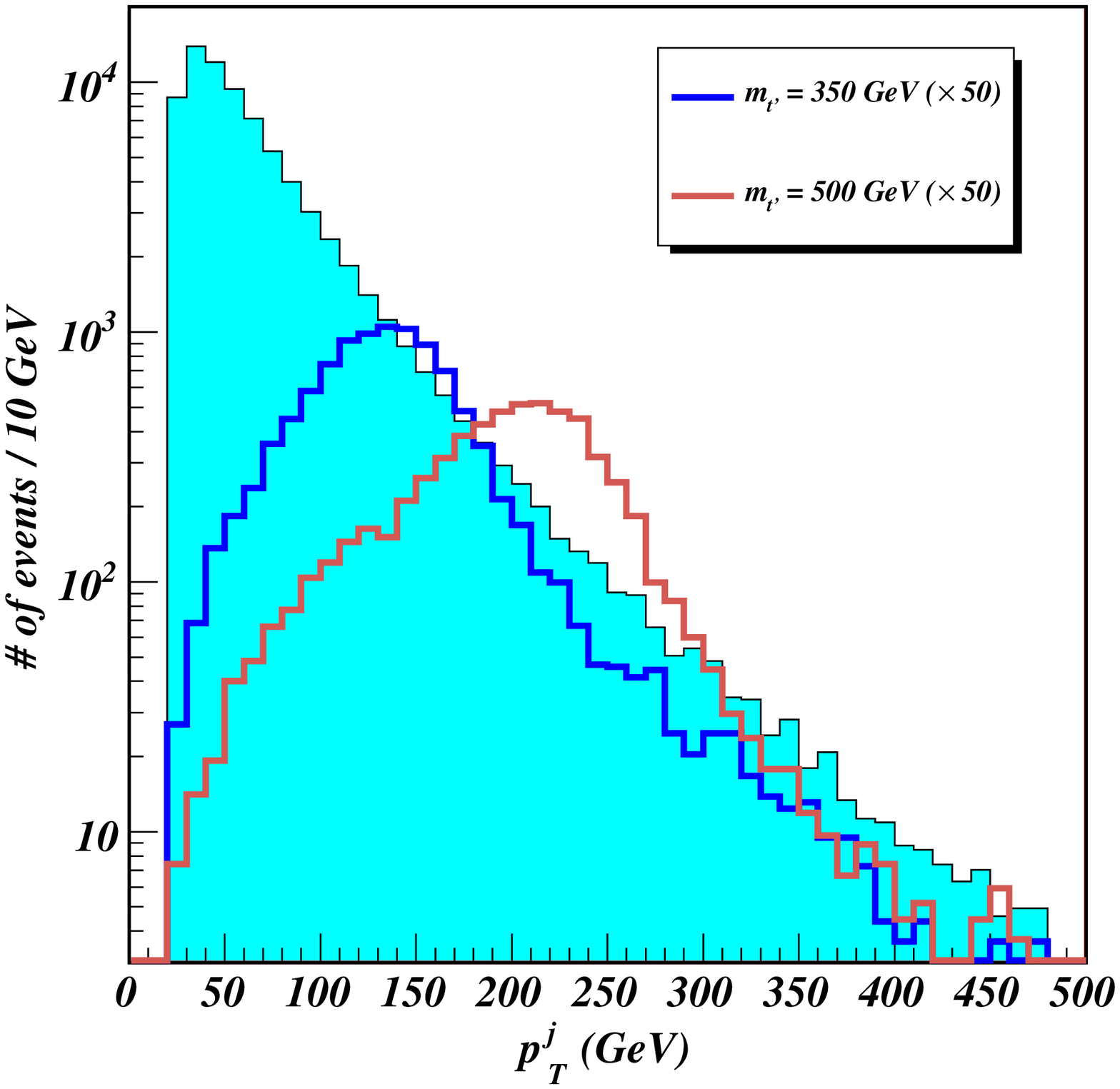,width=0.47\textwidth}
\caption{\sl \underline{\jjbb} : $p_T$ of the higest $p_T$ light jet
at the LHC with centre of mass energy of 7 TeV and lumionsity of $10 fb^{-1}$. }   
\label{fig:jet_str0} 
}

The situation is different only for $\sin \theta^R = 0$:  in this case, the heavy top $t'$ decouples from the top, therefore it can only decay 
into light quarks.
An important difference is the presence of the $h j$ channel, that has not been considered in  \cite{Atre:2011ae}.
In our specific case, namely a non-standard doublet, the decays mediated by the $W$ to light quarks are suppressed by the mass of 
the quark, therefore they are negligible.
Furthermore, as we can see from Figure~\ref{fig:br_tp_str0}, the mode $h j$ could be comparable to $Zj$ especially for relatively large 
$t'$ mass.
The production cross sections are very similar to the previous case under consideration, as it can be seen from 
Figure~\ref{fig:br_tp_str0}.

In the rest of this section, we will focus on the decay mode
$$
t' \to h + j\,.
$$
Considering the dominant
decay of $h \to b \bar{b}$, we will get a signature of $p p \to t' j
\to j j b \bar{b}$. 
To study the $b \bar{b} j j$ final state, we impose the following {\sl
pre-selection} cuts: 
\begin{itemize}
\item{} Two light jets with $p_T > 20$ GeV and $|\eta| <
5$.
\item{} Two b-tagged jets with $p_T > 20$ GeV and $|\eta| <
3$. 
\end{itemize}

\FIGURE[t]{
\epsfig{file=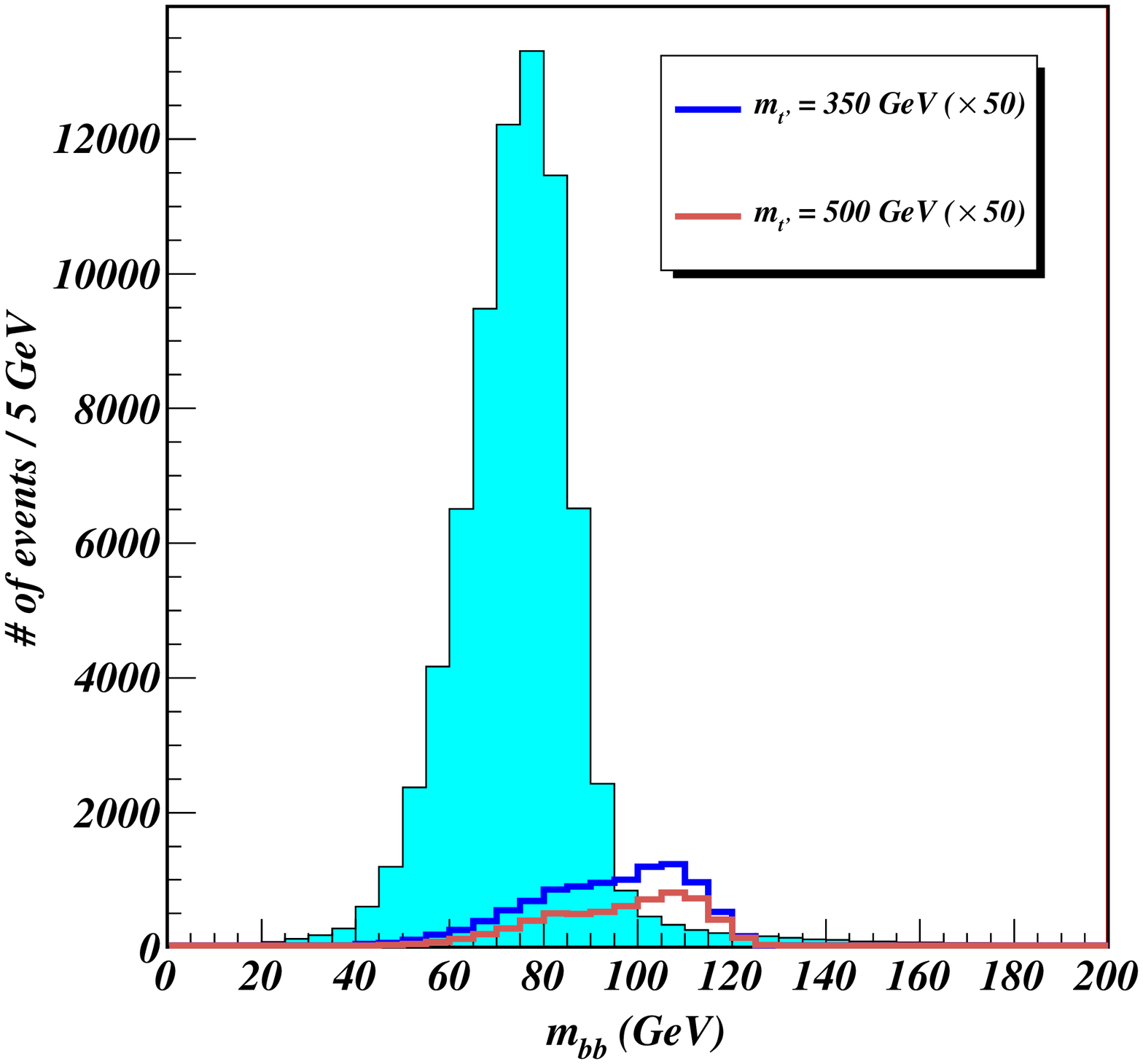,width=0.47\textwidth}
\epsfig{file=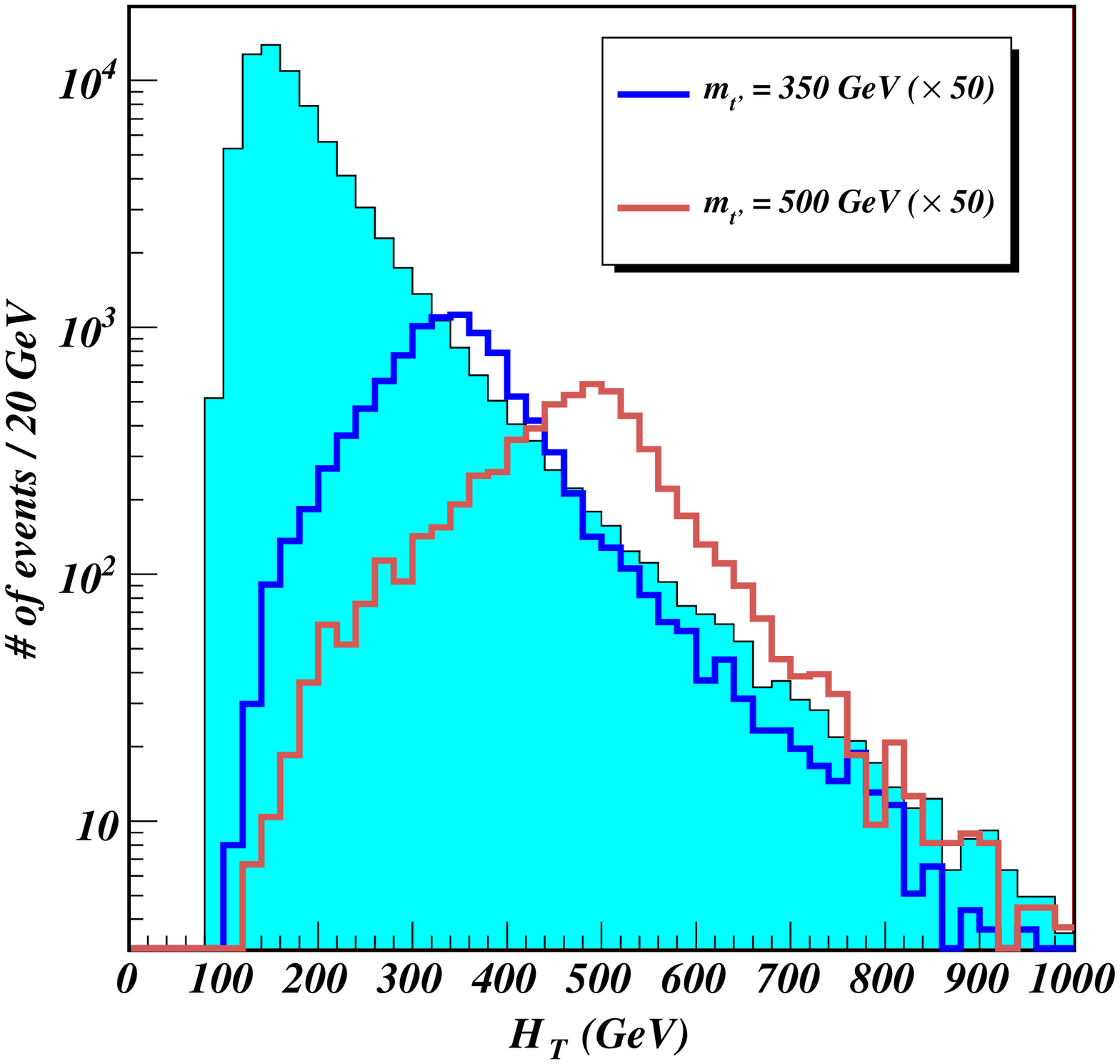,width=0.47\textwidth}
\caption{\sl \underline{\jjbb} : $m_{b \bar{b}}$ and $H_T$ distributions at the LHC with centre of mass energy of 7 TeV and lumionsity of $10 fb^{-1}$.  }  
\label{fig:bjet_str0} 
}

For this signature, we considered the background coming from $Z$ and $h$ + jets, with the bosons decaying into a $b \bar{b}$ pair, 
that has been generated with ALPGEN.
In Figure \ref{fig:jet_str0} we plotted the $p_T$ distribution of the
highest $p_T$ jet for signal and backgrounds. As can be seen from this
distribution one can reduce the backgrounds by using a
hard cut on the $p_T$ of the jet. 
In Figure \ref{fig:bjet_str0} we also plotted the invariant mass
$m_{b \bar{b}}$ and total transverse energy $H_T $ distributions, where $H_T$ is defined as the scalar sum of the transverse $p_T$ of the visible particles (jets and b-jets in our case). 
We can therefore improve the significance of this signature by imposing additional cuts:
\begin{itemize}
\item{} {\it Cut A : } $p_T^j > 100$ GeV. Hard cut on the $p_T$ of the
highest $p_T$ jet.
\item{} {\it Cut B : } $|m_{b \bar{b}} - m_h | < 25$ GeV. 
\item{} {\it Cut C : } Total transverse energy $H_T > 300$ GeV. 
\item{} {\it Cut D : } Total transverse energy $H_T > 350$ GeV. 
\end{itemize}

In Table \ref{table:lhcbbjj} we have shown the effects of the above cuts
on signal and backgrounds. 
These results show that the backgrounds can be reduced and a a good significance achieved on the signal.
\TABLE[t]{
\caption{\sl \underline{\jjbb} : No of signal events after the cuts
for LHC with CM energy of 7 TeV and Luminosity of 10 fb$^{-1}$. } 
\begin{tabular}{|c|c|c|c|} \hline 
Cut                & BP 1-A & BP 2-A &  SM       \\ \hline 
Pre-selection      & 205.3  & 124.2  & 74500   \\ 
$S/\sqrt{S+B}$ & 0.75 & 0.5 & - \\ \hline 
Cut A              & 164.5  & 116.7  & 11440   \\ 
$S/\sqrt{S+B}$ & 1.5  & 1.1 & -    \\ \hline
Cut B              & 82.5   & 65.3 &   498.4   \\ 
$S/\sqrt{S+B}$ & 3.4 & 2.8 & -  \\ \hline
Cut C              & 75.2   & 65.3  &   341.7   \\ 
$S/\sqrt{S+B}$ & 3.7 & 3.2 & - \\ \hline 
Cut D              & 52.2   & 60.8  &   219.1   \\ 
$S/\sqrt{S+B}$ & 3.2 & 3.6 & -  \\ \hline 
\end{tabular}
\label{table:lhcbbjj}
}
\section{Conclusions \label{sec:concl} }

We have discussed in detail new vector-like quarks which can mix with
the standard model quarks without conflicting  
with current experimental limits. 
In particular we focused on a non standard doublet that contains a top partner and an exotic quark with electric charge $5/3$.
This case is motivated from models of composite Higgs, where such fermions are the lightest top partners, and from the less severe 
bounds from precision observables on the general parameter space.
In this paper, we discussed in detail the full flavour structure of the model including mixing with the light generations, and we studied 
the bounds on such mixing from flavour observables in the D meson sector and atomic parity violation measurements.
We also studied their suppressed loop contribution to the mixing in the Kaon and B meson sectors, where large effects may appear 
especially in the CP violating observables.
While all other effects are smaller than the theoretical uncertainty on the form factors, we found a potentially large effect on the phase 
of the $B_s$ mixing amounting to a correction up to $\pm 150\%$ of the Standard Model phase.
We also considered the new bounds from the recent data at the LHC with a luminosity of up to 1 fb$^{-1}$: the bounds are lower than 
the nominal ones due to the reduced branching ratio in the final states analysed by the collaborations, therefore masses as low as 
$300$ or $380$ GeV are still allowed.

In the second part of the paper, we considered some LHC signatures of this scenario.
In particular, we focused on single production of the heavy top partner due to the flavour violation couplings of the $Z$ boson: this 
channel is dominated by up-quark initiated processes.
Notwithstanding the suppressed coupling, the cross section is enhanced by the large PDF's of the valence ups, and it can dominate 
over the pair production channel at large masses, about $\sim 400$ GeV.
We performed a detailed simulation of signals and backgrounds for a few benchmark points.
For generic parameters, the preferred decay final states involve $W$, $Z$ and Higgs plus third generation quarks.
The most promising channels contain at least one lepton in the final state from the top or $W$ decays.
The main background for these channels is given by events containing top quarks.
We identified the most promising final state as \bjlllet.
For light $t'$ masses, however, the top backgrounds have very similar kinematic structure as the $t'$ signal, therefore it is very hard to 
disentangle the signal events and an integrated luminosity of 10 fb$^{-1}$ at 7 TeV of centre of mass energy may not be enough for the 
discovery.
The LHC with a higher centre of mass energy and good
luminosity can improve the situation, especially by accessing higher mass cases where the distinction is possible if enough events are 
collected.

Finally, in the case of no mixing with the top quark, the $t'$ decays mostly into a light Higgs plus a jet: this channel would generate the 
final state $b \bar{b} j j$. By imposing cuts on the invariant mass of the b-tagged jet pair and on the total transverse energy, it is 
possible to reduce the backgrounds and discover this channel with luminosities below 10 fb$^{-1}$.

Our exploration showed that the phenomenology of vector-like fermions can be very rich, and many other channels, not considered 
here, deserve further detailed investigation. 
Moreover, a generic model independent analysis of the parameter space, not biased by any specific model, is crucial at this point of the 
LHC as it would allow us to fully exclude this possibility or, in case of excesses in the data, it would offer hints on the kind of new 
physics at the TeV scale.

\section*{Acknowledgements}
NG would like to thank IPNL, Lyon, France and National Central
University (NCU), Chung-Li, Taiwan for the hospitality where part of this work was done. 
Y.O. and L.P. would like to thank W.S. Hou and C.-P. Yuan for useful discussions
at "Focus Workshop on Heavy Quarks at LHC"  held at National Taiwan University. 
The research of Y.O. is supported in part by the Grant-in-Aid for Science Research, Japan Society for the Promotion of
Science (JSPS), No. 20244037 and No. 22244031.  
The work of N.G. is supported by a grants from Department of Science
\& Technology (DST), India under project no. SR/S2/HEP-09/10 and
University Grants Commission (UGC), India under project
No. 38-58/2009(SR) . 

\newpage
\appendix

\section{Appendix: Expansion of the CKM matrix}
\label{app:CKM}
\setcounter{equation}{0}

We give in the following a systematic expansion of the CKM matrix in our
modified framework, including the extra vector-like top partner. The mixing angles in the left-handed sector are small, 
and we can safely assume that the main structure of the CKM matrix is coming from the SM Yukawas, therefore 
$\tilde{V}_{CKM}$ is very close to the measured CKM matrix.
A simple way to estimate the impact of the mixing in the left-handed sector is to use the hierarchy in $\tilde{V}_{CKM}$ and 
the quark masses in terms of the Cabibbo angle $\lambda = \sin \theta_{12} = 0.2252$, where $\theta_{12}$ is the mixing angle between the first and second generation.
It is well known that the SM CKM matrix is hierarchical: we will use here the Wolfenstein parameterisation~\cite{pdg}
\beq
\tilde{V}_{CKM} = \left( \begin{array}{ccc}
1-\lambda^2/2 & \lambda & A \lambda^3 (\rho - i \eta) \\
-\lambda & 1-\lambda^2/2 & A \lambda^2 \\
A \lambda^3(1-\rho-i \eta) & -A \lambda^2 & 1
\end{array} \right) + \mathcal{O} (\lambda^4) \sim \left( \begin{array}{ccc}
\mathcal{O} (1) & \mathcal{O} (\lambda) & \mathcal{O} (\lambda^3) \\
\mathcal{O} (\lambda) & \mathcal{O} (1) & \mathcal{O} (\lambda^2) \\
\mathcal{O} (\lambda^3) & \mathcal{O} (\lambda^2) & \mathcal{O} (1)
\end{array} \right)\,. \nonumber
\eeq
The hierarchy in the quark masses can also be related to the Cabibbo angle, $\frac{m_c}{m_t} \sim \mathcal{O} (\lambda^3)$ and $\frac{m_u}{m_t} \sim \mathcal{O} (\lambda^7)$, therefore one can also determine a hierarchy in the matrix $V_L$:
\beq
{V}_{L} \sim \left( \begin{array}{ccc}
\mathcal{O} (1) & \mathcal{O} (\lambda^4) & \mathcal{O} (\lambda^7) \\
\mathcal{O} (\lambda^4) & \mathcal{O} (1) & \mathcal{O} (\lambda^3) \\
\mathcal{O} (\lambda^7) & \mathcal{O} (\lambda^3) & \mathcal{O} (1)
\end{array} \right)\,.
\eeq
Note that the powers of $\lambda$ only take into account the suppression coming from the quark masses, and that additional suppressions will come from the small mixing with the vector fermions, as discussed in Section~\ref{subsec:CKMmatrix}.
Using the expansion in powers of $\lambda$, we can order the contributions to the elements of $V_{CKM} = V_L^\dagger \cdot V'_{CKM}$. 
Moreover, we will explicitly indicate the phases of the new Yukawas $x_1$ and $x_2$ to make clear where the new phases enter while the SM phase is hidden in the standard CKM elements $V_{qq'}$:
\beq
\begin{array}{l}
V_{CKM}^{ud} = V_{ud}  [\lambda^0] + e^{i (\beta_2-\beta_1)} |V_L^{21}|  V_{cd} [\lambda^5] + 
e^{-i \beta_1} |V_L^{31}|  V_{td} [\lambda^{10}]\,, \\
V_{CKM}^{us} = V_{us}  [\lambda^2] + e^{i (\beta_2-\beta_1)} |V_L^{21}|  V_{cs} [\lambda^4] +  
e^{-i \beta_1} |V_L^{31}| V_{ts} [\lambda^9]\,,  \\
V_{CKM}^{ub} = V_{ub}  [\lambda^3] + e^{i (\beta_2-\beta_1)} |V_L^{21}|  V_{cb} [\lambda^6] + 
e^{-i \beta_1} |V_L^{31}|  V_{tb} [\lambda^7]\,, \\
\\
V_{CKM}^{cd} = V_{cd}  [\lambda^2] - e^{- i (\beta_2-\beta_1)} |V_L^{12}| V_{ud} [\lambda^4] +  
e^{-i \beta_2} |V_L^{32}| V_{td} [\lambda^6]\,,  \\
V_{CKM}^{cs} = V_{cs}  [\lambda^0] + (-e^{- i (\beta_2-\beta_1)} |V_L^{12}| V_{us} + 
e^{-i \beta_2} |V_L^{32}| V_{ts} ) [\lambda^4] \,, \\
V_{CKM}^{cb} = V_{cb}  [\lambda^2] + e^{-i \beta_2} |V_L^{32}| V_{tb} [\lambda^3] -  
e^{- i (\beta_2-\beta_1)} |V_L^{12}| V_{ub} [\lambda^7]\,, \\ 
\\
V_{CKM}^{td} = V_L^{33} V_{td}  [\lambda^3] - e^{i \beta_2} |V_L^{23}| V_{cd} [\lambda^4] - 
e^{i \beta_1} |V_L^{13}| V_{ud} [\lambda^7] \,, \\
V_{CKM}^{ts} = V_L^{33} V_{ts}  [\lambda^2] -  e^{i \beta_2} |V_L^{23}| V_{cs} [\lambda^3] - 
e^{i \beta_1} |V_L^{13}| V_{us} [\lambda^8] \,, \\
V_{CKM}^{tb} = V_L^{33} V_{tb}  [\lambda^0] - e^{i \beta_2} |V_L^{23}| V_{cb} [\lambda^5] - 
e^{i \beta_1} |V_L^{13}| V_{td} [\lambda^{10}]\,;
\end{array}
\eeq
in square brackets we give the power counting in $\lambda$ for each term.
Analogously, one can calculate the couplings of $t'$ to the $W$ (taking into account that 
$V_L^{14} \sim \mathcal{O} (\lambda^7)$, $V_L^{24} \sim \mathcal{O} (\lambda^3)$ and 
$V_L^{34} \sim \mathcal{O} (1)$): 
\beq
\begin{array}{l}
V_{CKM}^{t'd} =  V_L^{34} V_{td}  [\lambda^3] + e^{i \beta_2} |V_L^{24}| V_{cd} [\lambda^4] +
 e^{i \beta_1} |V_L^{14}| V_{ud} [\lambda^7] \,, \\
V_{CKM}^{t's} =  V_L^{34} V_{ts}  [\lambda^2] + e^{i \beta_2} |V_L^{24}| V_{cs} [\lambda^3] + 
e^{i \beta_1} |V_L^{14}|  V_{us} [\lambda^8] \,, \\
V_{CKM}^{t'b} =  V_L^{34} V_{tb}  [\lambda^0] + e^{i \beta_2} |V_L^{24}| V_{cb} [\lambda^5] + 
e^{i \beta_1} |V_L^{14}|  V_{td} [\lambda^{10}]\,.
\end{array}
\eeq

\section{Appendix: Notation for the meson mixing}
\label{app:mesonmixing}

The $K^{0} - \bar{K}^{0}$ mixing can be calculated in our model with an extra non-standard doublet by a simple generalisation 
of the usual formulas, by including the effect of the $t'$ quark as shown in Figure~\ref{k0k0barmixing}.
The effective Hamiltonian describing $K^{0} - \bar{K}^{0}$ mixing is a simple extension of the standard one and it is given by
\begin{equation}
{\cal H}_{eff} = \frac{G_{F}}{\sqrt{2}}\frac{\alpha}{4\pi\sin^{2}\theta_{W}} \sum_{i,j=c,t,t^{\prime}} \eta_{ij}\xi^{i}\xi^{j} E(x_{i},x_{j}) 
[\bar{s}\gamma_{\mu}P_{L}d] [\bar{s}\gamma^{\mu}P_{L}d] \,,
\end{equation}
with
\begin{eqnarray}
E(x_{i},x_{j}) &=& x_{i}x_{j} \Bigl[  -\frac{3}{4(1-x_{i})(1-x_{j})} + \frac{\log x_{i}}{(x_{i}-x_{j})(1-x_{i})^{2}} 
\left( 1 - 2x_{i} + \frac{x_{i}^{2}}{4} \right) \nonumber \\
&&+ \frac{\log x_{j}}{(x_{j}-x_{i})(1-x_{j})^{2}} \left( 1 - 2x_{j} + \frac{x_{j}^{2}}{4} \right) \Bigr]
\end{eqnarray}
for $i \neq j$. For $i = j$ the Inami-Lim function $E(x_{i},x_{j})$ becomes
\begin{equation}
E(x_{i},x_{i}) = \frac{x_{i}}{(1-x_{i})^{2}} \left[ 1 - \frac{11x_{i}}{4} + \frac{x_{i}^{2}}{4} - \frac{3x_{i}^{2}\log x_{i}}{2(1-x_{i})} \right]
\end{equation}
where $\eta_{ij}$ are QCD corrections to the Inami-Lim functions, $\xi^{i}=V^{*,is}_{CKM}V^{id}_{CKM}$ and $x_{i} = m_{i}^{2}/m_{W}^{2}$;  $V_{CKM}^{\dagger} V_{CKM}$ leads 
to the quadrangle condition $\xi^{u} + \xi^{c} + \xi^{t} + \xi^{t^{\prime}}=0$.
We checked these formulas by calculating the loop box diagrams in Feynman-'t Hooft gauge: in the mass eigenstate 
basis, the couplings of the $W$ bosons are given in Eq.~(\ref{eq:wcoupling}).

The hadronic matrix element entering the $\bar{K}^0$--$K^0$ mixing is
\begin{equation}
\bra{K^{0}} [\bar{s}\gamma_{\mu}P_{L}d] [\bar{s}\gamma^{\mu}P_{L}d] \ket{\bar{K}^{0}}
= \frac{2}{3}f_{K}^{2}B_{K}m_{K}^{2}
\end{equation}
where $f_{K}\simeq160$ MeV is the Kaon decay constant, its mass is $m_{K}=497.614 \pm 0.024$ MeV and $B_{K}=0.725\pm0.026$~\cite{Laiho:2009eu} is the bag parameter.
In the non-SM doublet model, the mixing matrix element $M_{12}$ becomes
\begin{eqnarray}
M_{12} = \frac{1}{2m_{K}} \bra{K^{0}} {\cal H}_{eff} \ket{\bar{K}^{0}} = \frac{1}{3} f_{K}^{2}B_{K}m_{K}
\frac{G_{F}}{\sqrt{2}}\frac{\alpha}{4\pi\sin^{2}\theta_{W}} \sum_{i,j=c,t,t^{\prime}} \eta_{ij}\xi^{i}\xi^{j} E(x_{i},x_{j}) \,.
\end{eqnarray}

Here we are interested in two CP violating quantities: the mass difference $\Delta m_{K}$ between the two mass eigenstates $K_{L/S}$, and the direct CP violation parameter $\epsilon_K$, defined in terms of the matrix element as
\begin{eqnarray}
\Delta m_{K} &\equiv& m_{K_{L}} - m_{K_{S}} = 2|M_{12}| \simeq 2 Re M_{12}\,, \\
\epsilon_{K} &\simeq& \frac{e^{i\pi/4}}{\sqrt{2}\Delta m_{K}} Im M_{12}\,.
\end{eqnarray}
With the non-SM doublet we consider, $\Delta m_{K}$ is given by
\begin{eqnarray}
\Delta m_{K} =
\frac{G_{F}}{\sqrt{2}}\frac{\alpha f_{K}^{2}m_{K}}{6\pi\sin^{2}\theta_{W}} \sum_{i,j=c,t,t^{\prime}} \left| \eta_{ij}\xi^{i}\xi^{j} 
E(x_{i},x_{j}) \right|\,;
\end{eqnarray}
the $\xi^{i}$ can be expressed as (where we show the ordering in powers of $\lambda$)
\begin{eqnarray}
\xi^{c} &\sim& V_{cs}^{*}V_{cd}[\lambda^{2}] - e^{-i(\beta_{2}-\beta_{1})} \left| V_{L}^{12} \right| V_{cs}^{*}V_{ud} [\lambda^{4}] \,,
\nonumber \\
\xi^{t} &\sim& \left| V_{L}^{33} \right|^{2} V_{ts}^{*}V_{td}[\lambda^{5}] - V_{L}^{33} \left| V_{L}^{23} \right| \left( e^{i
\beta_{2}}V_{ts}^{*}V_{cd} + e^{-i\beta_{2}}V_{cs}^{*}V_{td} \right) [\lambda^{6}] + \left| V_{L}^{23} \right|^{2} V_{cs}^{*}V_{cd} 
[\lambda^{7}]\,, \nonumber \\
\xi^{t^{\prime}} &\sim& \left| V_{L}^{34} \right|^{2} V_{ts}^{*}V_{td}[\lambda^{5}] + e^{-i\beta_{2}}V_{L}^{34} \left| V_{L}^{24} \right| 
V_{cs}^{*}V_{td} [\lambda^{6}] + e^{i\beta_{2}} V_{L}^{34} \left| V_{L}^{24} \right| V_{ts}^{*}V_{cd} [\lambda^{6}]\,. \nonumber
\end{eqnarray}

The formulas in this section can be easily generalised to the case of $B_s$ and $B_d$ mixing: it is enough to replace the $s$ quark with a $b$ and the $d$ quark with a $s$ or $d$ respectively.
The relevant CKM entries in the two cases will be
\begin{equation}
\xi_{s}^i=V^{*,ib}_{CKM}V^{is}_{CKM} \qquad \mbox{and} \qquad \xi_{d}^i=V^{*,ib}_{CKM}V^{id}_{CKM}
\end{equation}
with $i=u,c,t,t'$.


\end{document}